\pdfoutput=1
\documentclass[12pt]{article}

\advance \topmargin by -\headheight
\advance \topmargin by -\headsep

\evensidemargin \oddsidemargin
\input{epsf}

\usepackage{graphicx}
\usepackage{comment}
\usepackage{amsmath,amsfonts,amsthm,amssymb}
\usepackage{subfigure}
\usepackage{setspace}
\usepackage{Tabbing}
\usepackage{lastpage}
\usepackage{extramarks}
\usepackage{chngpage}
\usepackage[usenames,dvipsnames]{color}
\usepackage{graphicx,float,wrapfig}
\usepackage{lineno}
\usepackage{xcolor}

\begin{document}

\topmargin -.6in

\def\rh{{\hat \rho}}
\def\alie{{\hat{\cal G}}}
\newcommand{\sect}[1]{\setcounter{equation}{0}\section{#1}}
\renewcommand{\theequation}{\thesection.\arabic{equation}}

 %%      MACROS.TEX
%               macros formatting and equations
\def\rf#1{(\ref{eq:#1})}
\def\lab#1{\label{eq:#1}}
\def\nonu{\nonumber}
\def\br{\begin{eqnarray}}
\def\er{\end{eqnarray}}
\def\be{\begin{equation}}
\def\ee{\end{equation}}
\def\eq{\!\!\!\! &=& \!\!\!\! }
\def\foot#1{\footnotemark\footnotetext{#1}}
\def\lb{\lbrack}
\def\rb{\rbrack}
\def\llangle{\left\langle}
\def\rrangle{\right\rangle}
\def\blangle{\Bigl\langle}
\def\brangle{\Bigr\rangle}
\def\llbrack{\left\lbrack}
\def\rrbrack{\right\rbrack}
\def\lcurl{\left\{}
\def\rcurl{\right\}}
\def\({\left(}
\def\){\right)}
\newcommand{\nit}{\noindent}
\newcommand{\ct}[1]{\cite{#1}}
\newcommand{\bi}[1]{\bibitem{#1}}
\def\lskip{\vskip\baselineskip\vskip-\parskip\noindent}
\relax

%                     common physics symbols
\def\tr{\mathop{\rm tr}}
\def\Tr{\mathop{\rm Tr}}
\def\v{\vert}
\def\bv{\bigm\vert}
\def\Bgv{\;\Bigg\vert}
\def\bgv{\bigg\vert}
\newcommand\partder[2]{{{\partial {#1}}\over{\partial {#2}}}}
\newcommand\funcder[2]{{{\delta {#1}}\over{\delta {#2}}}}
\newcommand\Bil[2]{\Bigl\langle {#1} \Bigg\vert {#2} \Bigr\rangle}  %% <.|.>
\newcommand\bil[2]{\left\langle {#1} \bigg\vert {#2} \right\rangle} %% <.|.>
\newcommand\me[2]{\left\langle {#1}\bv {#2} \right\rangle} %% <.|.>
\newcommand\sbr[2]{\left\lbrack\,{#1}\, ,\,{#2}\,\right\rbrack}
\newcommand\pbr[2]{\{\,{#1}\, ,\,{#2}\,\}}
\newcommand\pbbr[2]{\lcurl\,{#1}\, ,\,{#2}\,\rcurl}
%
%                    math symbols
\def\a{\alpha}
\def\at{{\tilde A}^R}
\def\atc{{\tilde {\cal A}}^R}
\def\atcm#1{{\tilde {\cal A}}^{(R,#1)}}
\def\b{\beta}
\def\dc{{\cal D}}
\def\d{\delta}
\def\D{\Delta}
\def\eps{\epsilon}
\def\ve{\varepsilon}
\def\g{\gamma}
\def\G{\Gamma}
\def\grad{\nabla}
\def\h{{1\over 2}}
\def\l{\lambda}
\def\L{\Lambda}
\def\m{\mu}
\def\n{\nu}
\def\o{\over}
\def\om{\omega}
\def\O{\Omega}
\def\p{\phi}
\def\P{\Phi}
\def\pa{\partial}
\def\pr{\prime}
\def\pt{{\tilde \Phi}}
\def\qs{Q_{\bf s}}
\def\ra{\rightarrow}
\def\s{\sigma}
\def\S{\Sigma}
\def\t{\tau}
\def\th{\theta}
\def\Th{\Theta}
\def\tpp{\Theta_{+}}
\def\tmm{\Theta_{-}}
\def\tpg{\Theta_{+}^{>}}
\def\tms{\Theta_{-}^{<}}
\def\tp0{\Theta_{+}^{(0)}}
\def\tm0{\Theta_{-}^{(0)}}
\def\ti{\tilde}
\def\wti{\widetilde}
\def\jc{J^C}
\def\bj{{\bar J}}
\def\sj{{\jmath}}
\def\bsj{{\bar \jmath}}
\def\bp{{\bar \p}}
\def\vp{\varphi}
\def\vt{{\tilde \varphi}}
\def\faa{Fa\'a di Bruno~}
\def\ca{{\cal A}}
\def\cb{{\cal B}}
\def\ce{{\cal E}}
\def\cg{{\cal G}}
\def\cgh{{\hat {\cal G}}}
\def\ch{{\cal H}}
\def\chh{{\hat {\cal H}}}
\def\cl{{\cal L}}
\def\cm{{\cal M}}
\def\cn{{\cal N}}
\newcommand\sumi[1]{\sum_{#1}^{\infty}}   %% summation till infinity
\newcommand\fourmat[4]{\left(\begin{array}{cc}  %%   2x2 matrix
{#1} & {#2} \\ {#3} & {#4} \end{array} \right)}

%
%%%                    macros for Lie algebras
\def\lie{{\cal G}}
\def\kmlie{{\hat{\cal G}}}
\def\dlie{{\cal G}^{\ast}}
\def\elie{{\widetilde \lie}}
\def\edlie{{\elie}^{\ast}}
\def\hlie{{\cal H}}
\def\flie{{\cal F}}
\def\wlie{{\widetilde \lie}}
\def\f#1#2#3 {f^{#1#2}_{#3}}
\def\winf{{\sf w_\infty}}
\def\win1{{\sf w_{1+\infty}}}
\def\hwinf{{\sf {\hat w}_{\infty}}}
\def\Winf{{\sf W_\infty}}
\def\Win1{{\sf W_{1+\infty}}}
\def\hWinf{{\sf {\hat W}_{\infty}}}
\def\Rm#1#2{r(\vec{#1},\vec{#2})}          % r-operator kernel
\def\OR#1{{\cal O}(R_{#1})}           % R-coadjoint orbit
\def\ORti{{\cal O}({\widetilde R})}           % R-tilde-coadjoint orbit
\def\AdR#1{Ad_{R_{#1}}}              % R-adjoint group action
\def\dAdR#1{Ad_{R_{#1}^{\ast}}}      % R-coadjoint group action
\def\adR#1{ad_{R_{#1}^{\ast}}}       % R-coadjoint algebra action
\def\KP{${\rm \, KP\,}$}                 %% KP
\def\KPl{${\rm \,KP}_{\ell}\,$}         %% KP_l
\def\KPo{${\rm \,KP}_{\ell = 0}\,$}         %% KP_l=0
\def\mKPa{${\rm \,KP}_{\ell = 1}\,$}    %% modified KP-1
\def\mKPb{${\rm \,KP}_{\ell = 2}\,$}    %% modified KP-2
%
%       fake blackboard bold macros for reals, complex, etc.
\def\rlx{\relax\leavevmode}
\def\inbar{\vrule height1.5ex width.4pt depth0pt}
\def\IZ{\rlx\hbox{\sf Z\kern-.4em Z}}
\def\IR{\rlx\hbox{\rm I\kern-.18em R}}
\def\IC{\rlx\hbox{\,$\inbar\kern-.3em{\rm C}$}}
\def\IN{\rlx\hbox{\rm I\kern-.18em N}}
\def\IO{\rlx\hbox{\,$\inbar\kern-.3em{\rm O}$}}
\def\IP{\rlx\hbox{\rm I\kern-.18em P}}
\def\IQ{\rlx\hbox{\,$\inbar\kern-.3em{\rm Q}$}}
\def\IF{\rlx\hbox{\rm I\kern-.18em F}}
\def\IG{\rlx\hbox{\,$\inbar\kern-.3em{\rm G}$}}
\def\IH{\rlx\hbox{\rm I\kern-.18em H}}
\def\II{\rlx\hbox{\rm I\kern-.18em I}}
\def\IK{\rlx\hbox{\rm I\kern-.18em K}}
\def\IL{\rlx\hbox{\rm I\kern-.18em L}}
\def\one{\hbox{{1}\kern-.25em\hbox{l}}}
\def\0#1{\relax\ifmmode\mathaccent"7017{#1}%
B        \else\accent23#1\relax\fi}
\def\omz{\0 \omega}
%
%               \ltimes=\semiproduct
\def\ltimes{\mathrel{\vrule height1ex}\joinrel\mathrel\times}
\def\rtimes{\mathrel\times\joinrel\mathrel{\vrule height1ex}}
%
%               This defines remark, proposition etc.
\def\mark{\noindent{\bf Remark.}\quad}
\def\prop{\noindent{\bf Proposition.}\quad}
\def\theor{\noindent{\bf Theorem.}\quad}
\def\name{\noindent{\bf Definition.}\quad}
\def\exam{\noindent{\bf Example.}\quad}
\def\proof{\noindent{\bf Proof.}\quad}

\begin{titlepage}
\vspace*{-1cm}

\vskip 2cm

\vspace{.2in}
\begin{center}
{\large\bf Self-dual sectors for scalar field theories in $(1+1)$ dimensions }
\end{center}

\vspace{.5cm}

\begin{center}
L. A. Ferreira~$^{\star}$,
P. Klimas~$^{\nabla}$
 and Wojtek J. Zakrzewski~$^{\dagger}$

\small
\par \vskip .2in \noindent
$^{(\star)}$Instituto de F\'\i sica de S\~ao Carlos; IFSC/USP;\\
Universidade de S\~ao Paulo  \\ 
Caixa Postal 369, CEP 13560-970, S\~ao Carlos-SP, Brazil\\
email: laf@ifsc.usp.br
\small

 \par \vskip .2in \noindent
$^{(\nabla)}$Universidade Federal de Santa Catarina,\\
 Trindade, CEP 88040-900, Florian\'opolis-SC, Brazil\\
email: pawel.klimas@ufsc.br

\small 
\par \vskip .2in \noindent
$^{(\dagger)}$~Department of Mathematical Sciences,\\
 University of Durham, Durham DH1 3LE, U.K.\\
email: W.J.Zakrzewski@durham.ac.uk

\normalsize
\end{center}
%%%%%%%%%%%%%%%%%%%%%%%%%%%%%%%%%%%%%%%%%%%%%

\vspace{.5in}

\begin{abstract}

We use ideas of generalized self-duality conditions to construct real scalar field theories in $(1+1)$-dimensions with exact self dual sectors. The approach is based on a pre-potential $U$ that defines the topological charge and the potential energy of these theories. In our algebraic method to construct the required pre-potentials we use the representation theory of Lie groups. This approach leads naturally to an infinite set of degenerate vacua and so to topologically non-trivial self-dual solutions of  these models. We present explicit examples for the groups $SU(2)$, $SU(3)$ and $SO(5)$  and discuss some properties of these solutions.

\end{abstract} 
\end{titlepage}

%%%%%%%%%%%%%%%%%%%%%%%%%%%%%%%%%%%%%%%%%%%%%

\section{Introduction}
\label{sec:intro}
\setcounter{equation}{0}

Topological solitons are of great importance in many areas of  science  as they constitute non-trivial configurations of the degrees of freedom of the system, stabilized by topology. Among them there are instantons, magnetic monopoles and vortices in gauge theories, Skyrmions, baby-Skyrmions and kinks in scalar field theories \cite{mantonbook,shnirbook,wojtekbook}, and many other types of solitons with applications which range from high energy  and  condensed matter physics to fluid dynamics. The  spectrum of solutions, in a given theory, is split into disjoint classes characterized by their topological properties, which in many cases is labelled by the value of the topological charge. The topology introduces selection rules preventing solutions from one class to evolve, under the dynamics of the system, into another one.  Inside a given class solutions with the smallest possible value of energy (or Euclidean action) play the most prominent role since they are very stable under perturbations, as they cannot decay. 

In some special theories these solutions have  further interesting properties. They satisfy simpler differential equations, usually of first order in derivatives, that imply the full equations of motion of the system, which are usually of second order. In addition, these solutions saturate a lower bound on the energy (or Euclidean action) determined  the topological charge. Such solutions are called self-dual  or BPS, an abbreviation for the concept introduced by Bogomolny, Prasad and Sommerfield \cite{bpsoriginal} in the context of the theory of magnetic monopoles.  The fact that one can construct solutions of the system by performing one fewer integration is not related to the use of dynamically conserved quantities. The self-dual or BPS solutions arise in theories in which the topological charge has an integral representation and so it has a topological charge density. The fact that the topological charge is invariant under smooth deformations of the field configurations implies that its density satisfies identities that have the form of differential equations which, when combined with the (first-order) self-dual equations, imply the full equations of motion.

This intriguing interplay between topology and dynamics has been explored in \cite{bps} to construct, in a systematic way, theories with self-dual ({\it i.e.} BPS) sectors. The method put forward in \cite{bps} starts from a given topological charge with an integral representation, and involves splitting the density of that charge into the sum of products of pairs of some quantities, chosen in a convenient way. Different choices of the splitting lead to 
different theories associated to the same topological charge.  The self-duality equations are given by the equality (up to a sign) of these quantities in each pair. The static energy density (or Euclidean action density) is defined as the sum of squares of these quantities, in each pair. The self-duality equations together with the identities satisfied by the density of topological charge imply the Euler-Lagrange equations that follow from the static
 energy functionals (or Euclidean action) of such theories. The lower bound on such functionals, determined by the value of the topological charge, follows as a byproduct of the construction. In section \ref{sec:bps} we give more details of this method. 

Incidentally, let us point out that the scalar field theories possessing a self-dual sector that were constructed in \cite{bps}, see also \cite{santamaria}, were constructed in such a way that the number of real scalar fields had to be equal to the number of dimensions of the space in which each theory was defined.  Among the theories covered by such a method there were the sine-Gordon model in $(1+1)$-dimensions, the Belavin-Polyakov \cite{bpmodel} and baby-Skyrmion \cite{babyskyrme} models in $(2+1)$ models, various modifications of the $SU(2)$ Skyrme model in $(3+1)$-dimensions \cite{bpsskyrme,wojteklaf,shnirlaf,laf}, as well as generalizations of the Skyrme model to higher dimensions and higher target spaces \cite{yuki}.

The purpose of this paper is to extend the ideas of \cite{bps} to construct scalar field theories in $(1+1)$-dimensions  possessing an exact self-dual sector, and having more than one real scalar field. 
{Theories with two real scalar fields, and possessing a self-dual sector, have already been  constructed in \cite{bazeia} using a different approach and not considering  periodic potentials with infinitely degenerate vacua.} 
The basic ingredient in the construction is the pre-potencial $U\(\vp\)$, a functional of the real scalar fields of the theory $\vp_a$, $a=1,2,\dots r$, but not of their derivatives. This pre-potential allows us to define the topological charge as 
\be
Q=\int_{-\infty}^{\infty}dx\, \frac{d\, U}{d\,x}= \int_{-\infty}^{\infty}dx\,\frac{\delta\, U}{\delta\,\vp_a}\,\frac{d\,\vp_a}{d\,x}=
U\(\vp_a(x=\infty)\)-U\(\vp_a(x=-\infty)\).
\lab{topcharge}
\ee

The  action of the theories that we consider here have the form ($\mu=0,1$)
\be
S= \int d^2x\left[ \frac{1}{2}\,\eta_{ab}\,\partial_{\mu}\vp_a\,\partial^{\mu}\vp_b- V\(\vp\)\right],
\lab{actionintro}
\ee
where the potential is constructed from the pre-potential $U\(\vp\)$ as
\be
 V=\frac{1}{2}\,\eta^{-1}_{ab}\,\frac{\delta\, U}{\delta\,\vp_a}\,\frac{\delta\, U}{\delta\,\vp_b}
 \lab{potdef}
 \ee
and where $\eta_{ab}$ is a symmetric invertible matrix that can be quite general in its character as we discuss in section \ref{sec:bps}.  However, for the purposes of this paper we take it to be a  constant matrix with positive eigenvalues to keep the energy positive definite. The self-duality ({\it i.e.}BPS) equations are given by 
\be
 \eta_{ab}\,\frac{d\,\vp_b}{d\,x}=\pm \,  \frac{\delta\, U}{\delta\,\vp_a}.
 \lab{bpseqintro}
 \ee
Solutions of \rf{bpseqintro} are static solutions of the Euler-Lagrage equations that follow from \rf{actionintro}, and they saturate the bound $E\geq \mid Q\mid$, for the static energy $E$ of the theory \rf{actionintro}.  

Given the construction above there are basically two approaches to it. One can take a theory of the type \rf{actionintro}, with a given potential $V$ and matrix $\eta_{ab}$, and try to solve \rf{potdef} to find the corresponding pre-potential $U$ that leads to self-duality.  In general, that is not an easy task since the equation \rf{potdef} for the unknown functional $U$ is a non-linear equation and even the question of the existence of solutions might be non-trivial. 

 In this paper we have adopted the opposite approach; {\it i.e.} of constructing  pre-potentials $U$, and matrices $\eta_{ab}$, that lead to physically interesting theories of the type \rf{actionintro}. Thus, instead of taking \rf{potdef} as an equation to solve, we take it as the definition of the potential $V$. We are interested in theories with a definite positive energy, and so we take the matrix $\eta_{ab}$ to have only real and positive eigenvalues, and take the scalars fields $\vp_a$ to be real. In addition, we want the solutions to be of finite energy and so, as we show in section \ref{sec:prepotential} such solutions have to approach extrema of the pre-potential $U$ at spatial infinity, {\it i.e.} for $x\rightarrow \pm \infty$. Moreover, for the topological charge \rf{topcharge} to be non-trivial, the extrema of $U$ (vacua)  have to be as numerous as possible.   In order to achieve this we have decided 
to adopt the method of construction of pre-potentials based on representation theory of Lie groups, as explained in section  \ref{sec:prepotential}. Our approach leads to  infinite classes of scalar field theories with very interesting physical properties that may have applications is many areas of non-linear phenomena. We give some examples of such theories in section \ref{sec:examples}, where we specify our discussion to some representations of the Lie groups $SU(2)$, $SU(3)$ and $SO(5)$. 

The solutions of the self-duality equations \rf{bpseqintro} have a very nice geometrical interpretation as explained in section \ref{sec:geometry}. One can think of the space variable $x$ as being ''time'', and the fields $\vp_a$ the coordinates of a particle moving in the target space. For finite energy solutions, the trajectories of such a particle go from one given extremum of $U$ at the infinite past ($x\rightarrow -\infty$) to another one at the infinite future ($x\rightarrow -\infty$). In addition, for positive definite matrices $\eta_{ab}$, we show that, along a given path,  that is a solution of \rf{bpseqintro}, the pre-potential $U$ is a monotonic function of $x$, growing with $x$, for the the choice of the positive sign in  \rf{bpseqintro}, and decreasing with $x$ for the negative sign. According to \rf{bpseqintro} the velocity of the particle $\({\vec v}\)_a =\frac{d\,\vp_a}{d\,x}$, tangent to the trajectory, is parallel or anti-parallel to the gradient $\({\vec \nabla}_{\eta}U\)_a= \eta^{-1}_{ab}\frac{\delta\, U}{\delta\,\vp_b}$.  Thus, the finite energy solutions of the self-duality equations \rf{bpseqintro} correspond to trajectories in target space linking two extrema of the pre-potential, having ${\vec \nabla}_{\eta}U$ as its tangent vector, and leading to the variation of the pre-potential $U$ monotonically, either upwards or downwards. We show, in section \ref{sec:maxmin}, that all extrema of the pre-potential $U$ are minima of the potential $V$. So, it may be possible to construct finite energy and time dependent solutions from some sort of non-linear superposition of self-dual solutions. 

Despite the attractiveness of the method, the self-duality does not lead, in general, to integrable theories with exact and analytical methods for the construction of their solutions. 
So, most solutions of the equations of self-duality \rf{bpseqintro} have to be constructed numerically. 
The exception is the well-known sine-Gordon model that is integrable and admits a self-dual sector. The  generalizations of the sine-Gordon model, the so-called Affine Toda theories,  are integrable, but it is not certain if they possess a self-dual sector. In addition, for such theories to possess exact soliton solutions their scalar fields have to be taken as complex fields and so the energy is not only non-positive but it is complex. Since we are interested in positive definite energy theories we do not consider the Affine Toda models in this paper. However, there are modifications of the non-abelian version of Affine Toda theories that do possess positive definite energy \cite{miramontes}. It would be interesting to investigate if these theories can fit in our construction. 

Another interesting point to be analyzed is the interaction among the self-dual solutions that we construct numerically. This would involve time dependent simulations of the full equations of motion and it is beyond the scope of the present paper which deals only with the static self-dual solutions. There is vast literature about the interactions of kinks and solitons \cite{mantonbook,shnirbook,kink0,kink1,kink2,kink3,kink4,kink5,kink6,kink7,kink8} and it would be interesting to apply some of the techniques used in these papers to the models constructed in the present paper.  

 We present our numerical construction  of the self-dual solutions in section \ref{sec:numerical}, for the examples discussed in section \ref{sec:examples},  using the fourth order Runge-Kutta method. We have used  numerical simulations not only to check the expected properties of the self-dual solutions, but also to test their stability against small perturbations by letting them evolve under the full time dependent equations of motion.  

In section \ref{sec:conclusions} we present our conclusions and comments on possible extensions of our work. 

%%%%%%%%%%%%%%%%%%%%%%%%%%%%%%%%%%%%%%%%%%%%%
\section{The construction of self-dual sectors}
\label{sec:bps}
\setcounter{equation}{0}
%%%%%%%%%%%%%%%%%%%%%%%%%%%%%%%%%%%%%%%%%%%%%

The construction of self-dual sectors for scalar field theories in $(1+1)$-dimensions that we present in this paper is based on the methods of \cite{bps}, and can be summarized as follows: Suppose one has a topological charge $Q$ with an integral representation such that  its density  can be split into the sum of the products of two quantities as 
\be
Q=\int_{-\infty}^{\infty} dx\, {\cal A}_{\alpha}\,{\tilde {\cal A}}_{\alpha},
\lab{topchargegen}
\ee 
where ${\cal A}_{\alpha}$ and ${\tilde {\cal A}}_{\alpha}$ are functionals of the scalar fields $\vp_a$, $a=1,2,\ldots r$, and of their first space derivatives $\partial_{x}\vp_a$, but not of higher derivatives of these fields. The sub-index $\alpha$ stands for an index or a set of indices. The statement that $Q$ is a topological charge is equivalent to it being invariant under any smooth infinitesimal variation $\delta \vp_a$ of the fields. The fact that $\delta Q=0$ for any $\delta \vp_a$ leads to the following identities, which are second order in space derivatives of the fields,  
\be
\frac{d\;}{d\,x}\(\frac{\delta {\cal A}_{\alpha}}{\delta\, \partial_x\vp_a}\, {\tilde {\cal A}}_{\alpha}\)-\frac{\delta\, {\cal A}_{\alpha}}{\delta\, \vp_a}\, {\tilde {\cal A}}_{\alpha}+\frac{d\;}{d\,x}\(\frac{\delta\,{\tilde {\cal A}}_{\alpha}}{\delta\, \partial_x\vp_a}\, {\cal A}_{\alpha} \)-\frac{\delta\, {\tilde {\cal A}}_{\alpha}}{\delta\, \vp_a}\, {\cal A}_{\alpha}=0.
\lab{identitygen}
\ee

If one now imposes the following first order equations on the fields
\be
 {\cal A}_{\alpha}=\pm\, {\tilde {\cal A}}_{\alpha},
\lab{bpseqgen}
\ee
then it is easy to see that \rf{identitygen} combined with \rf{bpseqgen} imply the following second order equations
\be
\frac{d\;}{d\,x}\(\frac{\delta {\cal A}_{\alpha}}{\delta\, \partial_x\vp_a}\,  {\cal A}_{\alpha}\)-\frac{\delta\, {\cal A}_{\alpha}}{\delta\, \vp_a}\,  {\cal A}_{\alpha}+\frac{d\;}{d\,x}\(\frac{\delta\,{\tilde {\cal A}}_{\alpha}}{\delta\, \partial_x\vp_a}\, {\tilde {\cal A}}_{\alpha} \)-\frac{\delta\, {\tilde {\cal A}}_{\alpha}}{\delta\, \vp_a}\, {\tilde {\cal A}}_{\alpha}=0.
\lab{eleqgen}
\ee
However, \rf{eleqgen} are the Euler-Lagrange equations associated to the following static energy functional
\be
E=\frac{1}{2}\,\int_{-\infty}^{\infty} dx\, \left[ {\cal A}_{\alpha}^2+{\tilde {\cal A}}_{\alpha}^2\right].
\lab{energyfuncgen}
\ee
Thus, this clarifies  why the solutions of the first order self-duality equations \rf{bpseqgen} also solve the second-order Euler-Lagrange \rf{eleqgen} for the theory \rf{energyfuncgen}. The extra integration that would be needed to construct the solutions is provided by the identities \rf{identitygen} which
 follow from the homotopy properties of the topological charge  \rf{topchargegen}. As a by-product of our construction we see that if the static energy functional \rf{energyfuncgen} is positive definite then one obtains a lower bound on $E$, for each homotopy class of solutions, and this bound is saturated by the solutions of the self-duality equations \rf{bpseqgen}. The bound is obtained by rewriting $E$ as
 \be
 E=\frac{1}{2}\,\int_{-\infty}^{\infty} dx\, \left[ {\cal A}_{\alpha}\mp  {\tilde {\cal A}}_{\alpha}\right]^2 \pm \int_{-\infty}^{\infty} dx\, {\cal A}_{\alpha}\,{\tilde {\cal A}}_{\alpha}.
 \ee
For the (self-dual or anti-self-dual) solutions of \rf{bpseqgen} the topological charge can be written as
\be
Q_{\rm BPS}=\pm \int_{-\infty}^{\infty} dx\, {\cal A}_{\alpha}^2 = \pm\,\int_{-\infty}^{\infty} dx\, {\tilde {\cal A}}_{\alpha}^2.
\ee
So, if ${\cal A}_{\alpha}^2$ and ${\tilde {\cal A}}_{\alpha}^2$ are positive definite it follows that
\be
E\geq \mid Q\mid.
\lab{boundgen}
\ee
The bound is saturated for the self-dual solutions, and in such a case the energy becomes
\be
E_{\rm BPS}= \int_{-\infty}^{\infty} dx\,  {\cal A}_{\alpha}^2=\int_{-\infty}^{\infty} dx\, {\tilde {\cal A}}_{\alpha}^2=\mid Q\mid.
\ee
Note that for the self-dual solutions, the sign of the topological charge $Q$ is determined by the choice of sign in the  equations \rf{bpseqgen} with the opposite sign for the anti-self-dual ones. 

For the scalar field theories we consider in this paper, the topological charge is constructed simply from a pre-potential $U$,  as given in \rf{topcharge}. 
Note also that in order to apply the construction of self-dual sectors, explained above, we cannot allow the density of topological charge to depend upon the derivatives of the fields other than the first one. Therefore, the pre-potential $U$ can be a functional of the fields but not of their derivatives. In order to write \rf{topcharge} as in \rf{topchargegen} we take the quantities ${\cal A}_{\alpha}$ and ${\tilde {\cal A}}_{\alpha}$ as 
\be
{\cal A}_{\alpha}\equiv k_{ab}\, \frac{d\,\vp_b}{d\,x}\;;\qquad\qquad\qquad {\tilde {\cal A}}_{\alpha}\equiv \frac{\delta\, U}{\delta\,\vp_b}\,k^{-1}_{ba}\,, 
\lab{choiceofacal}
\ee
where $k_{ab}$ is an arbitrary invertible matrix that can be introduced into the theory due to the freedom one has as to the ways of splitting the density of the topological charge into the sum of products of terms in \rf{topchargegen}.
This matrix can be a constant matrix, depend on the fields $\vp_a$, or can even depend on new (external) fields.  With this choice the self-duality equations \rf{bpseqgen} then become 
\be
 \eta_{ab}\,\frac{d\,\vp_b}{d\,x}=\pm \,  \frac{\delta\, U}{\delta\,\vp_a},
 \lab{bpseq}
 \ee
 where $\eta_{ab}$ is an invertible symmetric matrix given by
 \be
 \eta= k^T\,k.
 \lab{hdef}
 \ee

 Furthermore, the energy functional \rf{energyfuncgen} then becomes
 \be
 E= \int_{-\infty}^{\infty} dx\,\left[ \frac{1}{2}\,\eta_{ab}\, \frac{d\,\vp_a}{d\,x}\, \frac{d\,\vp_b}{d\,x} + V\right],
 \lab{energyfunc}
 \ee
 where  the potential is given by \rf{potdef}. 

Let us now assume that the entries of the matrix $\eta_{ab}$ are functionals of the fields $\vp_a$, their first space derivatives and possibly of some extra independent fields  $\chi_{\beta}$ and their first space derivatives. From the self-duality equations \rf{bpseq} we then have
\br
\frac{d\;}{d\,x}\(\eta_{ab}\,\frac{d\,\vp_b}{d\,x}\)&=&\pm \,  \frac{\delta^2\, U}{\delta \vp_c\,\delta\,\vp_a}\,\frac{d\,\vp_c}{d\,x}= \frac{\delta^2\, U}{\delta \vp_c\,\delta\,\vp_a}\,\eta^{-1}_{cd}\,\frac{\delta\, U}{\delta\,\vp_d}
\nonumber\\
&=&\frac{\delta\;}{\delta\,\vp_a}\left[\frac{1}{2}\,\eta^{-1}_{cd}\,\frac{\delta\, U}{\delta\,\vp_c}\,\frac{\delta\, U}{\delta\,\vp_d}\right]-\frac{1}{2}\,\frac{\delta\, \eta^{-1}_{cd}}{\delta\,\vp_a}\,\frac{\delta\, U}{\delta\,\vp_c}\,\frac{\delta\, U}{\delta\,\vp_d}
\lab{nicerel1}\\
&=&\frac{\delta\, V}{\delta\,\vp_a}
+\frac{1}{2}\,\eta^{-1}_{ce}\, \frac{\delta\, \eta_{ef}}{\delta\,\vp_a}\,
\eta^{-1}_{fd}\,\frac{\delta\, U}{\delta\,\vp_c}\,\frac{\delta\, U}{\delta\,\vp_d}
=\frac{\delta\, V}{\delta\,\vp_a}+\frac{1}{2}\,\frac{\delta\, \eta_{ef}}{\delta\,\vp_a}\,\frac{d\,\vp_e}{d\,x}\,\frac{d\,\vp_f}{d\,x},
\nonumber
\er
where we have used the definition of the potential $V$ given in \rf{potdef}. Again using \rf{bpseq} one finds that
\be
\frac{\delta\, \eta_{ab}}{\delta\,\vartheta}\,\frac{d\,\vp_a}{d\,x}\,\frac{d\,\vp_b}{d\,x}=
\frac{\delta\, \eta_{ab}}{\delta\,\vartheta}\,\eta^{-1}_{ad}\,\frac{\delta\, U}{\delta\,\vp_d}\,\eta^{-1}_{be}\,\frac{\delta\, U}{\delta\,\vp_e}=-\frac{\delta\, \eta^{-1}_{ab}}{\delta\,\vartheta}\,\frac{\delta\, U}{\delta\,\vp_a}\,\frac{\delta\, U}{\delta\,\vp_b},
\lab{nicerel2}
\ee
where $\vartheta$ stands for anything that $\eta_{ab}$ can be a functional of. Thus we see that for any choice of
$\vartheta$ we have
\be
\frac{\delta\, \eta_{ab}}{\delta\,\vartheta}\,\frac{d\,\vp_a}{d\,x}\,\frac{d\,\vp_b}{d\,x}+\frac{\delta\, \eta^{-1}_{ab}}{\delta\,\vartheta}\,\frac{\delta\, U}{\delta\,\vp_a}\,\frac{\delta\, U}{\delta\,\vp_b}\,=\,0.
\lab{nicerel3}
\ee
Inserting \rf{nicerel3} into the first term in (under the derivative with respect to $x$) \rf{nicerel1} we conclude that the self-duality equations \rf{bpseq} alone imply the relation
%Combining \rf{nicerel1} and \rf{nicerel2} we can conclude that the self-duality equations \rf{bpseq} alone imply the relation
\br
\frac{d\;}{d\,x}\left[\eta_{ab}\,\frac{d\,\vp_b}{d\,x}+\frac{1}{2}\,\frac{\delta\, \eta_{cd}}{\delta\,\partial_x\vp_a}\,\frac{d\,\vp_c}{d\,x}\,\frac{d\,\vp_d}{d\,x}
+\frac{1}{2}\,\frac{\delta\, \eta^{-1}_{cd}}{\delta\,\partial_x\vp_a}\,\frac{\delta\, U}{\delta\,\vp_c}\,\frac{\delta\, U}{\delta\,\vp_d}\right]=\frac{\delta\, V}{\delta\,\vp_a}+\frac{1}{2}\,\frac{\delta\, \eta_{cd}}{\delta\,\vp_a}\,\frac{d\,\vp_c}{d\,x}\,\frac{d\,\vp_d}{d\,x}.
\lab{elforphi}
\er

Note that \rf{elforphi} are exactly the Euler-Lagrange equations for the fields $\vp_a$ coming from the functional $E$ given in \rf{energyfunc}. Moreover, taking $\vartheta=\partial_x\chi_{\beta}$ and and then $\vartheta=\chi_{\beta}$ we note that \rf{nicerel3} implies also that
\br
\frac{d\;}{d\,x}\left[\frac{\delta\, \eta_{cd}}{\delta\,\partial_x\chi_{\beta}}\,\frac{d\,\vp_c}{d\,x}\,\frac{d\,\vp_d}{d\,x}+\frac{\delta\, \eta^{-1}_{cd}}{\delta\,\partial_x\chi_{\beta}}\,\frac{\delta\, U}{\delta\,\vp_c}\,\frac{\delta\, U}{\delta\,\vp_d}\right]=
\frac{\delta\, \eta_{cd}}{\delta\,\chi_{\beta}}\,\frac{d\,\vp_c}{d\,x}\,\frac{d\,\vp_d}{d\,x}+\frac{\delta\, \eta^{-1}_{cd}}{\delta\,\chi_{\beta}}\,\frac{\delta\, U}{\delta\,\vp_c}\,\frac{\delta\, U}{\delta\,\vp_d}.
\lab{elforchi}
\er
This time, the obtained eqauations \rf{elforchi} are the Euler-Lagrange equations for the external field $\chi_{\beta}$ coming from the functional $E$ given in \rf{energyfunc}. Note that such extra fields could even be the entries of the matrix $\eta_{ab}$ themselves. 

Summarising, we see that the first order self-duality equations \rf{bpseq} alone imply the Euler-Lagrange equations corresponding to the static energy functional $E$ for the fields $\vp_a$ and any possible extra fields that the matrix $\eta_{ab}$ can depend on. Note that this fact had already been encoded in the construction presented above, between equations \rf{topchargegen} and  \rf{energyfuncgen}, since the fields $\vp_a$ which appear in \rf{identitygen} and \rf{eleqgen} can be any fields that the quantities ${\cal A}_{\alpha}$ and  ${\tilde{\cal A}}_{\alpha}$ depend on. With the choice we have made in \rf{choiceofacal}, the matrix $k$ and its inverse have become parts of these quantities and so they can depend on extra fields. 

Note also that the bound can be obtained by rewriting the energy functional $E$ given in \rf{energyfunc}  as 
\be
 E= \frac{1}{2}\,\int_{-\infty}^{\infty} dx\,\left[ k_{ab}\, \frac{d\,\vp_b}{d\,x}\mp 
 \frac{\delta\, U}{\delta\,\vp_b}\,k^{-1}_{ba}\right]^2\pm 
 \int_{-\infty}^{\infty} dx\,\frac{d\,\vp_a}{d\,x}\,\frac{\delta\, U}{\delta\,\vp_a}\geq \mid Q\mid,
 \lab{energyfuncbound}
 \ee
 which is, in fact, the same as \rf{boundgen}, and the bound is saturated by the self-dual solutions of \rf{bpseq}.

 %%%%%%%%%%%%%%%%%%%%%%%%%%%%%%%%%%%%%%%%%%%%%
 \section{The construction of the pre-potential $U$}
\label{sec:prepotential}
\setcounter{equation}{0}
%%%%%%%%%%%%%%%%%%%%%%%%%%%%%%%%%%%%%%%%%%%%%

 As we are interested in deriving physically relevant theories, from now one, we restrict our discussion to the cases where the scalar fields $\vp_a$, the pre-potential $U$, and the matrix $\eta_{ab}$ are real. In addition, we are interested in the cases for which the static energy functional $E$, given in \rf{energyfunc}, is positive definite. Thus we need to restrict our discussion to cases in which all the eigenvalues of $\eta_{ab}$ are positive definite. In order for  the self-dual solutions of \rf{bpseq} to possess finite energy $E$,  we need the energy density to vanish at spatial infinities when evaluated on such solutions, and so, given our restrictions, we require that   
\be
\frac{d\,\vp_b}{d\,x}\rightarrow 0\;;\qquad\qquad   \frac{\delta\, U}{\delta\,\vp_a}\rightarrow 0\;;\qquad\qquad {\rm as}\qquad x\rightarrow \pm \infty.
\lab{vacuaconditions}
\ee
Thus, the self-duality equations \rf{bpseq} should possess constant vacua solutions $\vp_a^{\rm (vac.)}$ that are zeros of all the first derivatives of the pre-potential, {\it i.e.} 
\be
\frac{\delta\, U}{\delta\,\vp_b}\mid_{\vp_a=\vp_a^{\rm (vac.)}}=0.
\lab{vacuau}
\ee 

We then see from \rf{potdef} that such vacua are also zeros of the potential $V$ and of its first derivatives, {\it i.e.}
\be
V\(\vp_a^{\rm (vac.)}\)=0\;;\qquad\qquad\qquad 
\frac{\delta\, V}{\delta\,\vp_b}\mid_{\vp_a=\vp_a^{\rm (vac.)}}=0.
\ee
Moreover, we would like the theories we are constructing to possess various soliton type solutions, and we know that, in general, the total topological charges of such solutions are obtained by additions, under some finite or infinite abelian group, of the charges of the constituent one-solitons. Thus, we would like to have systems of vacua as degenerate as possible. Certainly there are numerous ways  of achieving this goal. In this paper we use a group theoretical approach to the construction of  the prepotentials $U$.

\subsection{Details of the construction} 

 Consider a Lie algebra ${\cal G}$ and let ${\vec \alpha}_a$, $a=1,2,\ldots r\equiv {\rm rank}\,{\cal G}$, be the set of its simple roots.  We use the scalar fields $\vp_a$ to construct our basic vector in the root space: 
\be
 {\vec \vp}\equiv \sum_{a=1}^r \vp_a\, \frac{2\,{\vec \alpha}_a}{{\vec \alpha}_a^2}.
 \lab{phivectorb}
 \ee
Next we choose a representation  ${\cal R}$ (irreducible or not) of the Lie algebra ${\cal G}$, and we denote by ${\vec \mu}_k$  the set of weights of ${\cal R}$. We take the pre-potential $U$ to be of the form
\be
 U\equiv \sum_{{\vec \mu}_k\in{\cal R}} c_{{\vec \mu}_k}\, e^{i\, {\vec \mu}_k\cdot {\vec \vp}},
 \lab{prepotconstruct}
 \ee
where $c_{{\vec \mu}_k}$ are some (complex) constant coefficients. Note from section \ref{sec:bps}, that $U$ enters in our construction of the self-dual sectors only through its derivatives {\it w.r.t.} the fields $\vp_a$, and so any constant additive in $U$ is irrelevant. Therefore, we see from \rf{prepotconstruct} that the zero weights of  ${\cal R}$ play no role in our construction. Since we want $U$ to be real, we need for our definition of $U$ to consider representations for which, if  ${\vec \mu}_k$  is a weight of ${\cal R}$, so is its negative $-{\vec \mu}_k$. Some irreducible representations, like the adjoint, have this property. However, we can also consider ${\cal R}$ to have as many irreducible components as necessary to fulfill this reality requirement. For instance, in the case of $SU(N)$ one can take ${\cal R}$ to be the direct sum of the $N$ and ${\bar N}$ fundamental representations. In addition, for the reality of $U$ we need the coefficients $c_{{\vec \mu}_k}$ to satisfy $c_{-{\vec \mu}_k}=c_{{\vec \mu}_k}^*$. Writing $c_{{\vec \mu}_k}=\frac{1}{2}\(\gamma_{{\vec \mu}_k}- i\,\delta_{{\vec \mu}_k}\)$, we find that \rf{prepotconstruct} then takes the form:
\be
U\equiv \sum_{{\vec \mu}_k\in{\cal R}^{(+)}}\left[ \gamma_{{\vec \mu}_k}\, \cos\({\vec \mu}_k\cdot {\vec \vp}\)+
\delta_{{\vec \mu}_k}\, \sin\({\vec \mu}_k\cdot {\vec \vp}\) \right],
\lab{prepotconstruct2}
\ee
where the superscript $+$ in ${\cal R}^{(+)}$  denotes that we are taking just one weight of each pair $\({\vec \mu}_k\,,\,-{\vec \mu}_k\)$. For instance, in the case where  ${\cal R}$ is the direct sum of the $N$ and ${\bar N}$ fundamental representations of $SU(N)$, ${\cal R}^{(+)}$ would be either the $N$, or the ${\bar N}$, component. In the case where  ${\cal R}$ is the adjoint representation, ${\cal R}^{(+)}$ would contain only the positive roots. From \rf{prepotconstruct2} we then have that
\be
\frac{\delta\, U}{\delta\,\vp_a}= \frac{2\,{\vec \alpha}_a}{{\vec \alpha}_a^2}\cdot 
\sum_{{\vec \mu}_k\in{\cal R}^{(+)}} {\vec \mu}_k\, \left[- \gamma_{{\vec \mu}_k}\, \sin\({\vec \mu}_k\cdot {\vec \vp}\)+
\delta_{{\vec \mu}_k}\, \cos\({\vec \mu}_k\cdot {\vec \vp}\) \right].
\lab{prepotder}
\ee

There are several ways of satisfying \rf{vacuau}, and the vacuum structure of our theories can be quite complicated. Let us first mention various possibilities: 
\begin{enumerate}
\item In the highest weight irreducible representation of a Lie algebra ${\cal G}$, the weights are of the form ${\vec \mu}_k={\vec\lambda}-{\vec w}_k$, where ${\vec\lambda}$ is the highest weight and ${\vec w}_k$ is a sum of positive roots of ${\cal G}$. So, if one takes ${\vec \vp}^{({\rm vac.})}$ to be $2\,\pi$ times a vector in the  co-weight lattice of ${\cal G}$, then ${\vec \vp}^{({\rm vac.})}\cdot {\vec w}_k=2\,\pi\, n_k$, with $n_k$ being an integer. Thus, for any irreducible component ${\cal R}^{(+)}_{{\vec \lambda}}$ of ${\cal R}^{(+)}$  the coefficients $\gamma_{{\vec \mu}_k}$ and $\delta_{{\vec \mu}_k}$ should be taken such that  
\be
\sin\({\vec \lambda}\cdot {\vec \vp}^{({\rm vac.})}\)\,\sum_{{\vec \mu}_k\in{\cal R}^{(+)}_{{\vec \lambda}}} {\vec \mu}_k\,  \gamma_{{\vec \mu}_k}-
\cos\({\vec \lambda}\cdot {\vec \vp}^{({\rm vac.})}\)\sum_{{\vec \mu}_k\in{\cal R}^{(+)}_{{\vec \lambda}}} {\vec \mu}_k\,\delta_{{\vec \mu}_k}\,  =0.
\lab{quitepossible}
\ee
\item The weights ${\mu}$ of a Lie algebra ${\cal G}$ are defined as the vectors  which satisfy the condition $2{\vec \alpha}\cdot {\vec\mu}/{\vec \alpha}^2\in \IZ$, for any root ${\vec \alpha}$ of ${\cal G}$. So, from \rf{phivectorb} and \rf{prepotder}, one sees that \rf{vacuau} can be satisfied if 
\be
\vp_a^{({\rm vac.})}= \pi\, n_a\;;\qquad n_a \in \IZ\;\qquad {\rm and} \qquad  \delta_{{\vec \mu}_k}=0.
\lab{concretechoice}
\ee
In such a case we have that ${\vec \vp}_a^{({\rm vac.})}$ is $\pi$ times a vector in the co-root lattice of ${\cal G}$.  
In the cases where $\sum_{a=1}^r 2{\vec \alpha}_a\cdot {\vec\mu}_k/ {\vec \alpha}_a^2$ is an odd number,  one can also satisfy \rf{vacuau} if 
\be
\vp_a^{({\rm vac.})}= \pi\, \(n_a+\frac{1}{2}\)\;;\qquad n_a \in \IZ\;\qquad {\rm and} \qquad  \gamma_{{\vec \mu}_k}=0.
\lab{funnychoice}
\ee
\item The third possibility is provided by the cases that involve special vectors ${\vec \vp}^{({\rm vac.})}$ such that 
\be
\sum_{{\vec \mu}_k\in{\cal R}^{(+)}} \frac{2\,{\vec \alpha}_a}{{\vec \alpha}_a^2}\cdot{\vec \mu}_k\, \left[- \gamma_{{\vec \mu}_k}\, \sin\({\vec \mu}_k\cdot {\vec \vp}^{({\rm vac.})}\)+
\delta_{{\vec \mu}_k}\, \cos\({\vec \mu}_k\cdot {\vec \vp}^{({\rm vac.})}\) \right]=0
\lab{rarechoice}
\ee
even when the sines or cosines do not vanish individually. 
We will show below that such a possibility exists,  for instance, when ${\cal R}$ is the direct sum of the triplet and anti-triplet of $SU(3)$, and when $\delta_{{\vec \mu}_k}=0$. 

\end{enumerate} 

In most of the examples that we discuss in this paper we consider pre-potentials $U$ of the form \rf{prepotconstruct2} with $\delta_{{\vec \mu}_k}=0$.
Then, the possibility \rf{concretechoice} is always there and this guarantees that we have  infinitely many degenerate vacua.

%%%%%%%%%%%%%%%%%%%%%%%%%%%%%%%%%%%%%%%%%%%%%
 \section{Examples}
\label{sec:examples}
\setcounter{equation}{0}
%%%%%%%%%%%%%%%%%%%%%%%%%%%%%%%%%%%%%%%%%%%%%

In this section we present some concrete examples of the construction presented in   sections \ref{sec:bps} and \ref{sec:prepotential}. As shown there, the matrix $\eta_{ab}$ can depend on the fields $\vp_a$ and  their first derivatives as well as  on extra fields. The dependence of the $\eta$  matrix on derivatives of the fields  would not allow a kinetic term which is quadratic in field derivatives. So, such  cases are probably not of much interest. The cases in which $\eta_{ab}$ depends on the fields $\vp_a$ only and not on their derivatives are important if one considers field theories possessing a target space with non-trivial metric like non-linear sigma models, non-abelian Toda theories, {\it etc}. In this paper we consider only the cases where the matrix $\eta_{ab}$ is constant, real and positive definite, since the corresponding examples are already rich enough and lead to interesting theories. We leave the generalizations to more complicated theories to further studies.

%%%%%%%%%%%%%%%%%%%%%%%%%%%%%%%%%%%%%%%%%%%%%
\subsection{$SU(2)$}
%%%%%%%%%%%%%%%%%%%%%%%%%%%%%%%%%%%%%%%%%%%%%

The rank of $SU(2)$ is unity and so we have just one scalar field that we denote by $\vp$. The matrix $\eta$ is just a number that we take to be unity. In order to have the usual notation where the weights are integers or semi-integers, we normalize the only simple root to have its squared modulus  equal to one. For all irreducible representations of $SU(2)$  the non-zero weights come in pairs, {\it i.e.} the weight and its negative, and so the pre-potentials given in \rf{prepotconstruct2} apply to all such representations. For the spinor (doublet) representation we have two possibilities. First we can take the $\delta$-term in \rf{prepotconstruct2} to vanish and so consider the following pre-potential and the self-duality equations (see  \rf{bpseq})
\be
U_{j=1/2}^{(1)}= -\cos \vp\;;\qquad\qquad\qquad \partial_x \vp=\pm\,\frac{\delta U_{j=1/2}^{(1)}}{\delta \vp}=\pm \sin\vp.
\lab{firstsg}
\ee

The corresponding vacua are then $\vp^{({\rm vac.})}= \pi\, n$, $n\in \IZ$, which correspond to the case \rf{concretechoice}. The vacua of types \rf{quitepossible} and \rf{rarechoice} do not exist in this case. 
By differentiating \rf{firstsg} {\it w.r.t.} $x$ and using it again, one finds that the solutions of \rf{firstsg}  solve the static sine-Gordon equation
\be
\partial_x^2\vp= \frac{1}{2}\,\sin\(2\, \vp\).
\lab{sgeq}
\ee
The solutions of \rf{firstsg} are the familiar kink  ``tunneling'' from $0$ to $\pi$, and anti-kink ``tunneling'' in the reverse direction, and given by  
\be
\vp=2\,{\rm ArcTan}\( e^{\pm x}\).
\ee

The second choice for the spinor representation, corresponds to the case when the $\gamma$-term in \rf{prepotconstruct2} is zero and when  the  pre-potential and self-duality equations take the form: 
\be
U_{j=1/2}^{(2)}= \sin \vp\;;\qquad\qquad\qquad \partial_x \vp=\pm\,\frac{\delta U_{j=1/2}^{(2)}}{\delta \vp}=\pm \cos\vp
\lab{secondsg}
\ee
The vacua in this case are $\vp^{({\rm vac.})}= \pi\, \(n+\frac{1}{2}\)$, $n\in \IZ$ and so correspond to the case \rf{funnychoice}. Again the vacua of types \rf{quitepossible} and \rf{rarechoice} do not exist in this case. By differentiating \rf{secondsg} one finds that its solutions  satisfy the inverted sine-Gordon equation
\be
\partial_x^2\vp= -\frac{1}{2}\,\sin\(2\, \vp\).
\lab{sgeq2}
\ee
The solutions of \rf{secondsg} are
\be
\vp=2 \,{\rm ArcTan}\left[{\rm Tanh}\(\pm \frac{x}{2}\)\right],
\ee
which are also kink and anti-kink solutions but ``tunneling''  from $-\frac{\pi}{2}$ to $\frac{\pi}{2}$ for the kink, and vice-versa for the anti-kink. Note that redefining the field as $\vp={\tilde \vp}-\pi/2$, the equation \rf{sgeq2} becomes the usual sine-Gordon equation for the field ${\tilde \vp}$, which now ``tunnels'' from $0$ to $\pi$ for the kink, and vice-versa for the anti-kink.

For the triplet representation we do not get anything new since the zero weight term in the pre-potential leads to a constant term and so is irrelevant. We just get the same equations as in the doublet representation but with the fields rescaled by a factor $2$.  However, we can take a representation which is reducible and being given by the sum of the $j=1/2$ and $j=1$ representations. Then  we can set the $\delta$-term in \rf{prepotconstruct2} to vanish and consider the following pre-potential 
\be
U_{1+1/2}^{(1)}= -\gamma_1\, \cos \vp - \gamma_2\, \cos\(2 \vp\) 
\lab{fakedouble1}
\ee
 and the corresponding self-duality equations now become
\be
\partial_x \vp=\pm\,\frac{\delta U_{1+1/2}^{(1)}}{\delta \vp}=\pm \left[\gamma_1\,\sin\vp+2\,\gamma_2\,\sin\(2 \vp\)\right]=\pm\,\sin\vp\,\left[\gamma_1\,+4\,\gamma_2\,\cos\vp\right].\lab{eqsu2}
\ee 
The vacua are now:
\be
\vp^{({\rm vac.})}= \pi\, n\;;\qquad {\rm and}\qquad 
\vp^{({\rm vac.})}= {\rm ArcCos}\(-\frac{\gamma_1}{4\,\gamma_2}\)+2\,\pi\,m\;;\qquad n,m\in \IZ
\lab{vacuasu2first}
\ee
The first class is of the type \rf{concretechoice} and the second of the type \rf{rarechoice}. Of course, we need $\mid\frac{\gamma_1}{\gamma_2}\mid \leq 4$ for the second type of vacua to exist. 

For the case $1+1/2$, we can also set the $\gamma$-term in \rf{prepotconstruct2} to zero and so consider the following pre-potential 
\be
U_{1+1/2}^{(2)}= \delta_1\, \sin \vp + \delta_2\, \sin\(2 \vp\) 
\lab{fakedouble2}
\ee
leading to the self-duality equations:
\be
\partial_x \vp=\pm\,\frac{\delta U_{1+1/2}^{(2)}}{\delta \vp}=\pm \left[\delta_1\,\cos\vp+2\,\delta_2\,\cos\(2 \vp\)\right]=\pm \left[4\,\delta_2\,\cos^2\vp+\delta_1\,\cos\vp-2\,\delta_2\right].
\ee 
The vacua are now
\be
\vp^{({\rm vac.})}={\rm ArcCos}\left[\frac{-\delta_1\pm\sqrt{\delta_1^2+32\,\delta_2^2}}{8\,\delta_2}\right]+2\,\pi\,n\;;\qquad n\in \IZ.
\ee

Below we present examples of solutions of equation \rf{eqsu2} (the case with upper sign). In the new variable $x\rightarrow \gamma_1 x$ \rf{eqsu2} takes the form $\partial_x\varphi=\sin\vp[1+b\cos\vp]$, in which the ratio $4\frac{\gamma_2}{\gamma_1}$ is denoted by $b$. So
\be
x=x_0+\int \frac{d\vp}{\sin\vp[1+b\cos\vp]}.\lab{inteq}
\ee
The form of the integral on the right hand side of \rf{inteq} depends on the value of the parameter $b$. For $|b|\le1$ the pre-potential has only the vacua $\vp^{({\rm vac.})}= \pi\, n$ and for $|b|>1$ the second type of vacua in \rf{vacuasu2first} appear. Taking the constant $x_0$ such that $\vp(0)=\frac{\pi}{2}$ one gets the solution
\begin{eqnarray}
x=\left\{\begin{array}{lcl}
\frac{1}{4}\big[\tan^2\left(\frac{\vp}{2}\right)-1+2\ln\left[\tan\left(\frac{\vp}{2}\right)\right]\big]&{\rm for}&b=1,\\
\frac{1}{4}\big[1-\cot^2\left(\frac{\vp}{2}\right)+2\ln\left[\tan\left(\frac{\vp}{2}\right)\right]\big]&{\rm for}&b=-1,\\
\frac{1}{1-b^2}\ln\left[\tan\left(\frac{\vp}{2}\right)\left(\frac{1+b+(1-b)\tan^2\left(\frac{\vp}{2}\right)}{2\tan\left(\frac{\vp}{2}\right)}\right)^b\right]&{\rm for}&b\neq \pm1.\\
\end{array}\right.\lab{solanayt}
\end{eqnarray}

\begin{figure}[h!]
  \centering
  \includegraphics[width=0.9\textwidth,height=0.4\textwidth, angle =0]{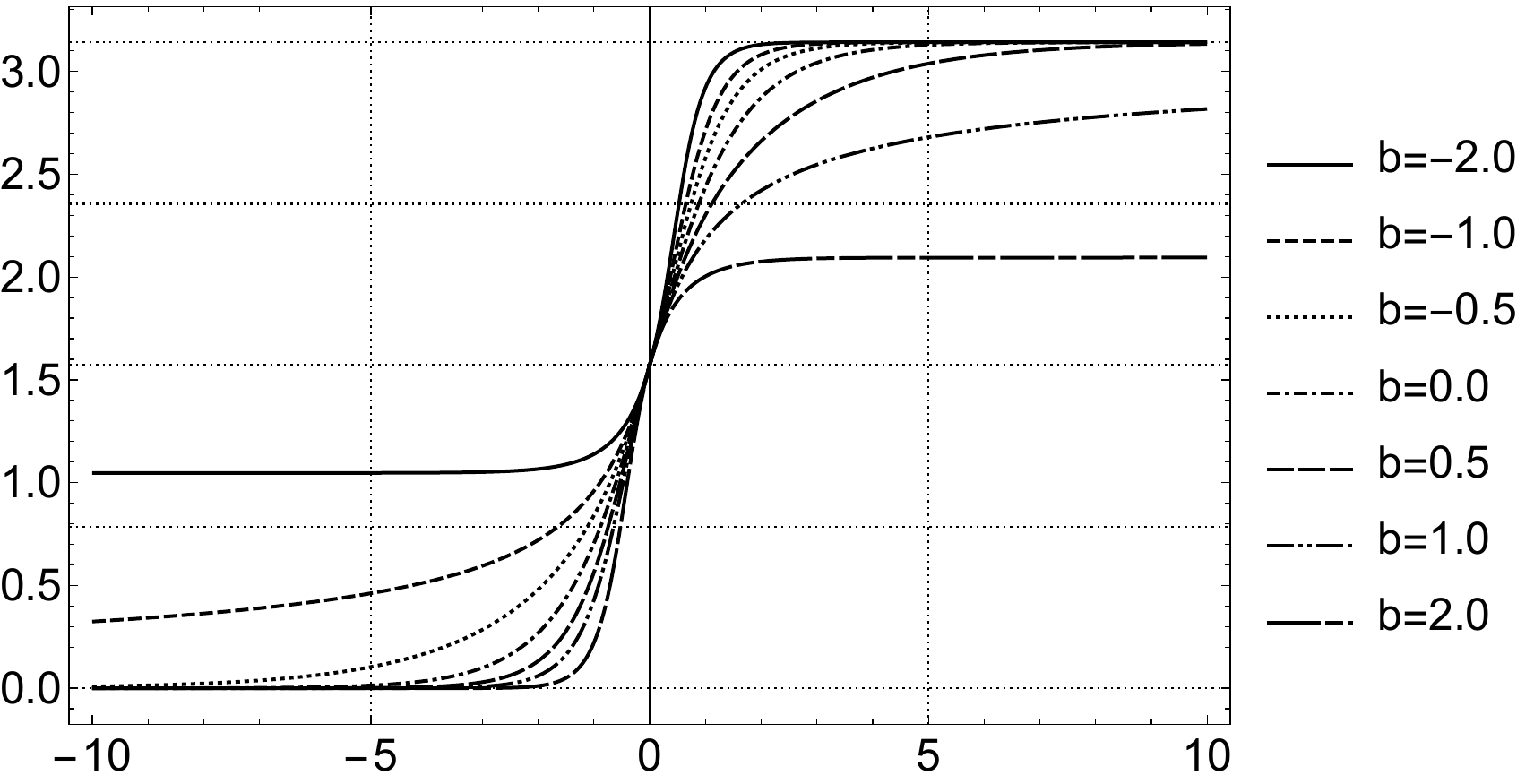}               
  \caption{Plots of  $\varphi(x)$ against $x$ for solutions of the equation \rf{eqsu2} with the upper sign and with $x$ rescaled as $x\rightarrow \gamma_1\,x$. The curves correspond to solutions of  \rf{eqsu2}  for different values of the ratio $b=4\frac{\gamma_2}{\gamma_1}$ and the initial condition $\varphi(0)=\frac{\pi}{2}$.}
  \label{su2plot}
\end{figure}

Note that the solutions given by the last formula in \rf{solanayt} interpolate between different vacua for different values of $b$. In the case of $|b|<1$ the solution describes the tunneling between the vacua $\vp^{({\rm vac.})}=0$ and $\vp^{({\rm vac.})}=\pi$. For $|b|>1$ the solution connects $\vp^{({\rm vac.})}={\rm ArcCos}[-b^{-1}]$ and $\vp^{({\rm vac.})}=\pi$ when $b<-1$ and the vacua $\vp^{({\rm vac.})}=0$ and $\vp^{({\rm vac.})}={\rm ArcCos}[-b^{-1}]$ when $b>1$. In Figure \ref{su2plot} we present solutions that correspond to the cases $b=\{0, \pm \frac{1}{2}, \pm 1,\pm 2\}$. 
%%%%%%%%%%%%%%%%%%%%%%%%%%%%%%%%%%%%%%%%%%%%%
%							FIGURE 1								      %
%%%%%%%%%%%%%%%%%%%%%%%%%%%%%%%%%%%%%%%%%%%%%

Note that the pre-potentials \rf{fakedouble1} and \rf{fakedouble2} contain sines and cosines of the field $\vp$ and its double $2\,\vp$, but they do not correspond to the usual double sine-Gordon model \cite{doublesg} since the potentials one gets from \rf{potdef}, with $\eta=1$ for instance, do not correspond to the double sine-Gordon potential.  However, the procedure for finding self-duality equations for theories in $(1+1)$-dimensions with just one scalar field is very well known.  Indeed, from \rf{potdef} one notes that the pre-potential for the double sine-Gordon potential $V_{\rm 2-SG}$ can be obtained by integrating the equation \rf{potdef}  (with $\eta=1$) 
\be
\frac{d\,U_{\rm 2-SG}}{d\, \vp}= \sqrt{2\, V_{\rm 2-SG}}\;;\qquad\quad 
{\rm with }\qquad\quad V_{\rm 2-SG}=1- a\,\cos \vp -\(1-a\)\, \cos\(2\,\vp\)
\lab{2sgprepotential}
\ee
where $0\leq a\leq 1$.  However, note that the pre-potential $U_{\rm 2-SG}$ one obtains from \rf{2sgprepotential} is not of the form \rf{prepotconstruct} that we have used in our construction for theories with several scalar fields.

%%%%%%%%%%%%%%%%%%%%%%%%%%%%%%%%%%%%%%%%%%%%%
\subsection{$SU(3)$}
%%%%%%%%%%%%%%%%%%%%%%%%%%%%%%%%%%%%%%%%%%%%%

The rank of $SU(3)$ is two and so we have two fields, $\vp_1$ and $\vp_2$, in this case. We take the matrix $\eta_{ab}$ to be of the form\footnote{Note that $\eta_{ab}\mid_{\lambda=1}=K_{ab}$, with $K_{ab}$ being the Cartan matrix of $SU(3)$, given in \rf{cartansu3}.}
\br
\eta=\(
\begin{array}{cc}
2&-\,\lambda\\
-\,\lambda &2
\end{array}\),
\qquad\qquad\qquad
\eta^{-1}=\frac{1}{4-\lambda^2}\(
\begin{array}{cc}
2&\lambda\\
\lambda&2
\end{array}\),
\lab{etamatricessu3lambda}
\er
where we have introduced a real parameter $\lambda$. The eigenvalues of $\eta$ are $2\pm \lambda$, and so we have to keep $\lambda$ in the interval $-2<\lambda<2$, to have $\eta$ positive definite and invertible.  The weights of the triplet representation of $SU(3)$ are given by
\be
{\vec \mu}_1={\vec \lambda}_1, \qquad\qquad {\vec \mu}_2={\vec \lambda}_1-{\vec \alpha}_1,
\qquad\qquad {\vec \mu}_3={\vec \lambda}_1-{\vec \alpha}_1-{\vec \alpha}_2
\ee
and those of the anti-triplet by
\be
{\vec {\bar\mu}}_1={\vec \lambda}_2=- {\vec \mu}_3,\qquad\qquad {\vec {\bar \mu}}_2={\vec \lambda}_2-{\vec \alpha}_2=-{\vec \mu}_2,
\qquad\qquad {\vec {\bar \mu}}_3={\vec \lambda}_2-{\vec \alpha}_1-{\vec \alpha}_2=-{\vec \mu}_1,
\ee
where $\lambda_a$, and $\alpha_a$, $a=1,2$ are, respectively, the fundamental weights and simple roots of $SU(3)$. They satisfy
\br
\frac{2\,{\vec \alpha}_a\cdot{\vec \lambda_b}}{{\vec \alpha_a}^2}=\delta_{ab}\;; \qquad\qquad\hbox{and}\qquad\qquad
K_{ab}=\frac{2\,{\vec \alpha}_a\cdot{\vec \alpha_b}}{{\vec \alpha_b}^2}=\(
\begin{array}{cc}
2&-1\\
-1 &2
\end{array}\),
\lab{cartansu3}
\er
where $K_{ab}$ is the Cartan matrix of $SU(3)$. Note that the weights of the anti-triplet representation are the negatives of those of the triplet. Therefore, if we take the representation ${\cal R}$ in  \rf{prepotconstruct} to be the direct sum of the triplet and anti-triplet we satisfy the conditions for the reality of the pre-potential $U$. Thus the set of weights ${\cal R}^{(+)}$ can be taken to be those of the triplet representation and so from \rf{prepotconstruct2} we get the pre-potential as
\be
U=\gamma_1\,\cos \vp_1+\gamma_2\,\cos\vp_2+\gamma_3\,\cos\(\vp_1-\vp_2\),
\lab{usu3tripletantitriplet}
\ee
where we have chosen the $\delta$-terms in \rf{prepotconstruct2}  to vanish. 

%%%%%%%%%%%%%%%%%%%%%%%%%%%%%%%%%%%%%%%%%%%%%
%							FIGURE 2							      %
%%%%%%%%%%%%%%%%%%%%%%%%%%%%%%%%%%%%%%%%%%%%%
 \begin{figure}[h!]
  \centering
  \subfigure[]{\includegraphics[width=0.48\textwidth,height=0.38\textwidth, angle =0]{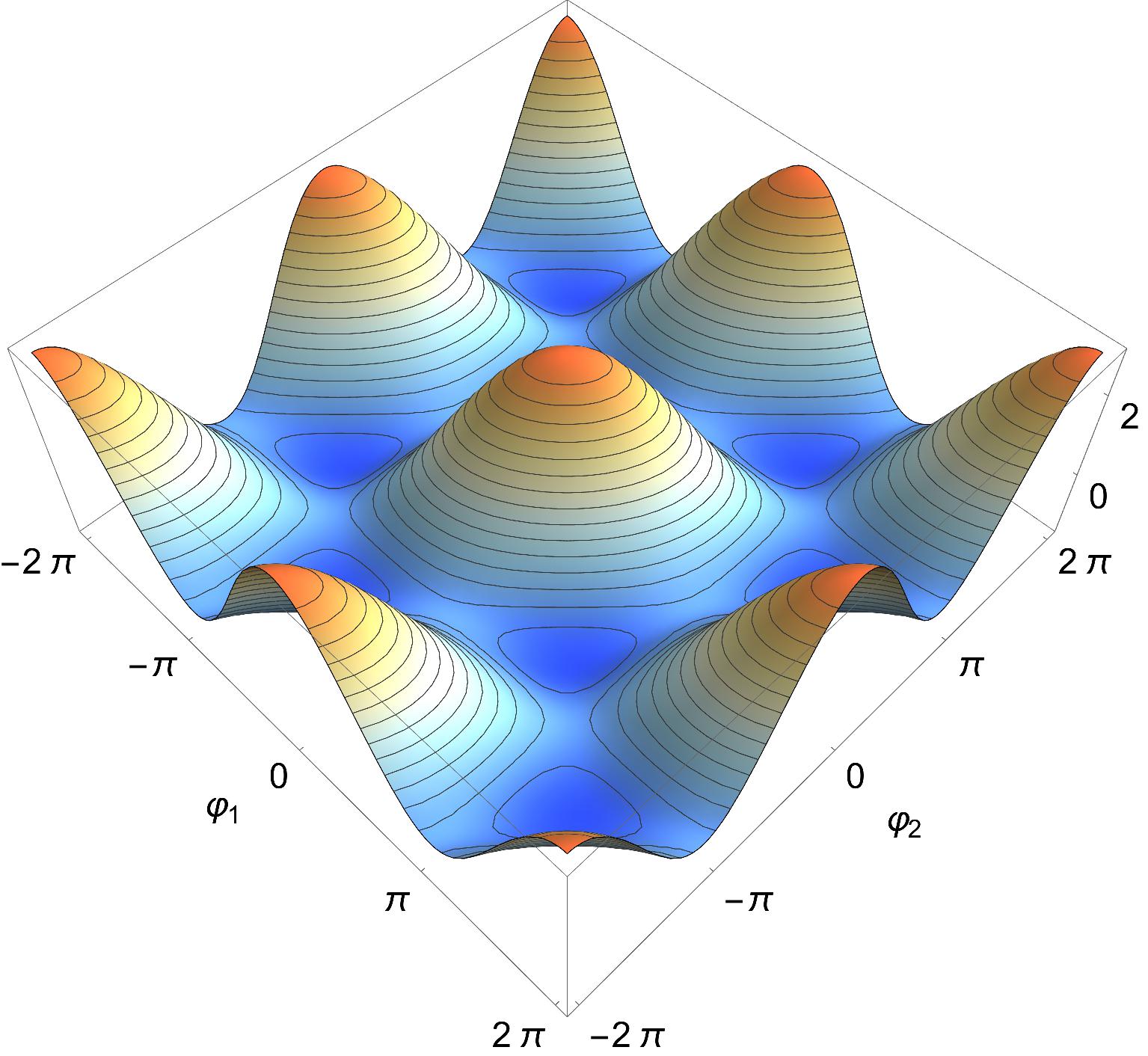}}               
  \subfigure[]{\includegraphics[width=0.48\textwidth,height=0.38\textwidth, angle =0]{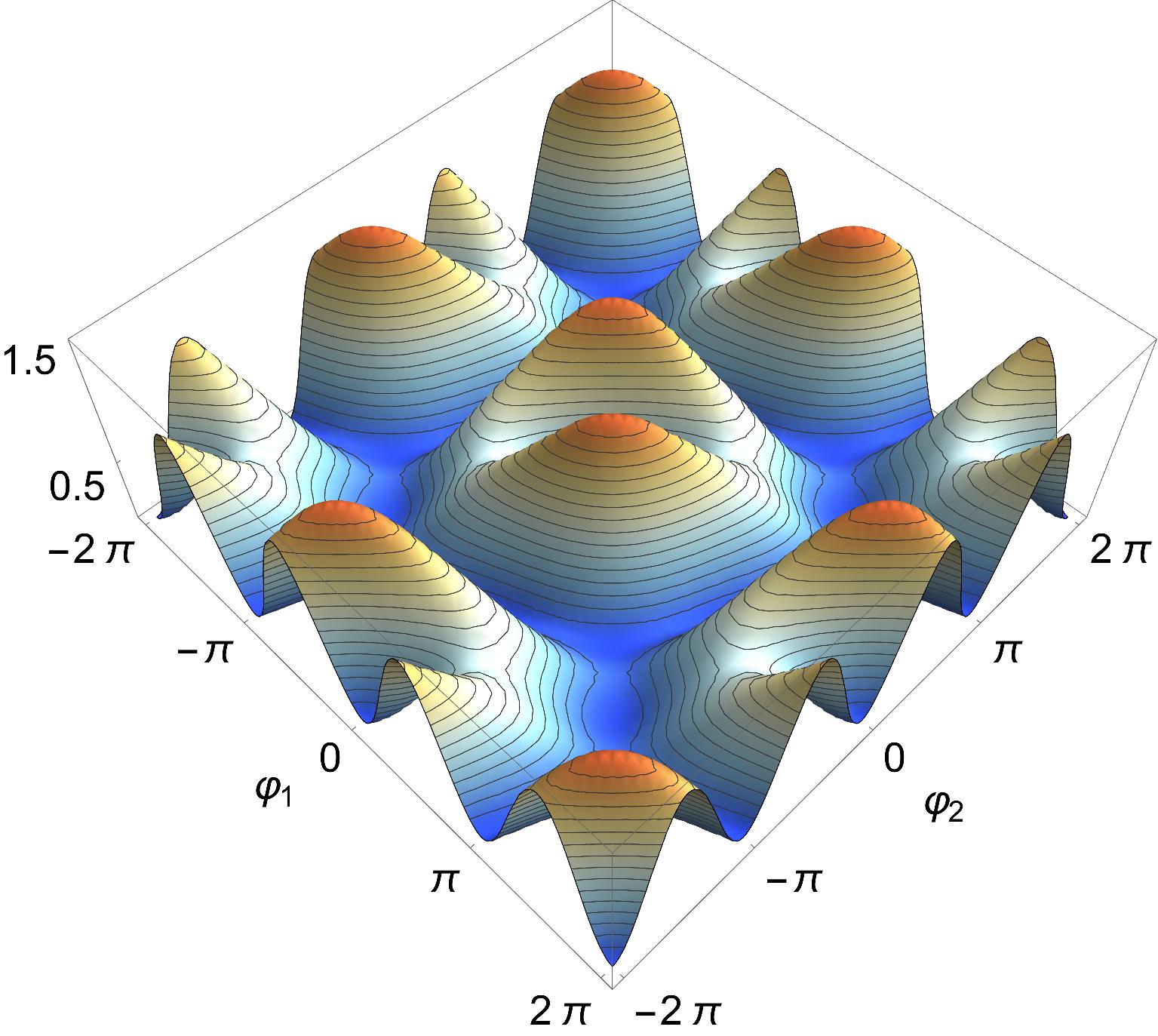}}
  \subfigure[]{\includegraphics[width=0.48\textwidth, angle =0]{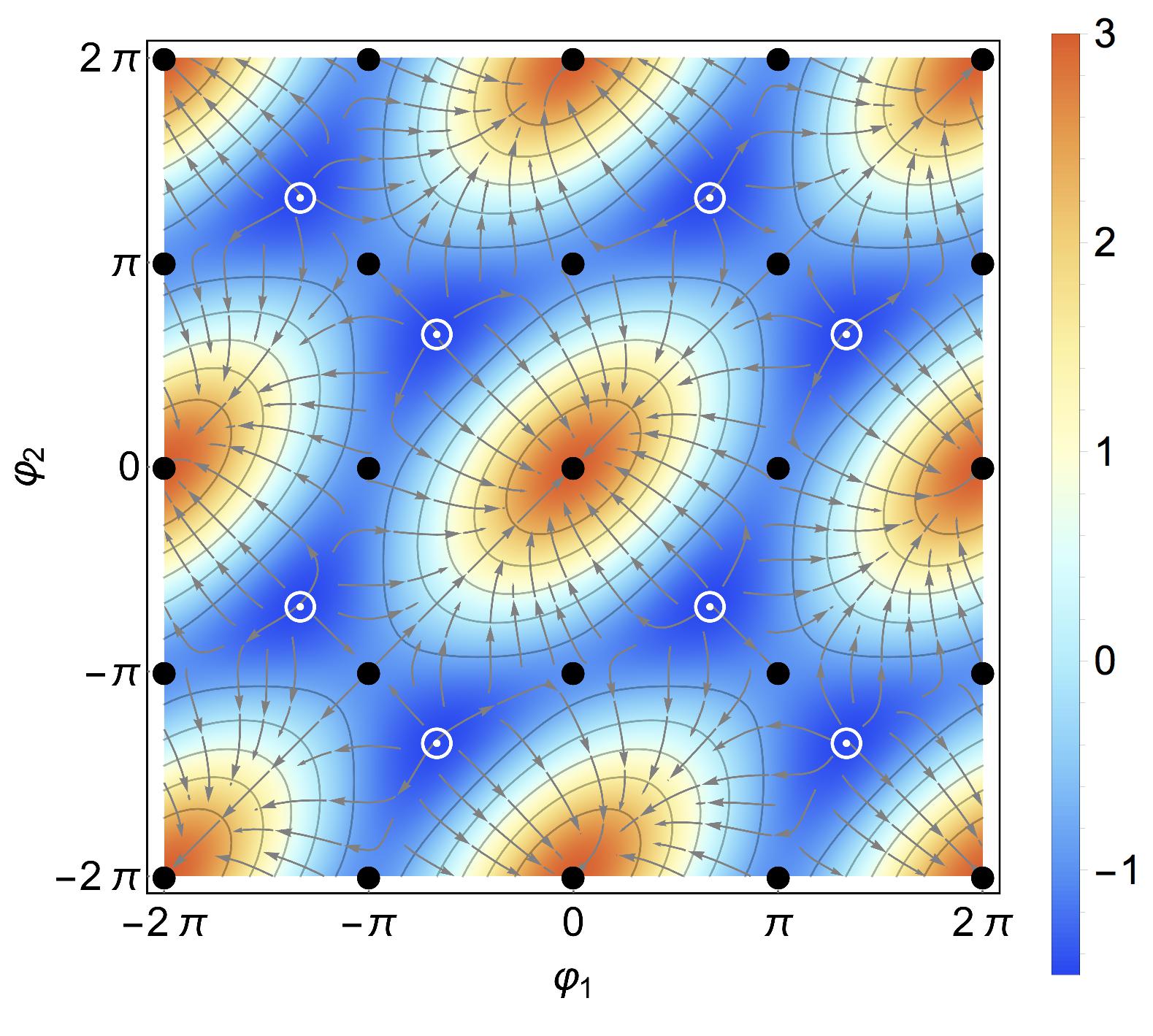}}
  \subfigure[]{\includegraphics[width=0.48\textwidth, angle =0]{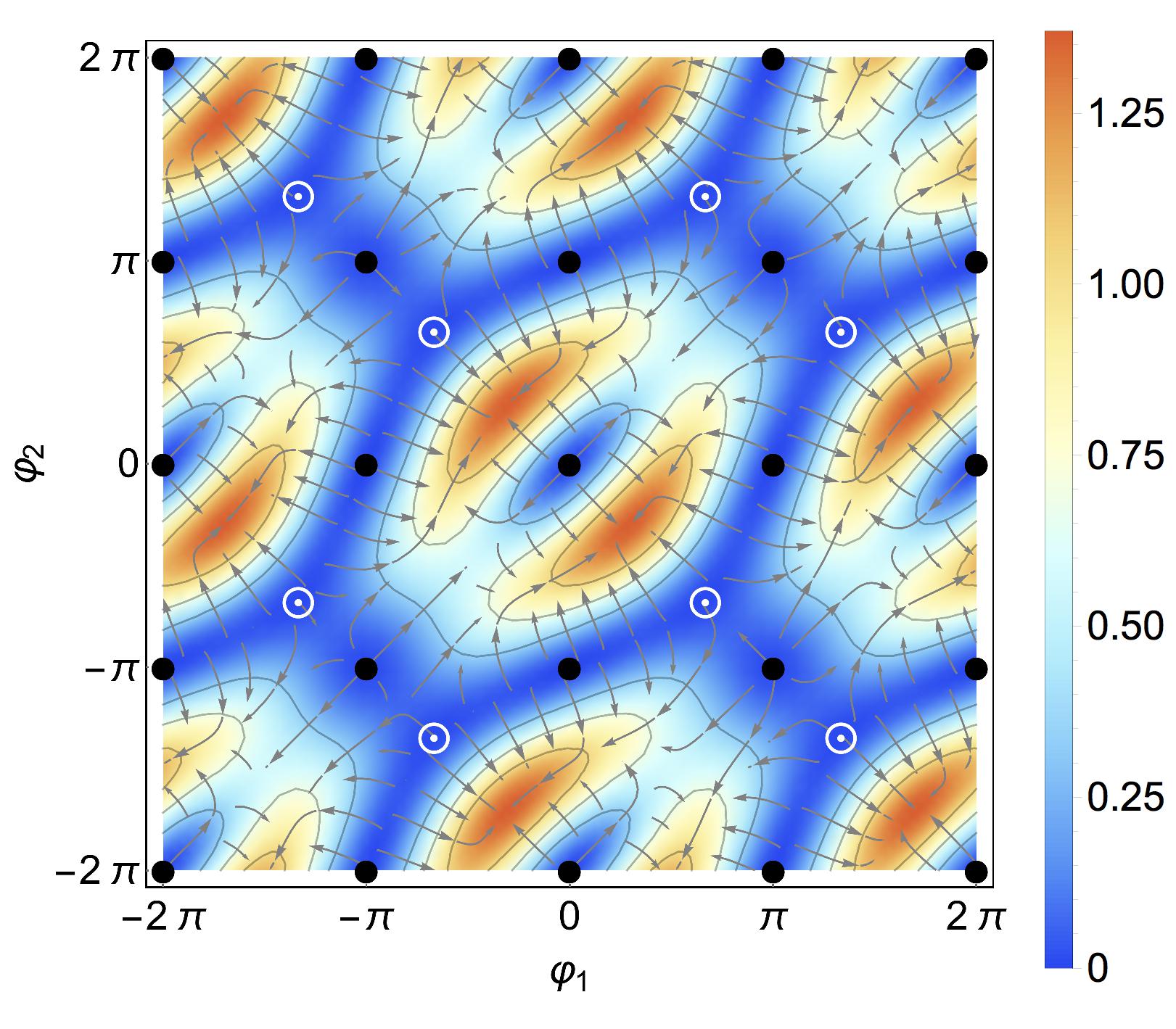}}
  \caption{(a), (c) The pre-potential \rf{usu3tripletantitriplet} and (b), (d) the potential \rf{vsu3tripletantitriplet}. The vacua \rf{vacuasu3b} are given by the black dots, \rf{othervacuasu3} by $\odot$ and the lines stand for the gradient flow of $\vec\nabla U$ and $\vec\nabla V$, where $\vec\nabla=(\partial_{\varphi_1},\partial_{\varphi_2})$. Here $\lambda=\frac{1}{4}$ and  $\gamma_1=\gamma_2=\gamma_3=1$. }
  \label{}
\end{figure}
%%%%%%%%%%%%%%%%%%%%%%%%%%%%%%%%%%%%%%%%%%%%%

The static energy \rf{energyfunc} now becomes  
\be
E=\int_{-\infty}^{\infty}dx\,\left[\(\partial_x\vp_1\)^2+\(\partial_x\vp_2\)^2-\lambda\,\partial_x\vp_1\,\partial_x\vp_2 + V\(\vp_1,\vp_2\) \right],
\lab{energy}
\ee
where the potential \rf{potdef} is given by
\br
V&=&
\frac{1}{\lambda^2-4}\left[-\gamma_1^2 \sin ^2(\vp_1)+\gamma_1 \sin (\vp_1) (\gamma_3
   (\lambda-2) \sin (\vp_1-\vp_2)-\gamma_2 \lambda \sin
   (\vp_2))
   \right. 
   \\
   &-&\left. \gamma_2^2 \sin ^2(\vp_2)-\gamma_2 \gamma_3 (\lambda-2) \sin
   (\vp_2) \sin (\vp_1-\vp_2)+\gamma_3^2 (\lambda-2) \sin
   ^2(\vp_1-\vp_2)\right].\lab{vsu3tripletantitriplet}
   \nonumber
   \er

 The self-duality equations \rf{bpseq} are  now of the form:
   \br
   \partial_x\vp_1&=&\pm 
   \frac{\left[2 \gamma_1 \sin (\vp_1)+\gamma_2 \lambda \sin (\vp_2)-\gamma_3
   (\lambda-2) \sin (\vp_1-\vp_2)\right]}{\lambda^2-4},
   \lab{su3bpseq33bar}\\
   \partial_x\vp_2&=&\pm 
   \frac{\left[\gamma_1 \lambda \sin (\vp_1)+2 \gamma_2 \sin (\vp_2)+\gamma_3
   (\lambda-2) \sin (\vp_1-\vp_2)\right]}{\lambda^2-4}.
   \nonumber
   \er
   The vacua are determined by the conditions \rf{vacuau} which in this case become
    \br
\frac{\partial U}{\partial \vp_1}\mid_{\vp_a=\vp_a^{\rm (vac.)}}&=&   
   -\gamma_1 \sin (\vp_1^{\rm (vac.)})-\gamma_3 \sin (\vp_1^{\rm (vac.)}-\vp_2^{\rm (vac.)})=0,
   \lab{vacuasu3}\\
   \frac{\partial U}{\partial \vp_2}\mid_{\vp_a=\vp_a^{\rm (vac.)}}&=&  
   \gamma_3 \sin (\vp_1^{\rm (vac.)}-\vp_2^{\rm (vac.)})-\gamma_2 \sin (\vp_2^{\rm (vac.)})=0,
   \nonumber
   \er
  and these conditions imply that
   \be
   \gamma_1 \sin (\vp_1^{\rm (vac.)})=-\gamma_3 \sin (\vp_1^{\rm (vac.)}-\vp_2^{\rm (vac.)})=-\gamma_2 \sin (\vp_2^{\rm (vac.)}).
   \lab{vacuasu3b}
   \ee
Certainly \rf{vacuasu3b} are satisfied if 
\be
\vp_a^{\rm (vac.)}=\pi\,n_a\;,\qquad\qquad n_a\in \IZ, \qquad\qquad \qquad   a=1,2\qquad\qquad 
\mbox{\rm any values of the $\gamma$'s}\lab{firstkindvacuasu3}
\ee
and these are the vacua of the type \rf{concretechoice}. However, we also have vacua of the type \rf{rarechoice} that depend upon the particular values of the $\gamma$-constants that we are free to choose. For instance, one finds that  \rf{vacuasu3b} are satisfied if
\br
\(\vp_1^{\rm (vac.)}\,,\, \vp_2^{\rm (vac.)}\)&=&\(\frac{2\pi}{3}+2\,\pi\,n_1\,,\, \frac{4\pi}{3}+2\,\pi\,n_2\)\;;\qquad \qquad \gamma_1=\gamma_2=\gamma_3=1,
\nonumber\\ 
\(\vp_1^{\rm (vac.)}\,,\, \vp_2^{\rm (vac.)}\)&=&\(\frac{4\pi}{3}+2\,\pi\,n_1\,,\, \frac{2\pi}{3}+2\,\pi\,n_2\)\;;\qquad \qquad n_1\,,\,n_2\in \IZ.\lab{othervacuasu3}
\er

%%%%%%%%%%%%%%%%%%%%%%%%%%%%%%%%%%%%%%%%%%%%%
\subsection{$SO(5)$}
%%%%%%%%%%%%%%%%%%%%%%%%%%%%%%%%%%%%%%%%%%%%%

The rank of $SO(5)$ is also two and so again we have two fields, $\vp_1$ and $\vp_2$. In this case, we take the matrix $\eta_{ab}$ to be of the form\footnote{Note that $\eta_{ab}\mid_{\lambda=1}=K_{ab}\,2/{\vec \alpha}_a^2$, where $K_{ab}$ is the Cartan matrix for $SO(5)$, given in \rf{cartanso5}, and where we have normalized the roots as ${\vec \alpha}_1^2=1$ and ${\vec \alpha}_2^2=2$. }
\br
\eta=\(
\begin{array}{cc}
4&-2\,\lambda\\
-2\,\lambda &2
\end{array}\),
\qquad\qquad\qquad
\eta^{-1}=\frac{1}{2\(2-\lambda^2\)}\(
\begin{array}{cc}
1&\lambda\\
\lambda&2
\end{array}\).
\lab{etamatricesso5lambda}
\er

Note that the  eigenvalues of $\eta$ are $3\pm\sqrt{1+4\,\lambda^2}$. Thus, to keep $\eta$ positive definite and invertible, we have to restrict $\lambda$ to the interval $-\sqrt{2}< \lambda<\sqrt{2}$. 

We consider here the case of the adjoint representation of $SO(5)$ in which case we can use, as weights, all the 8 roots (positive and negative) plus the zero weight which is doubly degenerate. However, the zero weights are irrelevant because in our construction they lead to additive constants in the pre-potential $U$. The roots of $SO(5)$ are ${\vec \alpha}_1$, ${\vec \alpha}_2$, ${\vec \alpha}_2+{\vec \alpha}_1$, ${\vec \alpha}_2+2\,{\vec \alpha}_1$, and their negatives, where ${\vec \alpha}_1$ and ${\vec \alpha}_2$ are the simple roots, with ${\vec \alpha}_1$ being the shorter simple root. They satisfy
\br
K_{ab}=\frac{2\,{\vec \alpha}_a\cdot{\vec \alpha_b}}{{\vec \alpha_b}^2}=\(
\begin{array}{cc}
2&-1\\
-2 &2
\end{array}\).
\lab{cartanso5}
\er
The set of weights ${\cal R}^{(+)}$ in \rf{prepotconstruct2} are the positive roots of $SO(5)$, that we order as ${\vec \mu}_1={\vec \alpha}_2+2\,{\vec \alpha}_1$, ${\vec \mu}_2={\vec \alpha}_2+{\vec \alpha}_1$, 
${\vec \mu}_3={\vec \alpha}_1$ and ${\vec \mu}_4={\vec \alpha}_2$, and so, using \rf{phivectorb} we get
\be
{\vec \mu}_1\cdot{\vec \vp}=2\vp_1\;;\qquad 
{\vec \mu}_2\cdot{\vec \vp}=\vp_2\;;\qquad
{\vec \mu}_3\cdot{\vec \vp}=2\vp_1-\vp_2\;;\qquad
{\vec \mu}_4\cdot{\vec \vp}=2\vp_2-2\vp_1.
\ee

Therefore, taking the $\delta$-terms to vanish in \rf{prepotconstruct2}, we get the following pre-potential  
\be
U= \gamma_1\,\cos\(2\vp_1\)+\gamma_2\, \cos\(\vp_2\)+\gamma_3\, \cos\(2\vp_1-\vp_2\)+\gamma_4\,\cos\(2\vp_2-2\vp_1\).
\lab{adjointuso5new}
\ee
The corresponding self-duality equations \rf{bpseq} now become 
\br
\partial_x \vp_1&=&\pm \frac{1}{2\(2-\lambda^2\)}\left[\(\lambda-2\)\,\gamma_3\, \sin\(2\vp_1-\vp_2\)-2\,\(\lambda-1\)\,\gamma_4\,\sin\(2\vp_2-2\vp_1\)
\right. \nonumber\\
&-&\left. 2\,\gamma_1\,\sin\(2\vp_1\)-\lambda\,\gamma_2\, \sin\(\vp_2\)
\right],
\lab{bpseqso5adj}\\
\partial_x \vp_2&=&\pm \frac{1}{2\(2-\lambda^2\)}\left[-2\,\(\lambda-1\)\, \gamma_3\, \sin\(2\vp_1-\vp_2\)
+2\,\(\lambda-2\)\,\gamma_4\,\sin\(2\vp_2-2\vp_1\)
\right. \nonumber\\
&-& \left. 2\,\lambda\,\gamma_1\,\sin\(2\vp_1\)-2\,\gamma_2\, \sin\(\vp_2\)
\right].
\nonumber
\er
%%%%%%%%%%%%%%%%%%%%%%%%%%%%%%%%%%%%%%%%%%%%%
%							FIGURE 3								      %
%%%%%%%%%%%%%%%%%%%%%%%%%%%%%%%%%%%%%%%%%%%%%
\begin{figure}[h!]
  \centering
  \subfigure[]{\includegraphics[width=0.48\textwidth,height=0.38\textwidth, angle =0]{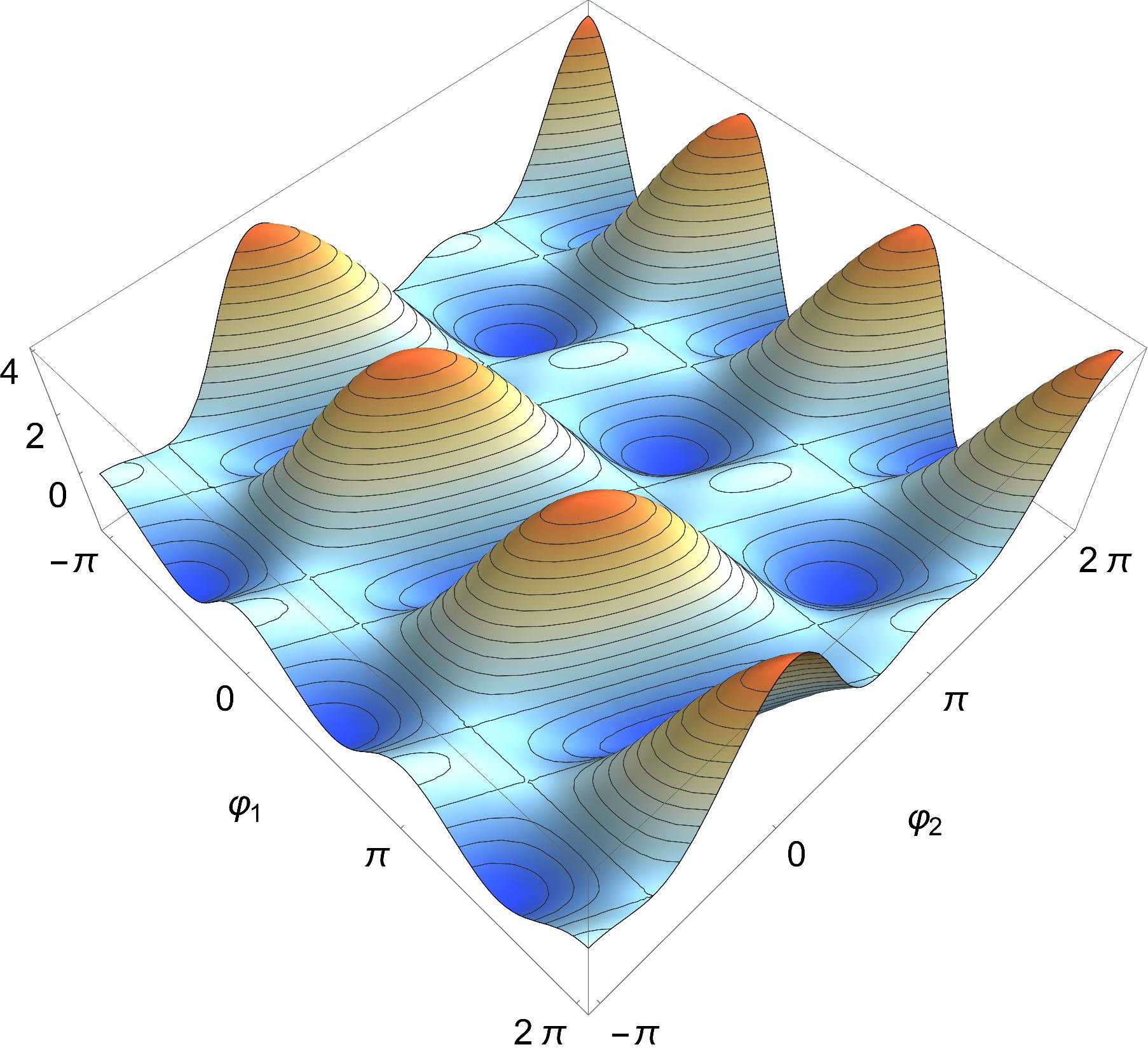}}               
  \subfigure[]{\includegraphics[width=0.48\textwidth,height=0.38\textwidth, angle =0]{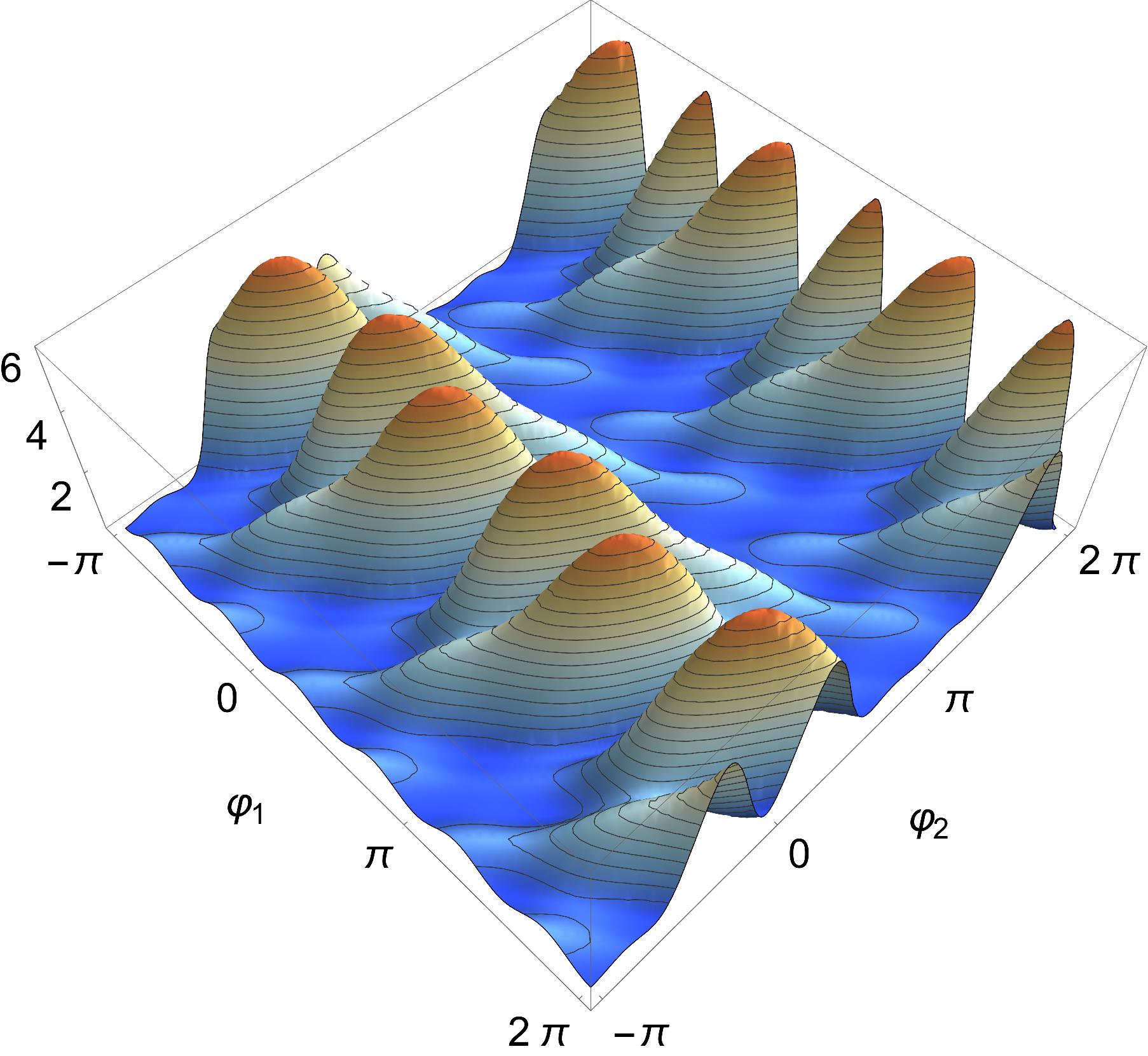}}
  \subfigure[]{\includegraphics[width=0.48\textwidth, angle =0]{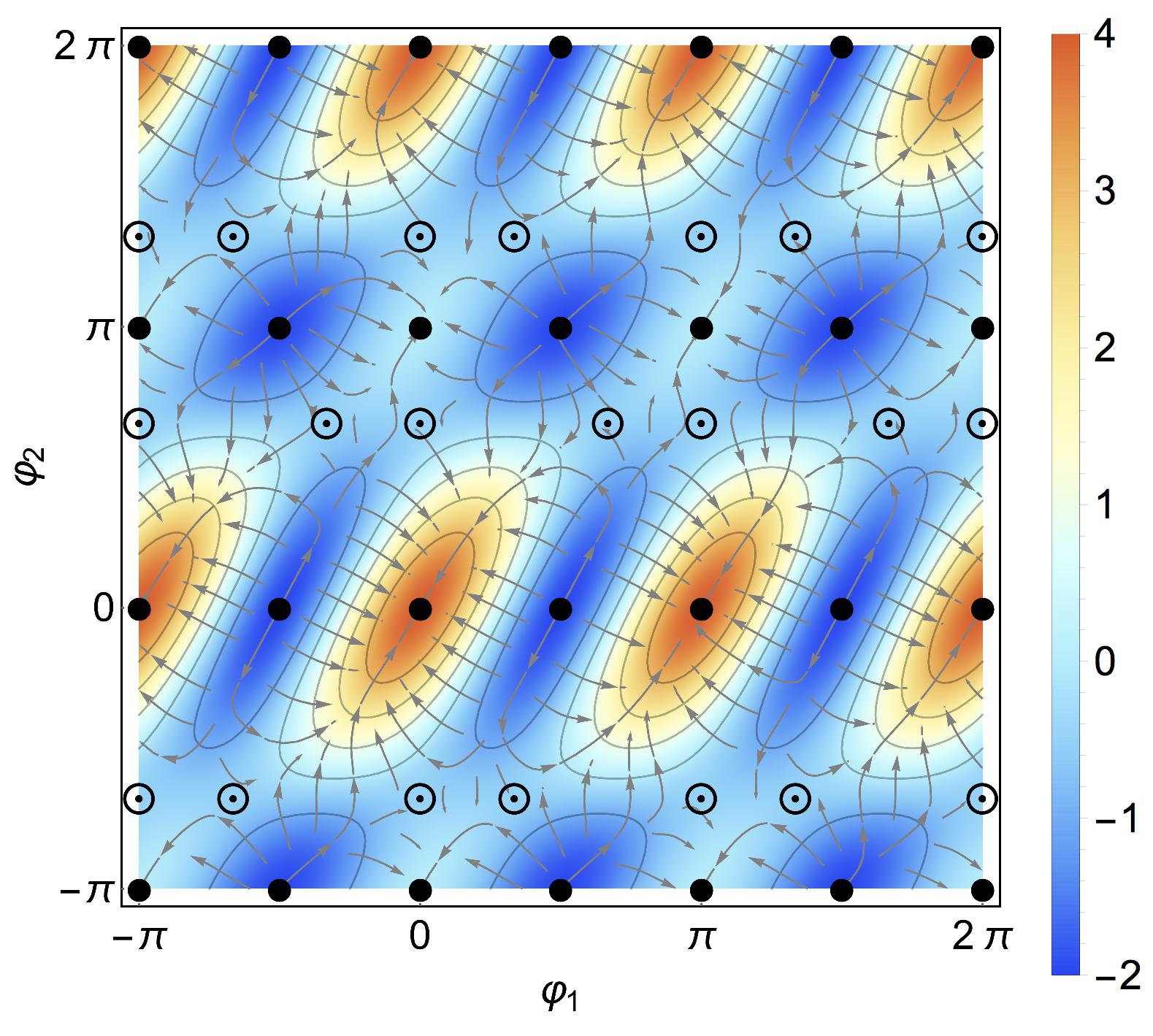}}
  \subfigure[]{\includegraphics[width=0.48\textwidth, angle =0]{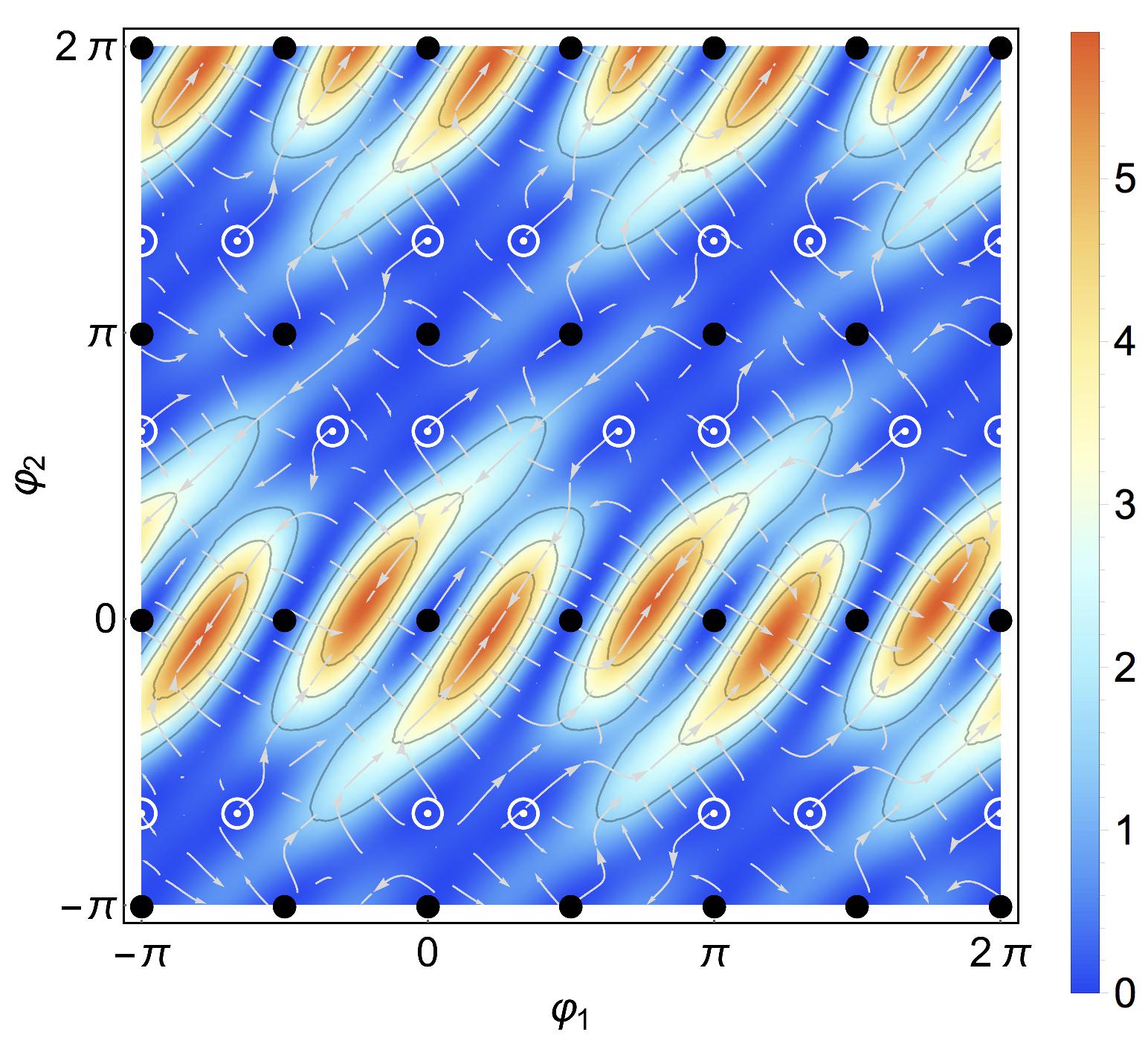}}
  \caption{(a), (c) The pre-potential \rf{adjointuso5new} and (b), (d) the potential \rf{potentialso5}. The vacua \rf{truevacuaso5} are given by the black dots, \rf{othervacuaso5} by $\odot$ and the lines stand for the gradient flow of $\vec\nabla U$ and $\vec\nabla V$, where $\vec\nabla=(\partial_{\varphi_1},\partial_{\varphi_2})$.
 Here $\gamma_1=\gamma_2=\gamma_3=\gamma_4=1$ and $\lambda=\frac{1}{4}$. }
  \lab{}
\end{figure}
%%%%%%%%%%%%%%%%%%%%%%%%%%%%%%%%%%%%%%%%%%%%%
The energy functional \rf{energyfunc} takes the form 
\be
E=\int_{-\infty}^{\infty} dx\, \left[2\,\(\partial_{x}\vp_1\)^2+\(\partial_{x}\vp_2\)^2-2\,\lambda \(\partial_{x}\vp_1\)\(\partial_{x}\vp_2\) + V\(\vp_1\,,\,\vp_2\) \right]
\ee
with the potential $V$ being given by
%%%%%%%%%%%%%%%%%%
\begin{comment}
\br
V&=&\frac{1}{4\(2-\lambda^2\)}\left[
4\gamma_1^2 \sin ^2(2 \vp_1)+4\gamma_1 \gamma_2 \lambda \sin (2 \vp_1)
   \sin (\vp_2)-4\gamma_1 \gamma_3 \lambda \sin (2 \vp_1) \sin (2
   \vp_1-\vp_2)
   \right.\nonumber\\
   &+& \left. 8\gamma_1 \gamma_3 \sin (2 \vp_1) \sin (2
   \vp_1-\vp_2)-8\gamma_1 \gamma_4 \lambda \sin (2 \vp_1) \sin (2
   \vp_1-2 \vp_2)
\right.\nonumber\\
   &+& \left.    
   8\gamma_1 \gamma_4 \sin (2 \vp_1) \sin (2 \vp_1-2
   \vp_2)+2 \gamma_2^2 \sin ^2(\vp_2)+4 \gamma_2 \gamma_3 \lambda \sin
   (\vp_2) \sin (2 \vp_1-\vp_2)
   \right.\nonumber\\
   &-& \left. 
   4 \gamma_2 \gamma_3 \sin (\vp_2) \sin
   (2 \vp_1-\vp_2)+4 \gamma_2 \gamma_4 \lambda \sin (\vp_2) \sin (2
   \vp_1-2 \vp_2)
   \right.\nonumber\\
   &-& \left. 
   8 \gamma_2 \gamma_4 \sin (\vp_2) \sin (2 \vp_1-2
   \vp_2)-4 \gamma_3^2 \lambda \sin ^2(2 \vp_1-\vp_2)+6 \gamma_3^2 \sin
   ^2(2 \vp_1-\vp_2)
   \right.\nonumber\\
   &-& \left. 
   12 \gamma_3 \gamma_4 \lambda \sin (2 \vp_1-2
   \vp_2) \sin (2 \vp_1-\vp_2)+16 \gamma_3 \gamma_4 \sin (2 \vp_1-2
   \vp_2) \sin (2 \vp_1-\vp_2)
   \right.\nonumber\\
   &-& \left. 
   8 \gamma_4^2 \lambda \sin ^2(2 \vp_1-2
   \vp_2)+12 \gamma_4^2 \sin ^2(2 \vp_1-2 \vp_2)\right].
   \lab{potentialso5}
\er
\end{comment}
%%%%%%%%%%%%%%%%%%
\begin{align}
V=&\frac{1}{4\(2-\lambda^2\)}\Big[4\gamma_1^2 \sin^2(2\vp_1)+2\gamma_2^2 \sin^2(\vp_2)+2(3-2\lambda)\gamma_3^2 \sin^2(2\vp_1-\vp_2)\nonumber\\
+&4(3-2\lambda)\gamma_4^2\sin^2(2\vp_1-2\vp_2)+4\lambda \gamma_1\gamma_2\sin(2\vp_1)\sin(\vp_2)\nonumber\\
+&4(2-\lambda)\gamma_1\gamma_3\sin(2\vp_1)\sin(2\vp_1-\vp_2)+8(1-\lambda)\gamma_1\gamma_4\sin(2\vp_1)\sin(2\vp_1-2\vp_2)\nonumber\\+&4(\lambda-1)\gamma_2\gamma_3\sin(\vp_2)\sin(2\vp_1-\vp_2)+4(\lambda-2)\gamma_2\gamma_4\sin(\vp_2)\sin(2\vp_1-2\vp_2)\nonumber\\+&
4(4-3\lambda)\gamma_3\gamma_4\sin(2\vp_1-\vp_2)\sin(2\vp_1-2\vp_2)\Big]. \lab{potentialso5}
\end{align}

The vacua conditions \rf{vacuau} in this case are given by
\br
\frac{\partial\,U}{\partial\,\vp_1}\mid_{\vp_a=\vp_a^{\rm (vac.)}}&=&-2\,\gamma_3\, \sin\(2\vp_1^{\rm (vac.)}-\vp_2^{\rm (vac.)}\)+2\,\gamma_4\,\sin\(2\vp_2^{\rm (vac.)}-2\vp_1^{\rm (vac.)}\)
\nonumber\\
&-&2\,\gamma_1\,\sin\(2\vp_1^{\rm (vac.)}\)=0,
\nonumber\\
\frac{\partial\,U}{\partial\,\vp_2}\mid_{\vp_a=\vp_a^{\rm (vac.)}}&=&\gamma_3\, \sin\(2\vp_1^{\rm (vac.)}-\vp_2^{\rm (vac.)}\)-2\,\gamma_4\,\sin\(2\vp_2^{\rm (vac.)}-2\vp_1^{\rm (vac.)}\)
\nonumber\\
&-&\gamma_2\, \sin\(\vp_2^{\rm (vac.)}\)=0.
\lab{uderivatives}
\er
Note that in  \rf{uderivatives}, $\vp_1^{\rm (vac.)}$ always appears multiplied by a factor $2$, and $\vp_2^{\rm (vac.)}$  never appears divided by any integer. Therefore, the following set of  values of fields are solutions of  \rf{uderivatives} 
\be
\vp^{{\rm vac.}}_1= m_1\, \frac{\pi}{2},\qquad\qquad\qquad 
\vp^{{\rm vac.}}_2= m_2\, \pi,\qquad\qquad\qquad m_a \in \IZ\qquad a=1,2
\lab{truevacuaso5}
\ee
for any values of the $\gamma$'s. These are the vacua of type \rf{concretechoice}, with the particularity that $2{\vec \mu}_j\cdot {\vec \alpha}_1/{\vec \alpha}_1^2$ is not odd for any weight, and so $\vp_1$ can be integer, as well as  half integer, multiples of $\pi$.

The vacua of the type \rf{rarechoice} depend upon the values of the $\gamma$'s. For instance, if we take all $\gamma$'s to be unity we have the following vacua:
\br
\(\vp_1^{{\rm vac.}}\,,\,\vp_2^{{\rm vac.}}\)&=& \(\frac{2\,\pi}{3}+2\,\pi\,n_1\,,\, \frac{2\,\pi}{3}+2\,\pi\,n_2\),
\nonumber\\
\(\vp_1^{{\rm vac.}}\,,\,\vp_2^{{\rm vac.}}\)&=&\(\pi+2\,\pi\,n_1\,,\, \frac{2\,\pi}{3}+2\,\pi\,n_2\),
\nonumber\\
\(\vp_1^{{\rm vac.}}\,,\,\vp_2^{{\rm vac.}}\)&=&\(2\,\pi\,n_1\,,\, \frac{2\,\pi}{3}+2\,\pi\,n_2\)
\qquad \qquad \qquad
\gamma_1=\gamma_2=\gamma_3=\gamma_4=1,
\nonumber\\
\(\vp_1^{{\rm vac.}}\,,\,\vp_2^{{\rm vac.}}\)&=&\(2\,\pi\,n_1\,,\, \frac{4\,\pi}{3}+2\,\pi\,n_2\)
\qquad\qquad \qquad n_1\,,\,n_2\in\IZ,
\nonumber\\
\(\vp_1^{{\rm vac.}}\,,\,\vp_2^{{\rm vac.}}\)&=&\(\frac{\pi}{3}+2\,\pi\,n_1\,,\, \frac{4\,\pi}{3}+2\,\pi\,n_2\),
\nonumber\\
\(\vp_1^{{\rm vac.}}\,,\,\vp_2^{{\rm vac.}}\)&=&\(\pi+2\,\pi\,n_1\,,\, \frac{4\,\pi}{3}+2\,\pi\,n_2\),
\nonumber\\
\(\vp_1^{{\rm vac.}}\,,\,\vp_2^{{\rm vac.}}\)&=&\(\frac{4\pi}{3}+2\,\pi\,n_1\,,\, \frac{4\,\pi}{3}+2\,\pi\,n_2\),
\nonumber\\
\(\vp_1^{{\rm vac.}}\,,\,\vp_2^{{\rm vac.}}\)&=&\(\frac{5\pi}{3}+2\,\pi\,n_1\,,\, \frac{2\,\pi}{3}+2\,\pi\,n_2\).\lab{othervacuaso5}
\er

%%%%%%%%%%%%%%%%%%%%%%%%%%%%%%%%%%%%%%%%%%%%%%
 \section{Geometric interpretation of the BPS solutions}
\label{sec:geometry}
\setcounter{equation}{0}
%%%%%%%%%%%%%%%%%%%%%%%%%%%%%%%%%%%%%%%%%%%%%%

As we have seen in \rf{vacuaconditions} and \rf{vacuau}, the finite energy solutions of the self-duality equations \rf{bpseq} have to go to  constant vacua solutions for $x\rightarrow \pm \infty$.  Therefore, each of these solutions connect two vacua of the theory. In order to have a geometric picture of these solutions let us write the self-duality equations \rf{bpseq} as 
\be
{\vec v}=\pm {\vec \nabla}_{\eta}U\;; \qquad\qquad {\rm with}\qquad \({\vec v}\)_a =\frac{d\,\vp_a}{d\,x}\;;\qquad 
\({\vec \nabla}_{\eta}U\)_a= \eta^{-1}_{ab}\frac{\delta\, U}{\delta\,\vp_b}.
 \lab{equiv_bpseq}
 \ee

 Given the pre-potential $U$ and the metric $\eta_{ab}$, which we assume real, constant and positive definite, the $\eta$-gradient of $U$ defines curves in the space of $\varphi_1,\ldots,\varphi_r$, with ${\vec \nabla}_{\eta}U$ being the tangent vector to these curves. The curves never intersect each other, since otherwise ${\vec \nabla}_{\eta}U$ would not be uniquely defined on a given point in $\vp$-space.  They can at most touch each other tangentially, or meet at points where ${\vec \nabla}_{\eta}U$ vanishes. The self-duality equation is a first order partial differential equation and so a given solution is determined by the values of the fields $\vp_a$ at a given point $x=x_0$.\footnote{We would like to thank 
Nick Manton for his suggestion, in our discussions with him, to look at the flow of the prepotential $U$.} Given the choice of values $\vp_a\(x_0\)$ one selects a point in the $\vp$-space and so a curve defined by the $\eta$-gradient of $U$. This choice of the curve is unique as long as the values of $\vp_a\(x_0\)$ do not correspond to a point where ${\vec \nabla}_{\eta}U$ vanishes, or to a point where two curves touch tangentially. The self-duality equation `says' that the solution `travels' along this curve with $x$-velocity ${\vec v}$ which is equal to the the $\eta$-gradient of $U$, or the negative of it. The geometric picture is therefore that of a particle traveling in the $\vp$-space with $x$-velocity ${\vec v}$, and with the space coordinate $x$ playing the role of time. Therefore, the problem of solving the self-duality equation \rf{bpseq} reduces to that of constructing the curves in the $\vp$-space determined by the $\eta$-gradient of $U$. Any particular solution corresponds to a particular curve  determined by the initial values $\vp_a\(x_0\)$. The finite energy solutions correspond to the curves that start and end at the extrema of the pre-potential $U$, {\it i.e.} at the points where ${\vec \nabla}_{\eta}U$ vanishes. 
 
Note that a given curve determined by the  $\eta$-gradient of $U$ cannot intersect itself, since otherwise the value of ${\vec \nabla}_{\eta}U$ would not be uniquely determined at the point of the intersection. A given curve can at most touch itself tangentially at a given point. However, as we show below, if $\eta$ is a positive definite matrix, a given curve can not close on itself at a point where ${\vec \nabla}_{\eta}U$ vanishes. For the case of a theory with just one field $\vp$, like the sine-Gordon model, the curves determined the $\eta$-gradient of $U$ live in a one dimensional space. Therefore, if a given curve starts (at $x=-\infty$) at a given vacuum, it either stays there all the 'time' and so is reduced to a point, or it is bound to end (at $x=\infty$) at a different vacuum.  Consequently, the profile function $\vp\(x\)$, that is a solution of the self-duality equation, has to be a monotonic function of $x$. %This explains why the two-soliton solution of such theories, corresponding to a soliton and an anti-soliton, and starting and ending at the same vacua, can not be self-dual solutions. 
This is indeed the case for the sine-Gordon model. For the case of a theory with several fields this is no longer the case. Indeed, the profile functions $\vp_a\(x\)$ are projections of the curves in $\vp$-space, determined by the $\eta$-gradient of $U$, onto the $\vp_a$-axis, and so they are not bound to be monotonic functions of $x$. Indeed, this is what we have observed in our numerical simulations. 

Consider now a given curve $\gamma$ in the $\vp$-space, parameterized by $x$, {\it i.e.} $\vp_a\(x\)$, which is a solution of the self-duality equation \rf{bpseq}, and associated to this curve define the quantity
\be
{\cal Q}\(\gamma\)=\int_{\gamma}dx\,{\vec v}\cdot {\vec \nabla}U= \int_{\gamma}dx\,\frac{d\,\vp_a}{d\,x}\,\frac{\delta\,U}{\delta\,\vp_a}=U\(x_f\)-U\(x_i\),
\ee
where $x_f$ and $x_i$ correspond to the final and initial points respectively, of the curve $\gamma$. Note that the tangent vector to this curve is ${\vec \nabla}_{\eta}U$ and not the ordinary gradient of $U$, {\it i.e.} ${\vec \nabla}U$, since the curve is a solution of the self-duality equations \rf{bpseq}. From these self-duality equations we see that
\be
{\cal Q}\(\gamma\)=\pm \int_{\gamma}dx\,\eta_{ab}\,\frac{d\,\vp_a}{d\,x}\,\frac{d\,\vp_b}{d\,x}=\pm \int_{\gamma}dx\,\omega_a\(\frac{d\,{\tilde \vp}_a}{d\,x}\)^2,
\lab{qgammadef}
\ee
where we have diagonalized the matrix $\eta$, {\it i.e.}
\be
\eta=\Lambda^{T}  \,\eta^D\, \Lambda\;;\qquad\qquad\Lambda^T\,\Lambda=\one\;; \qquad\qquad \eta^D_{ab}=\omega_a \,\delta_{ab}\;;\qquad\qquad \omega_a>0
\lab{diagonaletadef}
\ee
and have assumed that the eigenvalues of $\eta$ are all positive, and have defined ${\tilde \vp}_a=\Lambda_{ab}\,\vp_b$. Under the assumption that $\eta$ is positive definite, one observes that ${\cal Q}\(\gamma\)$ can only vanish  if the fields are constant along the whole curve, or in other words, if the curve is just a point. Therefore, the solutions of the self-duality equations cannot start and end on points in the $\vp$-space, where the the pre-potential $U$ has the same value.  In fact, there is more to this. As one progresses along the curve, the difference between the value of the  pre-potential $U$ at this particular point and at the initial point, only increases in modulus. This means that the curve, that is a solution of the self-duality equations \rf{bpseq}, climbs the pre-potential $U$, either upwards or downwards, without ever returning to an altitude that it has already passed through.  

One further observation one can make concerning this geometric picture of a particle moving in the $\vp$-space is that there is a quantity conserved in the `time' $x$, namely 
\be
{\cal E}=\frac{1}{2}\,\eta_{ab}\,\frac{d\,\vp_a}{d\,x}\, \frac{d\,\vp_b}{d\,x}-V
\ee
with $V$ given by \rf{potdef}. Indeed, assuming $\eta_{ab}$ to be constant, one gets from the self-duality equation \rf{bpseq} that $\frac{d^2\,\vp_a}{d\,x^2}=\eta_{ab}^{-1}\frac{\delta\,V}{\delta\,\vp_b}$.
Then using \rf{nicerel1} we see that
\be
\frac{d\,{\cal E}}{d\,x}=0.
\ee

However, the self-duality equation implies that such a quantity has to vanish on the self-dual solutions, {\it i.e.} ${\cal E}=0$. Such a result resembles what one has for the Euclidean Yang-Mills instanton solutions, that are also self-dual. The solutions of our self-duality equations \rf{bpseq} correspond to zero-energy `pseudo-particles' evolving in an imaginary time $\tau=i\,x$, and tunneling between vacua.

As an example of our  geometrical interpretation of the BPS solution we look at the model 
involving the $SU(3)$ triplet-anti-triplet case, in which the arbitrary potential parameters
 $\gamma_1$, $\gamma_2$ and $\gamma_3$ have been chosen to take the values $\gamma_1=\gamma_2=\gamma_3=1$.  Since the matrix $\eta^{-1}_{ab}$ is a function of the coupling parameter $\lambda$ we discuss here three different cases: $\lambda=0$ and $\lambda=\pm 1.8$. In all presented examples we have taken the upper sign in \rf{equiv_bpseq} {\it i.e.} ${\vec v}= {\vec \nabla}_{\eta}U$.

%%%%%%%%%%%%%%%%%%%%%%%%%%%%%%%%%%%%%%%%%%%%%
%							FIGURE 4							      %
%%%%%%%%%%%%%%%%%%%%%%%%%%%%%%%%%%%%%%%%%%%%%
\begin{figure}[h!]
  \centering
  \subfigure[]{\includegraphics[width=0.45\textwidth, angle =0]{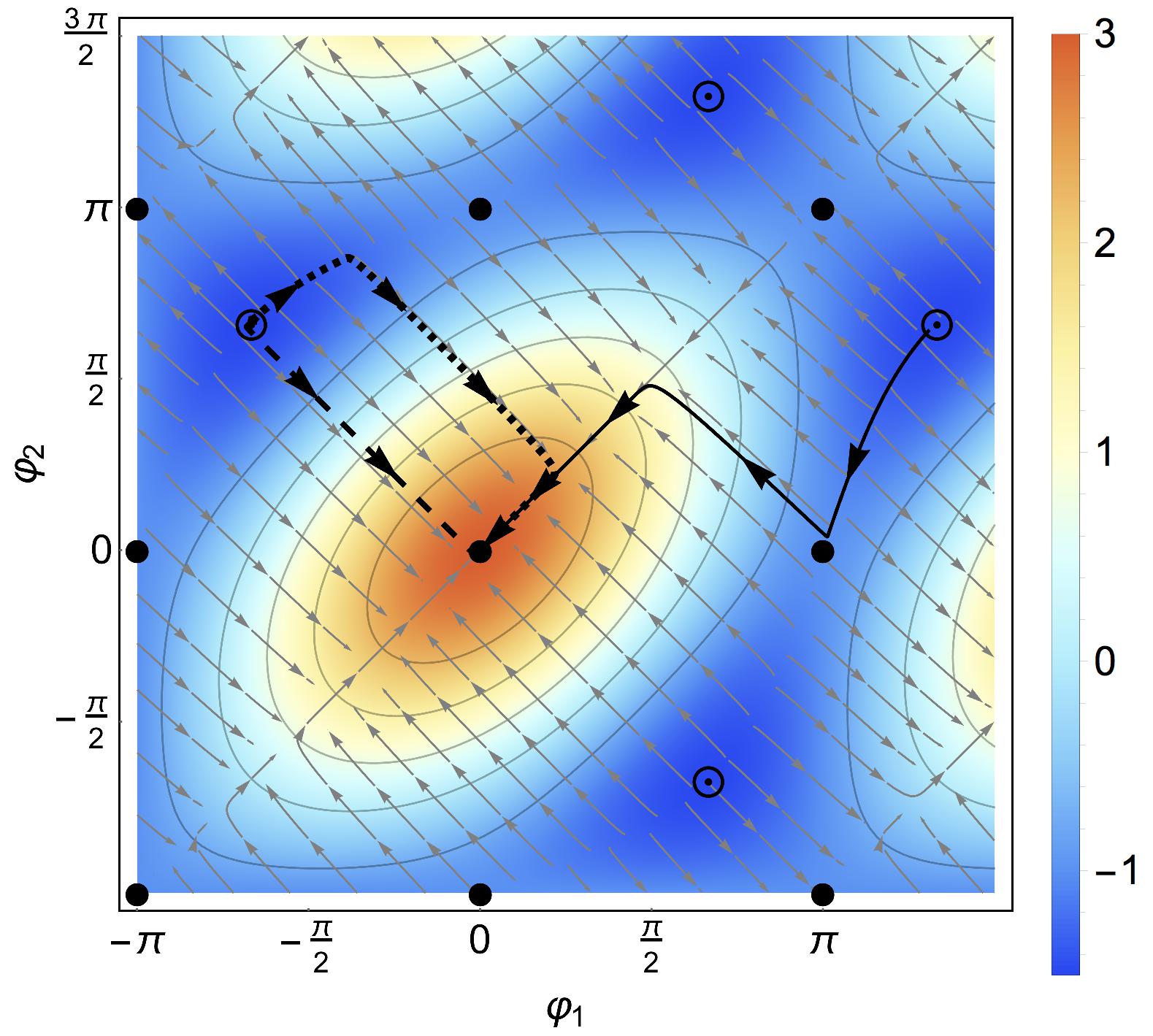}}               
  \subfigure[]{\includegraphics[width=0.45\textwidth, angle =0]{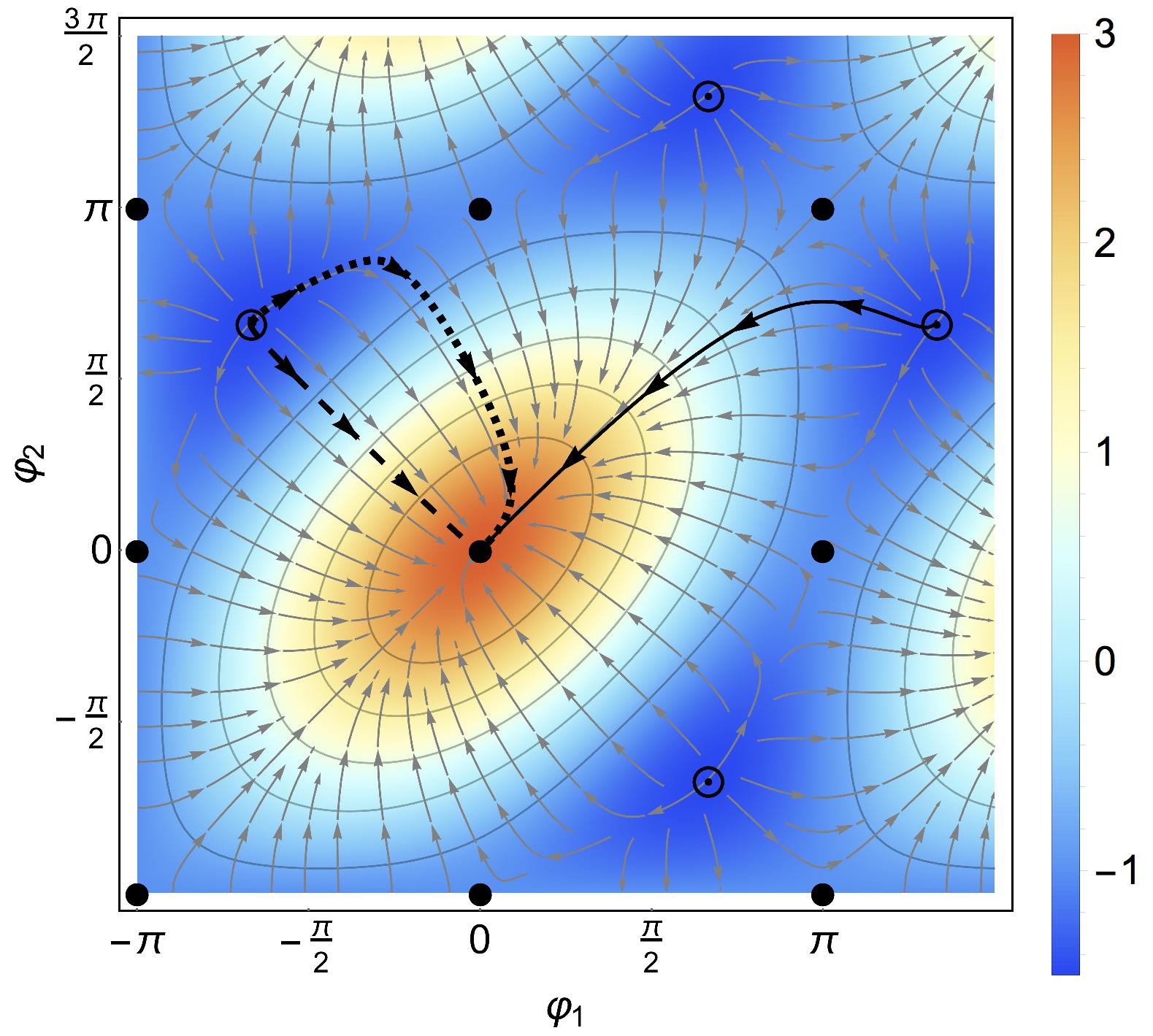}}
  \subfigure[]{\includegraphics[width=0.45\textwidth, angle =0]{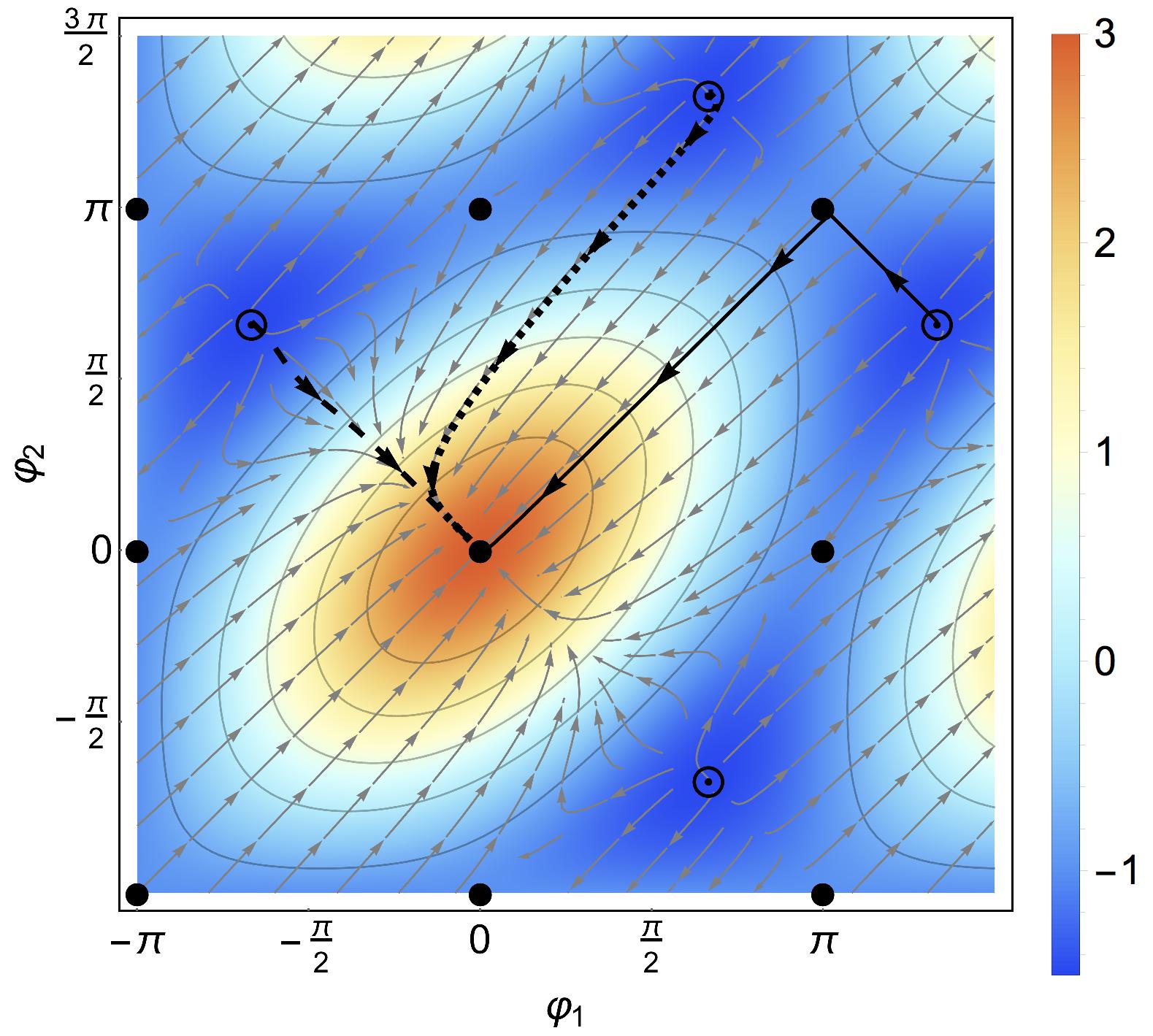}}
  \caption{The pre-potential $U(\vp_1,\vp_2)$ for the case of $SU(3)$ in which 
 $\gamma_1=\gamma_2=\gamma_3=1$ and the coupling parameter $\lambda$ takes values (a) $\lambda=-1.8$, 
(b) $\lambda=0$ and (c) $\lambda=1.8$. 
 These three examples of the numerical BPS solutions correspond to the curves that connect extrema 
of the pre-potential. The flow of ${\vec \nabla}_{\eta}U$ is depicted by tiny oriented lines.}
  \label{flowGamma}
\end{figure}
%%%%%%%%%%%%%%%%%%%%%%%%%%%%%%%%%%%%%%%%%%%%%

%%%%%%%%%%%%%%%%%%%%%%%%%%%%%%%%%%%%%%%%%%%%%
%							FIGURE 5							      %
%%%%%%%%%%%%%%%%%%%%%%%%%%%%%%%%%%%%%%%%%%%%%
\begin{figure}[h!]
  \centering
  \subfigure[]{\includegraphics[width=0.45\textwidth, angle =0]{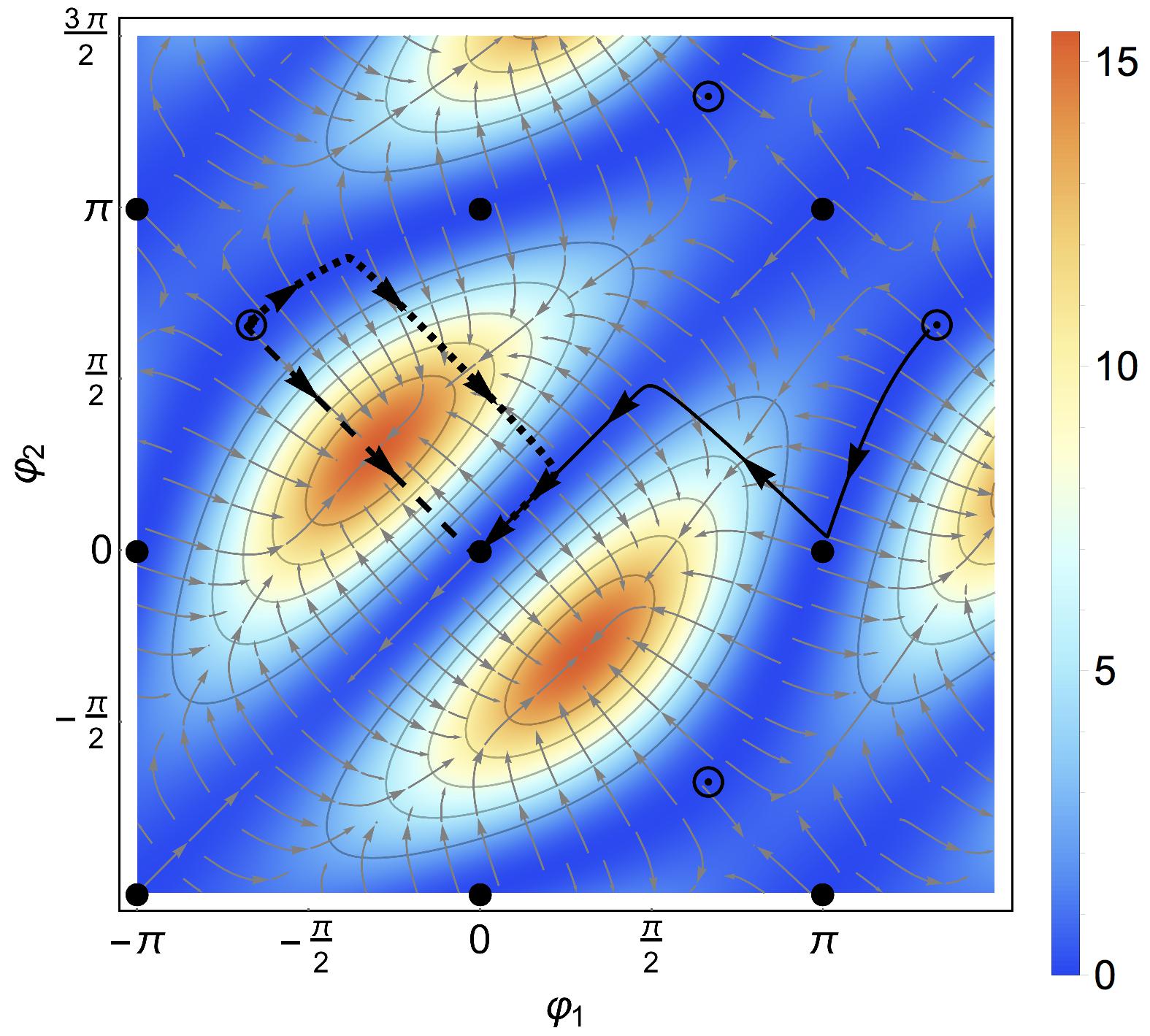}}               
  \subfigure[]{\includegraphics[width=0.45\textwidth, angle =0]{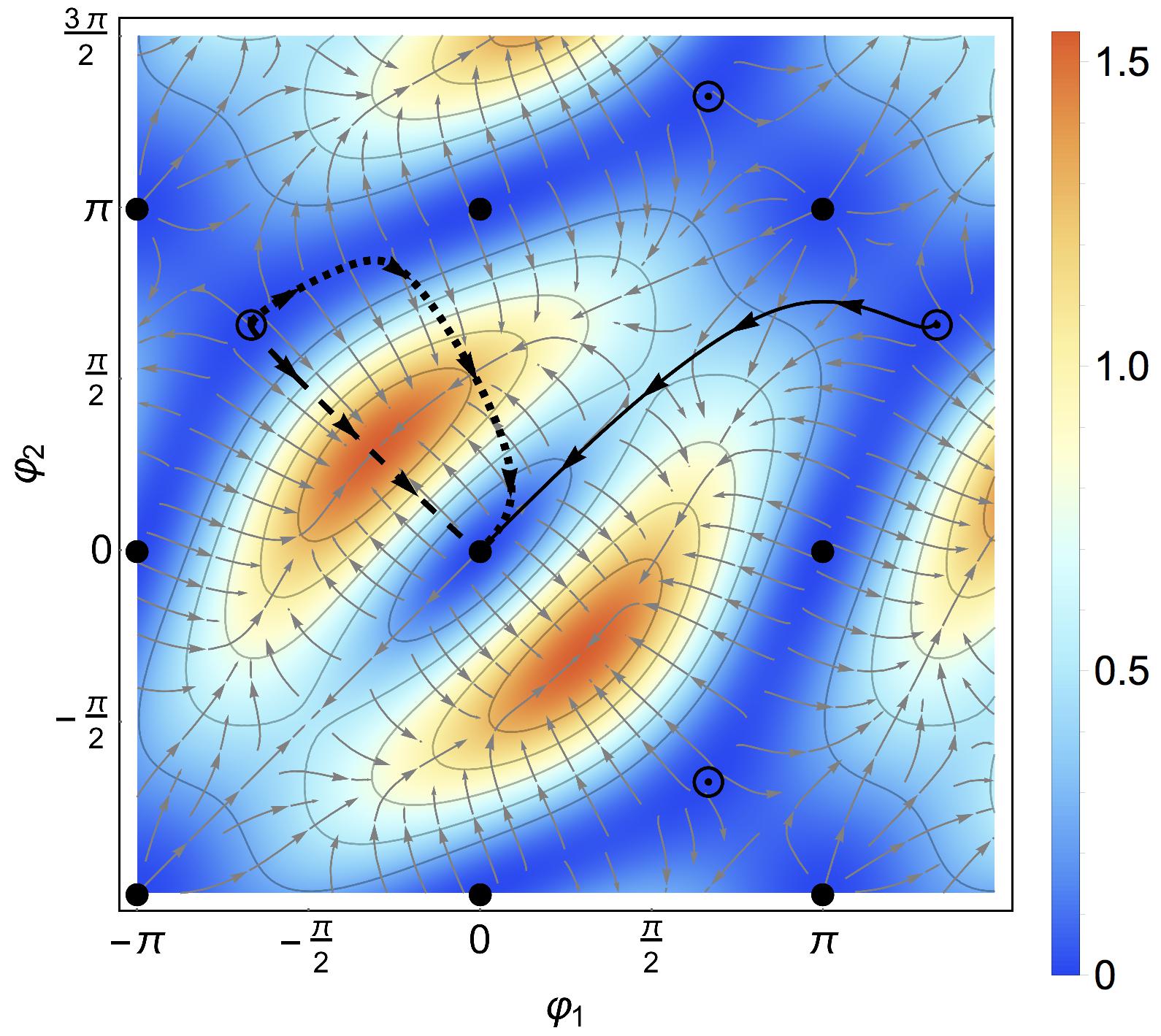}}
  \subfigure[]{\includegraphics[width=0.45\textwidth, angle =0]{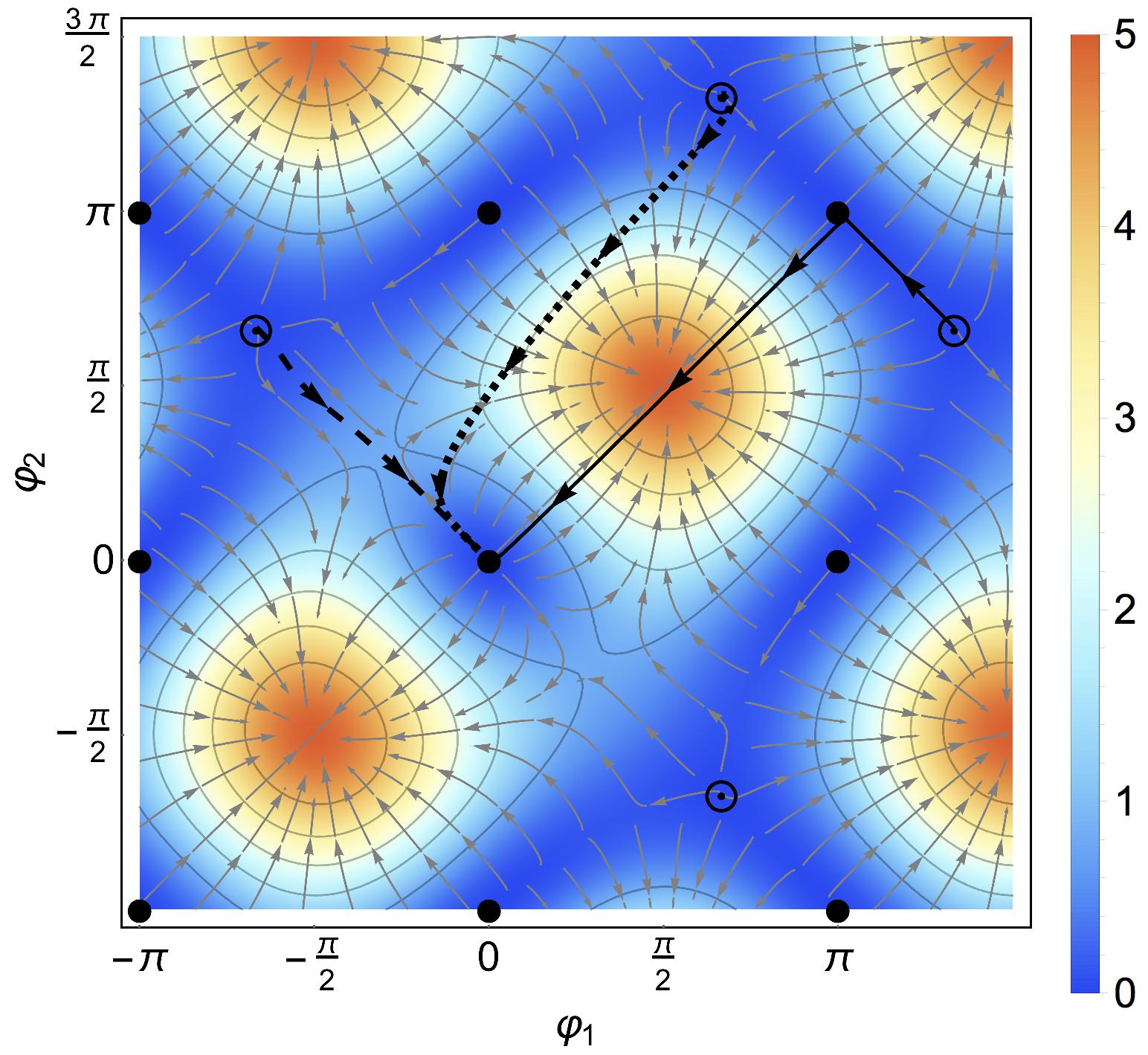}}
  \caption{The potential $V(\varphi_1,\varphi_2)$ and its gradient-flow $\vec\nabla V$ for
 the case of $SU(3)$ in which $\gamma_1=\gamma_2=\gamma_3=1$ and the coupling parameter $\lambda$ 
 takes the values (a) $\lambda=-1.8$, (b) $\lambda=0$ and (c) $\lambda=1.8$.}
  \label{gradflowV}
\end{figure}
%%%%%%%%%%%%%%%%%%%%%%%%%%%%%%%%%%%%%%%%%%%%%

In Fig.\ref{flowGamma} we present the plots, for these three values of $\lambda$, of the pre-potential $U$   and of the lines of the vector
 field ${\vec \nabla}_{\eta}U$. In each picture we have plotted three curves, each one for one numerical solution of the self-dual
 equations. Note that there is only one curve that passes through each point  not being an extremum 
(maximum, minimum or a saddle point) of the pre-potential.  The numerical curves follow very closely the lines
 of ${\vec \nabla}_{\eta}U$-flow in all three cases. The gradient flow $\vec\nabla U$ is the same in all
three cases. In the case of $\lambda=0$ one gets
 $\eta_{ab}^{-1}=\frac{1}{2}\delta_{ab}$ so both flows, gradient and  the ${\vec \nabla}_{\eta}U$-flow,
are proportional to each other and all is fine. However, for $\lambda\ne0$ the two flows are different, 
and one can easily see that the ${\vec \nabla}_{\eta}U$-flow is clearly different
 from the gradient flow  when $\lambda=\pm 1.8$.   In Fig.\ref{gradflowV} we present the picture of the potential $V$ 
and its gradient flow. 
It is quite clear from the pictures that the analysis of the potential $V$ and its gradient lines does not provide us with
 all the required information to determine the curves of the BPS solutions.

Another important point,  which can be immediately seen from ${\vec \nabla}_{\eta}U$-flow, is the presence of ``bumps'' 
in the solution $\varphi_a(x)$. The existence of such properties of the solutions has already been mentioned before and it will be 
discussed in more detail in Section \ref{sec:numerical}. Here we just note that if two vacua are connected
 by a curve which requires a non-monotonic change of fields then the ``bumps'' must necessarily occur. The number of ``bumps'' for a given BPS solution can be deduced directly from the form of ${\vec \nabla}_{\eta}U$-flow. For instance, let us look at the curves presented in Fig.\ref{flowGamma} (b). The initial point of each curve corresponds to $x=-\infty$ whereas the final one has $x=+\infty$. In the case of dashed curve the fields change monotonically with $x-$  $\vp_1$ increases and $\vp_2$ decreases - so in this case the solution has no bumps at all. On the other hand, the character of the dotted curve suggests that field $\vp_2$ has a ``bump'' with a local maximum and then $\vp_1$ has also a ``bump'' where its became positive-valued. A third (solid curve) is such that the bump occurs only for field $\vp_2$.

%%%%%%%%%%%%%%%%%%%%%%%%%%%%%%%%%%%%%%%%%%%%
 \section{The time dependent solutions}
\label{sec:maxmin}
\setcounter{equation}{0}
%%%%%%%%%%%%%%%%%%%%%%%%%%%%%%%%%%%%%%%%%%%%%

As we have shown in section \ref{sec:bps}, the solutions of the self-duality equation \rf{bpseq} are also solutions of the Euler-Lagrange equations associated to the static energy functional given in \rf{energyfunc}. Therefore, the solutions of \rf{bpseq} are static solutions of the $(1+1)$-dimensional theory defined by the action 
\be
 S= \int dt\,dx\,\left[ \frac{1}{2}\,\eta_{ab}\, \(\frac{d\,\vp_a}{d\,t}\, \frac{d\,\vp_b}{d\,t}-\frac{d\,\vp_a}{d\,x}\, \frac{d\,\vp_b}{d\,x}\)- V\right]
 \lab{lagrangiandef}
 \ee
with the potential $V$ given by \rf{potdef}. When studying non-self-dual and time dependent solutions it is important to know the properties of the potential $V$ and in particular its vacua structure. Since the potential $V$ is constructed from the pre-potential $U$ some of  these properties are easy to determine. Let us assume that $\eta$ is real, constant and positive definite, and let us diagonalize it as in \rf{diagonaletadef}. However, to make the notation clearer we absorb the eigenvalues $\omega_a$ of $\eta$ into the fields by redefining them as: 
\be
\phi_a\equiv  \sqrt{\omega_a}\,\Lambda_{ab} \,\vp_b.
\ee

 The self-duality equations \rf{bpseq} now become
\be
 \frac{d\,\phi_a}{d\,x}=\pm \,  \frac{\delta\, U}{\delta\,\phi_a}
 \lab{bpseqdiag}
 \ee
and the potential \rf{potdef} takes the form
\be
V=\frac{1}{2}\, \(\frac{\delta\, U}{\delta\,\phi_c}\)^2
\lab{potdiag}
\ee

Next we note that  
\be
\frac{\delta\, V}{\delta\,\phi_a}=\frac{\delta\, U}{\delta\,\phi_c}\,\frac{\delta^2\, U}{\delta\,\phi_c\,\delta\,\phi_a}\;;
\qquad\qquad
\frac{\delta^2\, V}{\delta\,\phi_a\,\delta\,\phi_b}=\frac{\delta^2\, U}{\delta\,\phi_b\,\delta\,\phi_c}\,\frac{\delta^2\, U}{\delta\,\phi_c\,\delta\,\phi_a}+
\frac{\delta\, U}{\delta\,\phi_c}\,\frac{\delta^3\, U}{\delta\,\phi_c\,\delta\,\phi_a\,\delta\,\phi_b}.
\ee
Thus, on the vacuum solutions, given by the extrema of $U$ (see \rf{vacuau}), we have  
\be
V\mid_{\phi_a=\phi_a^{\rm (vac.)}}=0\;;\qquad\qquad \qquad\qquad 
\frac{\delta\, V}{\delta\,\phi_a}\mid_{\phi_a=\phi_a^{\rm (vac.)}}=0
\ee
and
\be
\frac{\delta^2\, V}{\delta\,\phi_a\,\delta\,\phi_b}\mid_{\phi_a=\phi_a^{\rm (vac.)}}=\(M^2\)_{ab}\;;\qquad\qquad\qquad M_{ab}\equiv\frac{\delta^2\, U}{\delta\,\phi_a\,\delta\,\phi_b}\mid_{\phi_a=\phi_a^{\rm (vac.)}}.
\ee

Let us now Taylor expand the potential $V$ around an extrema $\phi_a^{\rm (vac.)}$ of the pre-potential $U$. Since $M$ is a real  and symmetric matrix, we find that
\be
V\(\phi\)\sim \frac{1}{2}\sum_{a,b}\left[M_{ab}\,\(\phi_b-\phi_b^{\rm (vac.)}\)\right]^2 +  O\(\(\phi-\phi^{\rm (vac.)}\)^3\)
\ee
and so, we see that $\phi_a^{\rm (vac.)}$ is a local minimum of the potential $V$. Consequently, we can  make the following statements about the potential $V$ and its relation to the pre-potential $U$: 
\begin{enumerate}
\item $V$ is non-negative, and it vanishes only at the extrema of $U$. 
\item The extrema of $U$ are extrema of $V$, but the converse may not be true. 
\item The extrema of $U$ are always minima of $V$, irrespective of being minima, maxima or saddle points of $U$. The maxima of $V$ are never extrema of $U$. 
\end{enumerate}

The self-dual solutions of \rf{bpseq} tunnel between extrema of $U$, and so between minima of the potential $V$. Therefore, we expect that there may exist finite energy, time dependent multi-soliton like solutions of the theory \rf{lagrangiandef}. %We plan to think about this further.
%Such solutions could perhaps be constructed numerically by taking as the initial configuration a superposition of self-dual solutions of \rf{bpseq} well separated from each other, and letting them evolve under the dynamics of  \rf{lagrangiandef}.
%I would not speculate here.

%%%%%%%%%%%%%%%%%%%%%%%%%%%%%%%%%%%%%%%%%%%%%%
 \section{Numerical support}
\label{sec:numerical}
\setcounter{equation}{0}
%%%%%%%%%%%%%%%%%%%%%%%%%%%%%%%%%%%%%%%%%%%%%%

In the preceding sections we have presented concrete procedures, based on representation theory of Lie groups, of constructing self-dual sectors of various real scalar field theories in $(1+1)$-dimensions. We have given examples for some representations of $SU(2)$, $SU(3)$ and $SO(5)$ groups but, with the exception of the $SU(2)$ case, we have not solved the self-duality equations \rf{bpseq} in analytical forms. The construction of analytical solutions becomes very difficult as the number of fields increases.
 Also, as it is not clear whether any of the models we have constructed are integrable, and we do
 not have analytical methods at hand to study this problem.

 Thus, in this section we present numerical solutions of the self-duality equations \rf{bpseq}. 
 The self-duality equations are first order in $x$-derivatives and so their solutions are determined
 by the initial values of the fields at a particular point in space. As we discussed
in the previous section, this point cannot be the extremum of $U$ as then the self-duality
equations do not 'evolve' the fields from their vacuum value. We have taken this point to be $x=0$, 
and solved \rf{bpseq}  first by propagating the solution along the positive $x$-axis and then along 
the negative $x$-axis. In each case we continued the solution until the fields did not change (and so 'effectively' reached a vacuum) and then glued the two branches of the evolved solutions to get the complete solution.  We have performed many such simulations, varying both the simulation step $dx$ and of the values of $x$ to which we carried the simulation (to check whether the fields really reached the vacua).  For small values of $dx$ ($dx<0.00001$) the results were essentially the same. In the plots that we include  in the next subsections, we present the results obtained for $dx=0.000002$. Moreover, in each case the solutions had essentially not changed much and so they  essentially `reached' the vacuum values. 

 We have also studied the stability of the BPS solutions. After constructing a given static self-dual solution we have used it as the initial static configuration for the Cauchy problem corresponding to the full (second order time dependent) equations of the model. The time variation of these solutions was simulated  using the 4th order Runge-Kutta method. Our simulations used double precision and were performed with absorbing boundary conditions but, in fact, the time variations of the fields at the boundaries were always extremely small and the absorption was always infinitesimal.

Of course, analytically, this was to be expected as our BPS fields were static solutions of the 
full equations and this was confirmed by the results of our simulations. However, small numerical errors
(inherent in any numerical work) could always alter any results and, in principle, they could lead
 to small evolution but we were genuinely surprised by the smallness of any changes 
(the errors had always been of the order
 of $10^{-3}\%$ and, effectively, they had not grown with the increase of the lattice). So, we have not seen any significant changes of the fields and we believe that we can trust our results. 
A bonus of these studies was the confirmation of the stability
of the solutions, at least with respect to small perturbations introduced
by the numerical errors. 

Thus, in the cases we have studied, we have found that the self-dual solutions are, as expected, stable, and do not send any radiation out, to lower their energies, confirming that they were, indeed, minima of the energy for the corresponding sector of the topological charge.

%%%%%%%%%%%%%%%%%%%%%%%%%%%%%%%%%%%%%%%%%%
\subsection{$SU(3)$ Simulations}
%%%%%%%%%%%%%%%%%%%%%%%%%%%%%%%%%%%%%%%%%%

In the $SU(3)$ case we have solved numerically the self-duality equations \rf{su3bpseq33bar}, corresponding to the pre-potential \rf{usu3tripletantitriplet}, constructed 
from the triplet and anti-triplet representations. We have performed simulations for the 
following sets of values of the $\gamma_i$ and $\lambda$ parameters: 
\begin{itemize}
\item Case I): $\(\gamma_1\,,\,\gamma_2\,,\,\gamma_3\,,\, \lambda\)=\(1\,,\,\sqrt{2}/2\,,\,1\,,\, 0.5\),$
\item Case II): $\(\gamma_1\,,\,\gamma_2\,,\,\gamma_3\,,\, \lambda\)=\(1\,,\,0.5\,,\,1\,,\, 0.5\),$
\item Case III): $\(\gamma_1\,,\,\gamma_2\,,\,\gamma_3\,,\, \lambda\)=\(0.1\,,\,0.5\,,\,0.5\,,\, 0.5\).$\newline
In this case we took three different values of pairs $(\vp_1\(0\), \vp_2\(0\))$ which lead to three different solutions.
\item Case IV): $\(\gamma_1\,,\,\gamma_2\,,\,\gamma_3\,,\, \lambda\)=\(1\,,\,\sqrt{2}/2\,,\,1\,,\, 1.8\).$
\end{itemize}
In cases I and IV the vacua (minima of the potential $V(\vp_1,\vp_2)$) are given by
\begin{align}
(\vp^{({\rm vac})}_1,\vp^{({\rm vac})}_2)&=(n_1\pi,n_2\pi),\label{vacua_su3a}\\
(\vp^{({\rm vac})}_1,\vp^{({\rm vac})}_2)&=\left(\pm\frac{3\pi}{4}+2\pi\, n_1,\mp\frac{\pi}{2}+2\pi\,n_2\right).\label{vacua_su3b}
\end{align}

%%%%%%%%%%%%%%%%%%%%%%%%%%%%%%%%%%%%%%%%%%%%%
%							FIGURE 6 NEW					      %
%%%%%%%%%%%%%%%%%%%%%%%%%%%%%%%%%%%%%%%%%%%%%
\begin{figure}[h!]
\centering
\subfigure[]{\includegraphics[width=0.45\textwidth,height=0.4\textwidth, angle =0]{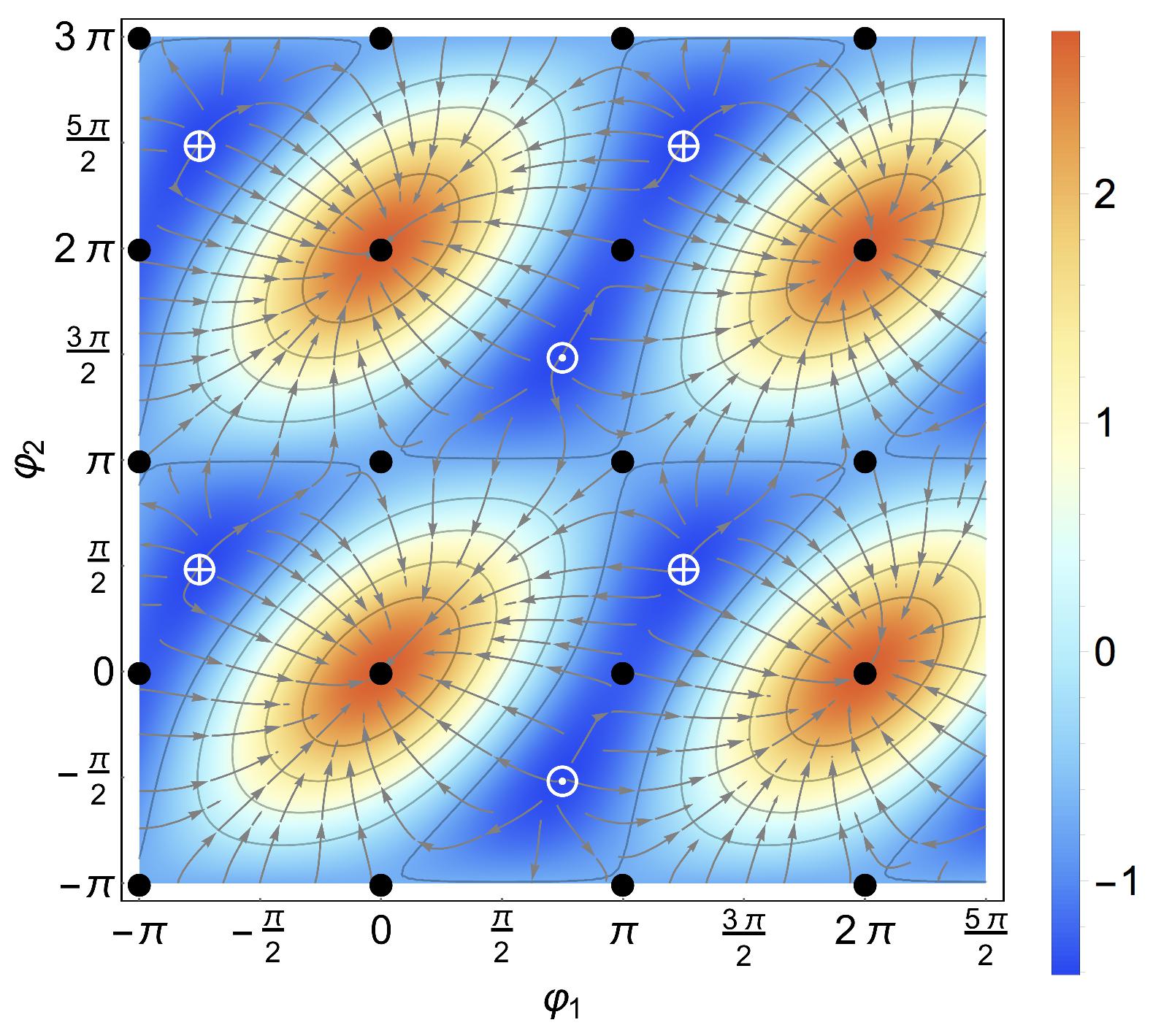}}   
\subfigure[]{\includegraphics[width=0.45\textwidth,height=0.4\textwidth, angle =0]{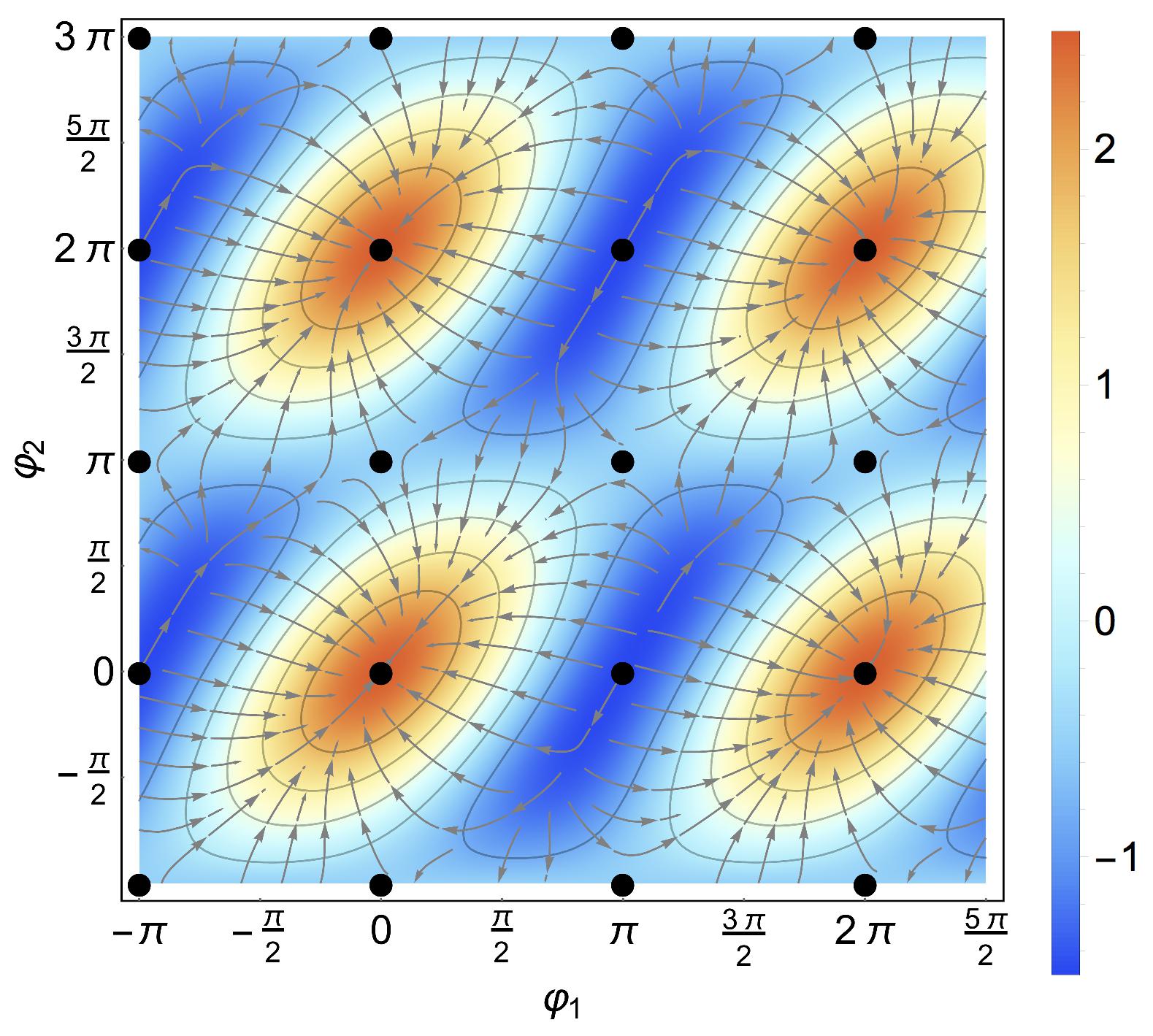}}
\caption{$SU(3)$ vacua and the pre-potential: (a) case I, (b) case  II. Black dots represent vacua \eqref{vacua_su3a}, $\odot$ stand for vacua \eqref{vacua_su3b} with the upper sign and $\oplus$ for vacua \eqref{vacua_su3b} with the lower sign. The arrows represent the  $\nabla_{\eta} U$ - flow.}\label{fig:su3vacua}
\end{figure}

\begin{figure}[h!]
\centering
\subfigure[]{\includegraphics[width=0.48\textwidth,height=0.4\textwidth, angle =0]{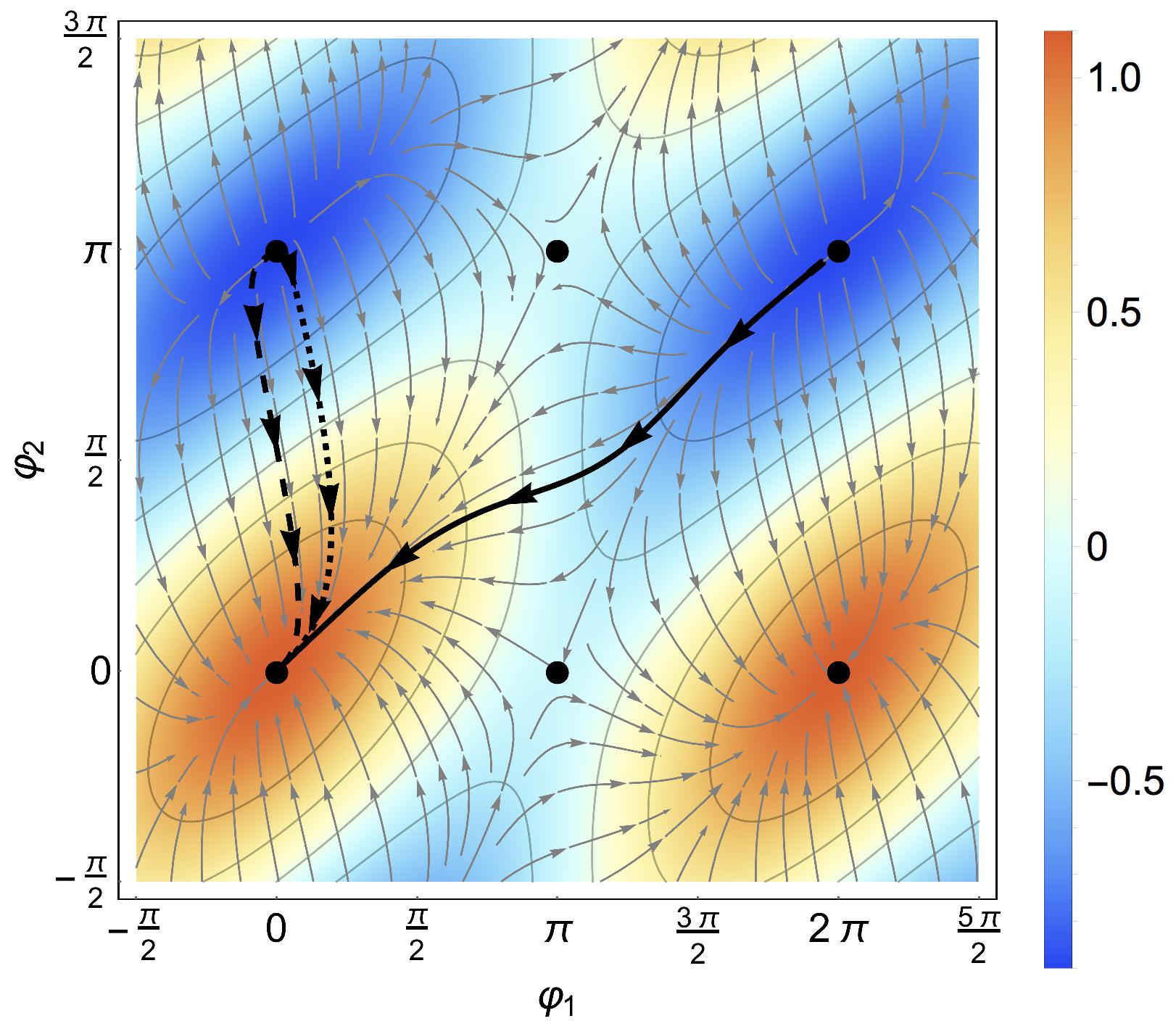}}   
\subfigure[]{\includegraphics[width=0.48\textwidth,height=0.4\textwidth, angle =0]{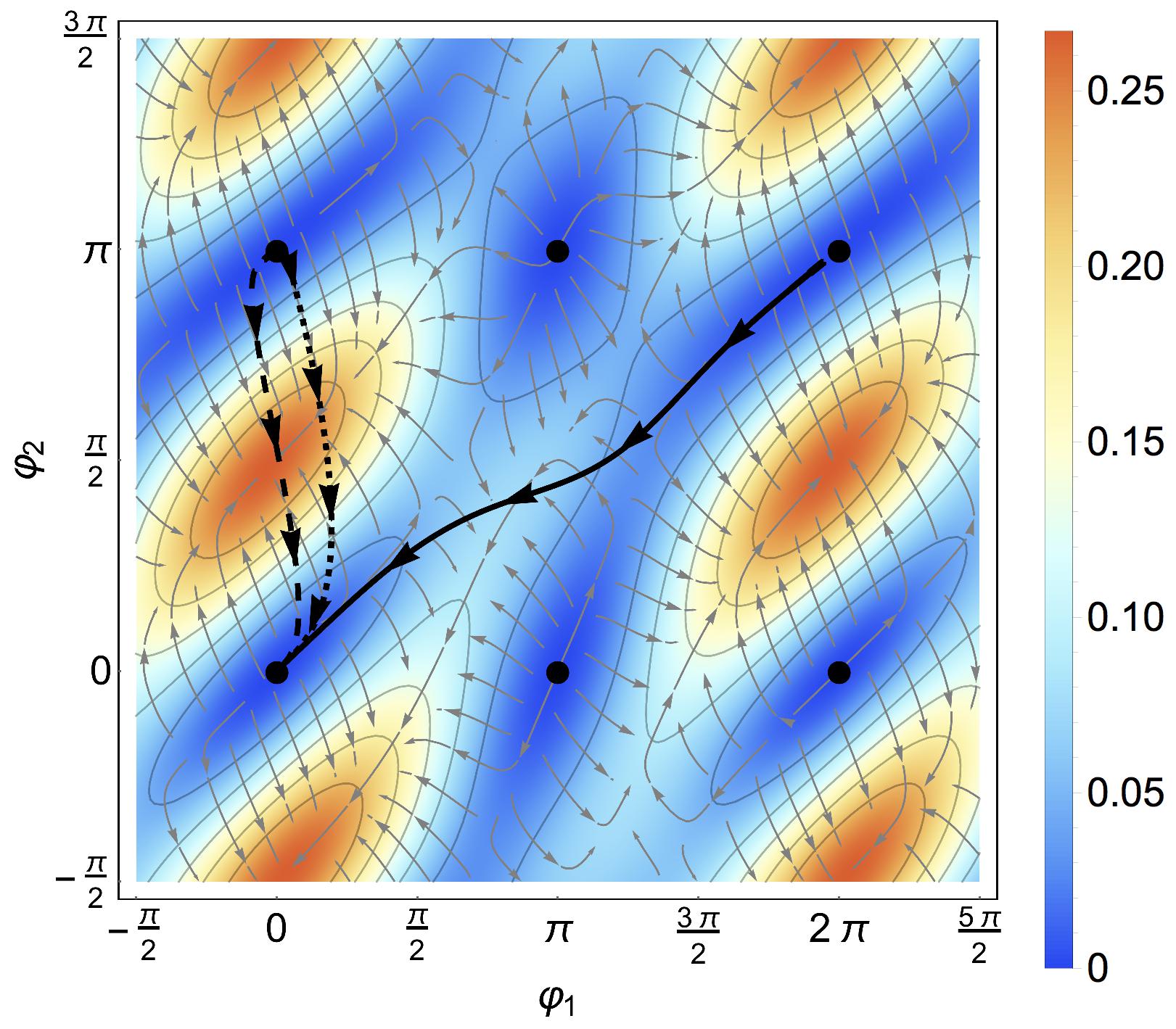}} 
\caption{$SU(3)$ fields - case III  (a) pre-potential $U$ and $\vec\nabla_{\eta}U-$flow and (b) potential $V$ and the gradient flow $\vec\nabla V$. A dotted curve describes the BPS solution obtained for the initial conditions $\vp_1(0)=0.1$, $\vp_2(0)=3.1095$, a dashed one for $\vp_1(0) =0.1$,  $\vp_2(0) = 1.3$ and a solid one for $\vp_1(0) = 2.75$, $\vp_2(0)= 1.3$.}\label{fig:su3case3UV}
\end{figure}

Expressions \eqref{vacua_su3a} result in the maxima and saddle points of the pre-potential $U$ for the cases I and IV and also in the minima for the cases II and III. For the cases I and IV the maxima $U_{\rm max}=2+\frac{1}{\sqrt{2}}$ occur when $n_1$ and $n_2$ are even numbers. For $n_2$ being odd and $n_1$ arbitrary the pre-potential has saddle points $U_{\rm s1}=-\frac{1}{\sqrt{2}}$ whereas for $n_1$ odd and $n_2$ even the pre-potential has saddle points $U_{\rm s2}=-2+\frac{1}{\sqrt{2}}$. Minima of the pre-potential for these two cases correspond to the vacua \eqref{vacua_su3b} where the pre-potential takes values $U_{\rm min}=-\sqrt{2}$. For the two other cases II and III all vacua are given by expressions \eqref{vacua_su3a}. For the case II  the pairs of numbers $(n_1,n_2)=$(even, even) give maxima of the pre-potential $U_{\rm max}=\frac{5}{2}$, $(n_1,n_2)=$(even, odd) and $(n_1,n_2)=$(odd, odd) give saddle points $U_{\rm s}=-\frac{1}{2}$ and finally $(n_1,n_2)=$(odd, even) give minima $U_{\rm min}=-\frac{3}{2}$. In the case III the maxima occur for $(n_1,n_2)=$(even, even) where the pre-potential takes value $U_{\rm max}=\frac{11}{10}$, the minima $U_{\rm min}=-\frac{9}{10}$ occur for $(n_1,n_2)=$(even, odd) and the saddle points $U_{\rm s}=-\frac{1}{10}$ for $(n_1,n_2)=$(odd, even) and $(n_1,n_2)=$(odd, odd). In Fig.\ref{fig:su3vacua} we present the vacua of the potential for the cases I and II. The vacua for the case III are shown in Fig.\ref{fig:su3case3UV}. The vacua for the case IV are the same as for the case I.

%%%%%%%%%%%%%%%%%%%%%%%%%%%%%%%%%%%%%%%%%%%%%
%							FIGURE 7					      %
%%%%%%%%%%%%%%%%%%%%%%%%%%%%%%%%%%%%%%%%%%%%%
\begin{figure}[h!]
\centering
\subfigure[]{\includegraphics[width=0.47\textwidth,height=0.47\textwidth, angle =0]{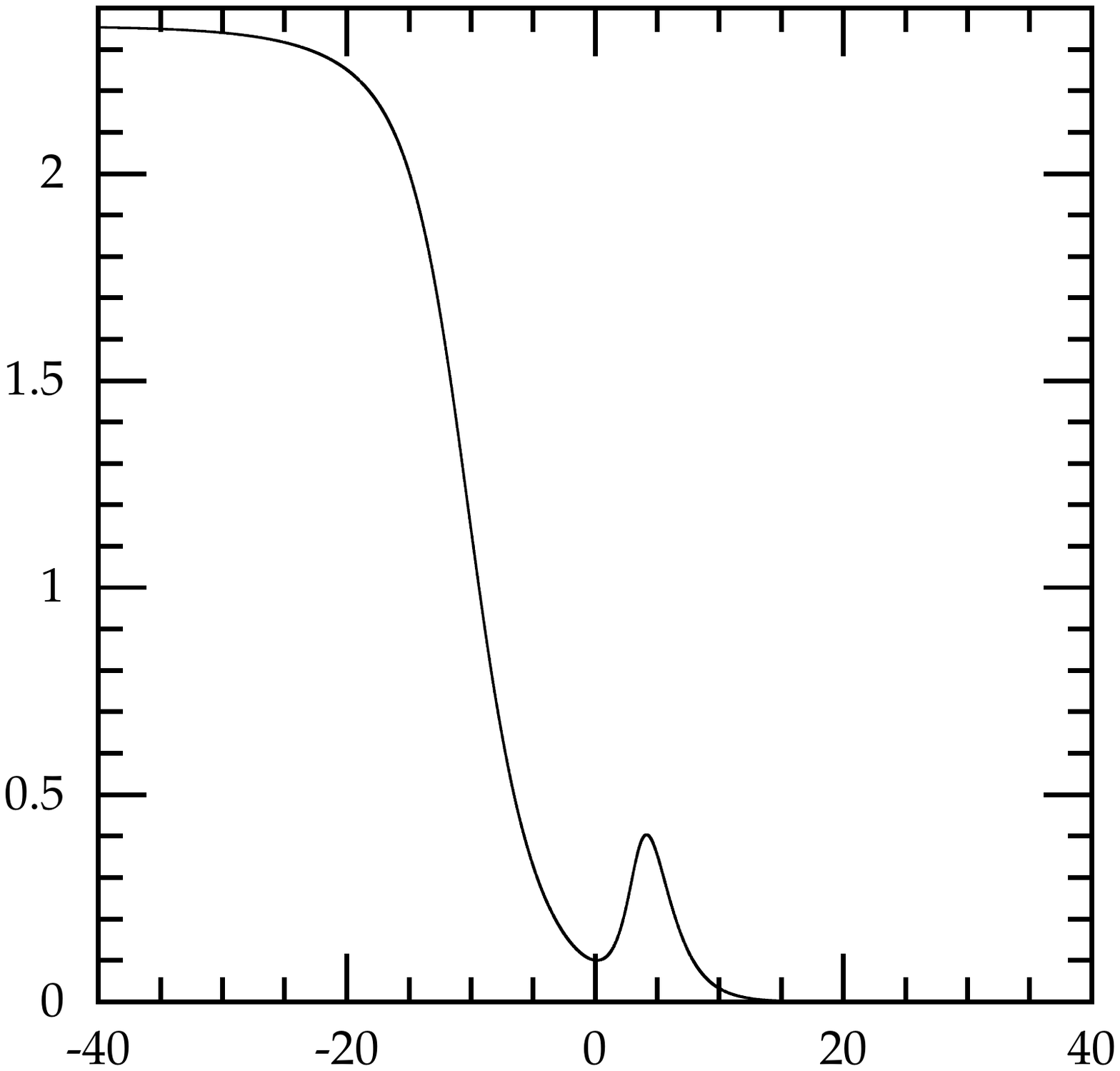}}   
\subfigure[]{\includegraphics[width=0.47\textwidth,height=0.47\textwidth, angle =0]{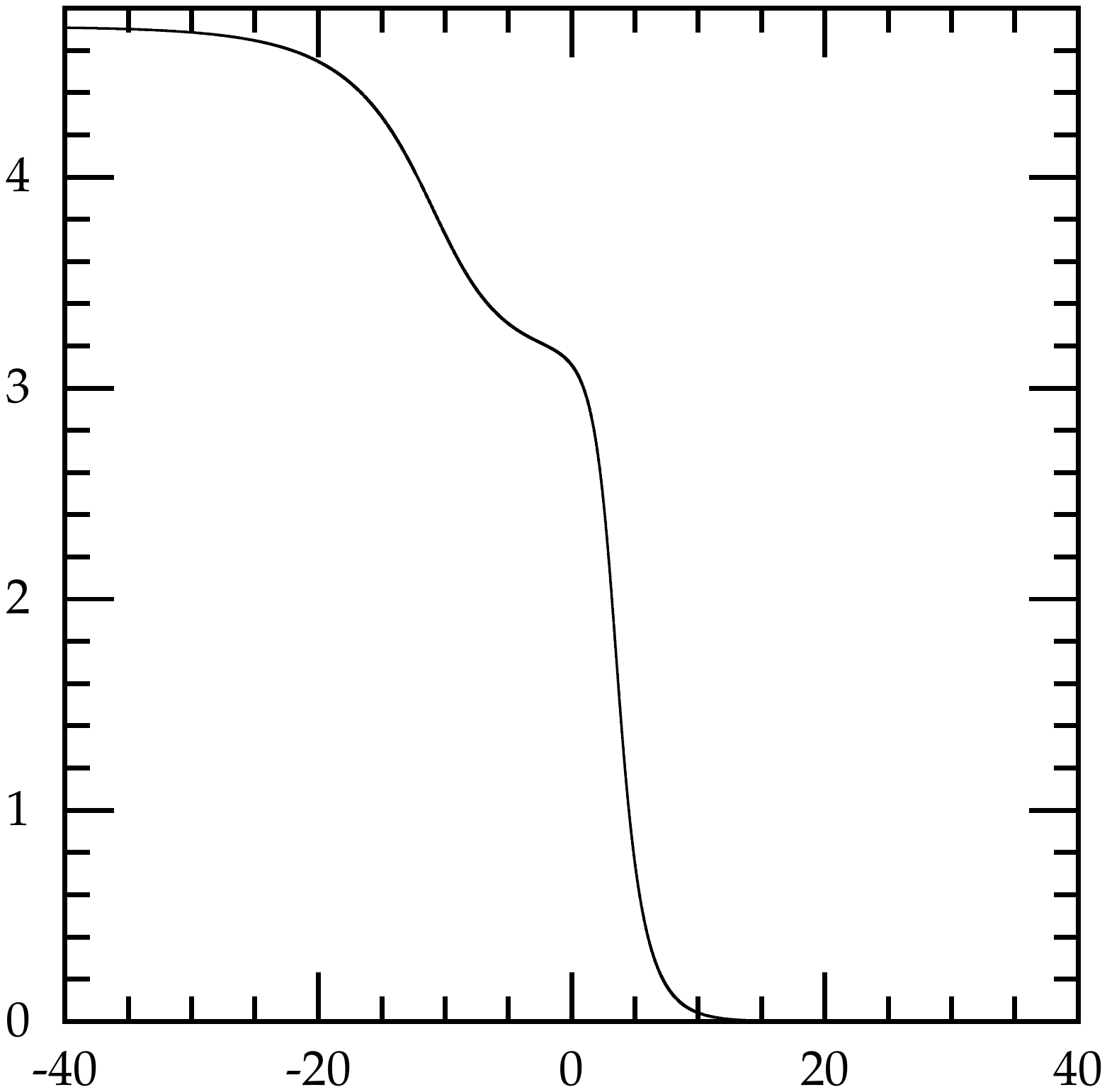} }
\caption{$SU(3)$ fields - case I; (a) $\vp_1$ and (b) $\vp_2$.}\label{fig:su3case1}
\end{figure}

%%%%%%%%%%%%%%%%%%%%%%%%%%%%%%%%%%%%%%%%%%%%%
%							FIGURE 8						               %
%%%%%%%%%%%%%%%%%%%%%%%%%%%%%%%%%%%%%%%%%%%%%
\begin{figure}[h!]
\centering
\subfigure[]{\includegraphics[width=0.47\textwidth,height=0.47\textwidth, angle =0]{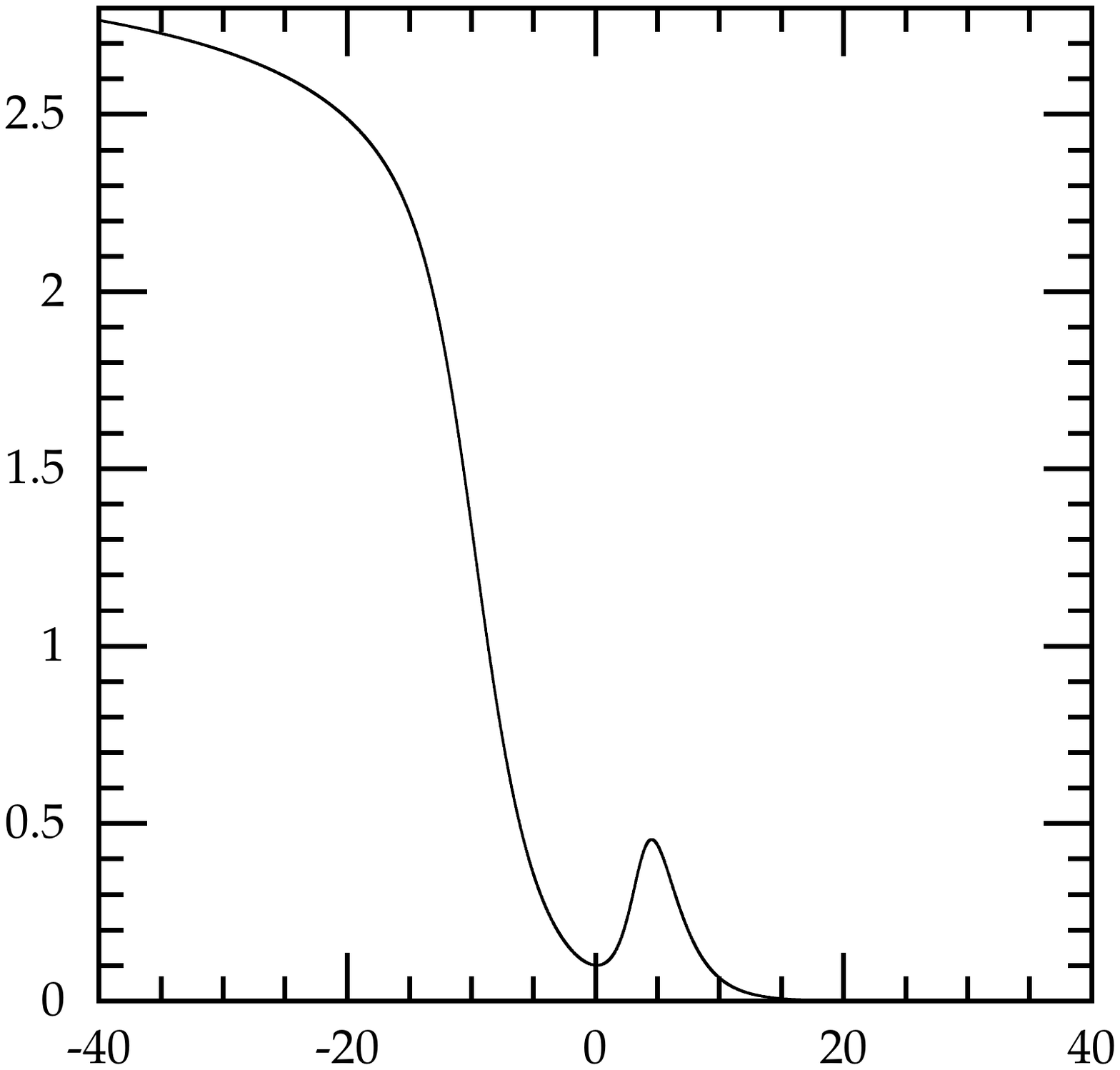}}   
\subfigure[]{\includegraphics[width=0.47\textwidth,height=0.47\textwidth, angle =0]{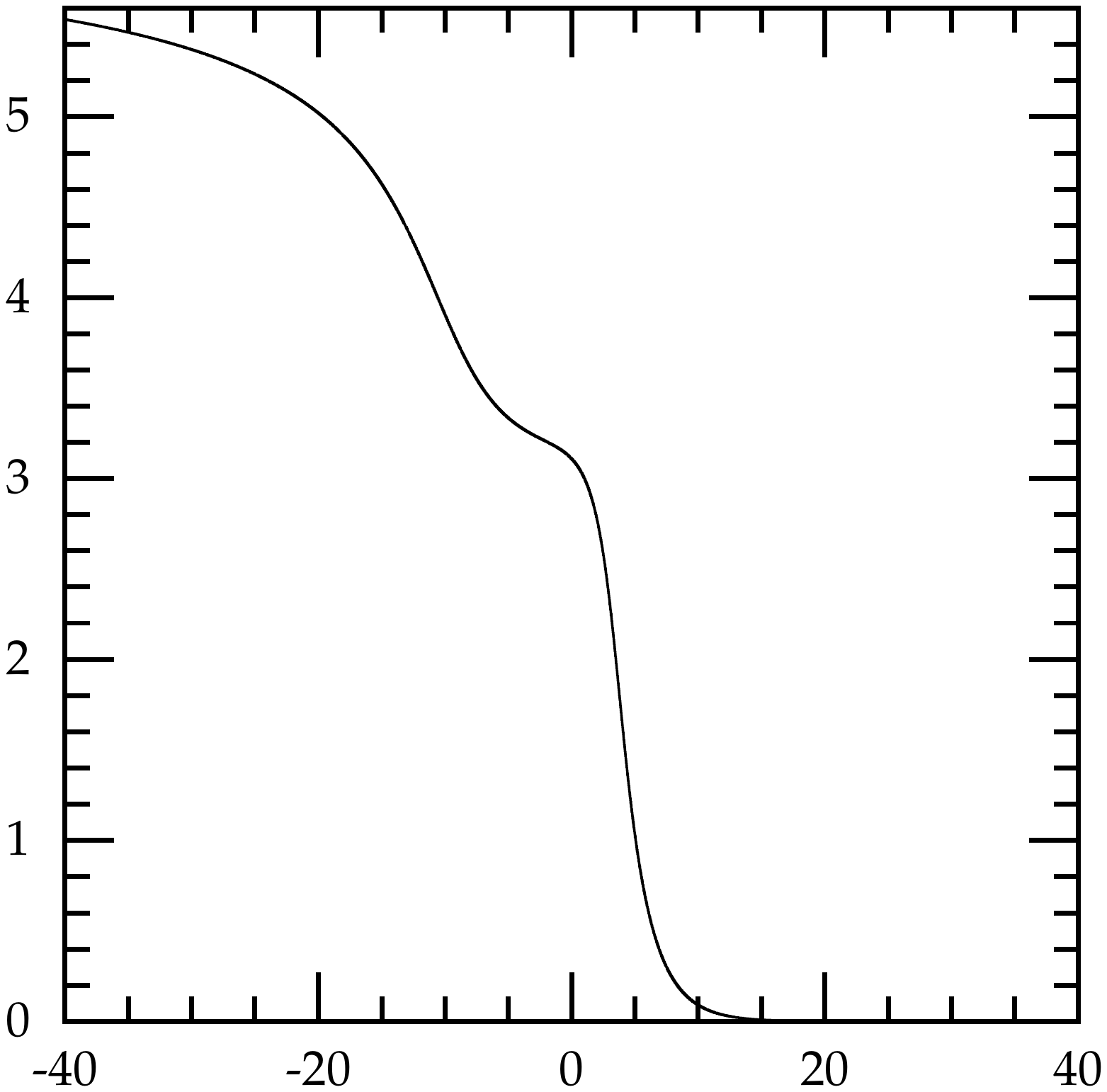}} 
\caption{$SU(3)$ fields - case II;  fields (a) $\vp_1$ and (b) $\vp_2$.}\label{fig:su3case2}
\end{figure}

%%%%%%%%%%%%%%%%%%%%%%%%%%%%%%%%%%%%%%%%%%%%%
%							FIGURE 9						               %
%%%%%%%%%%%%%%%%%%%%%%%%%%%%%%%%%%%%%%%%%%%%%
\begin{figure}[h!]
\centering
\subfigure[]{\includegraphics[width=0.32\textwidth,height=0.36\textwidth, angle =0]{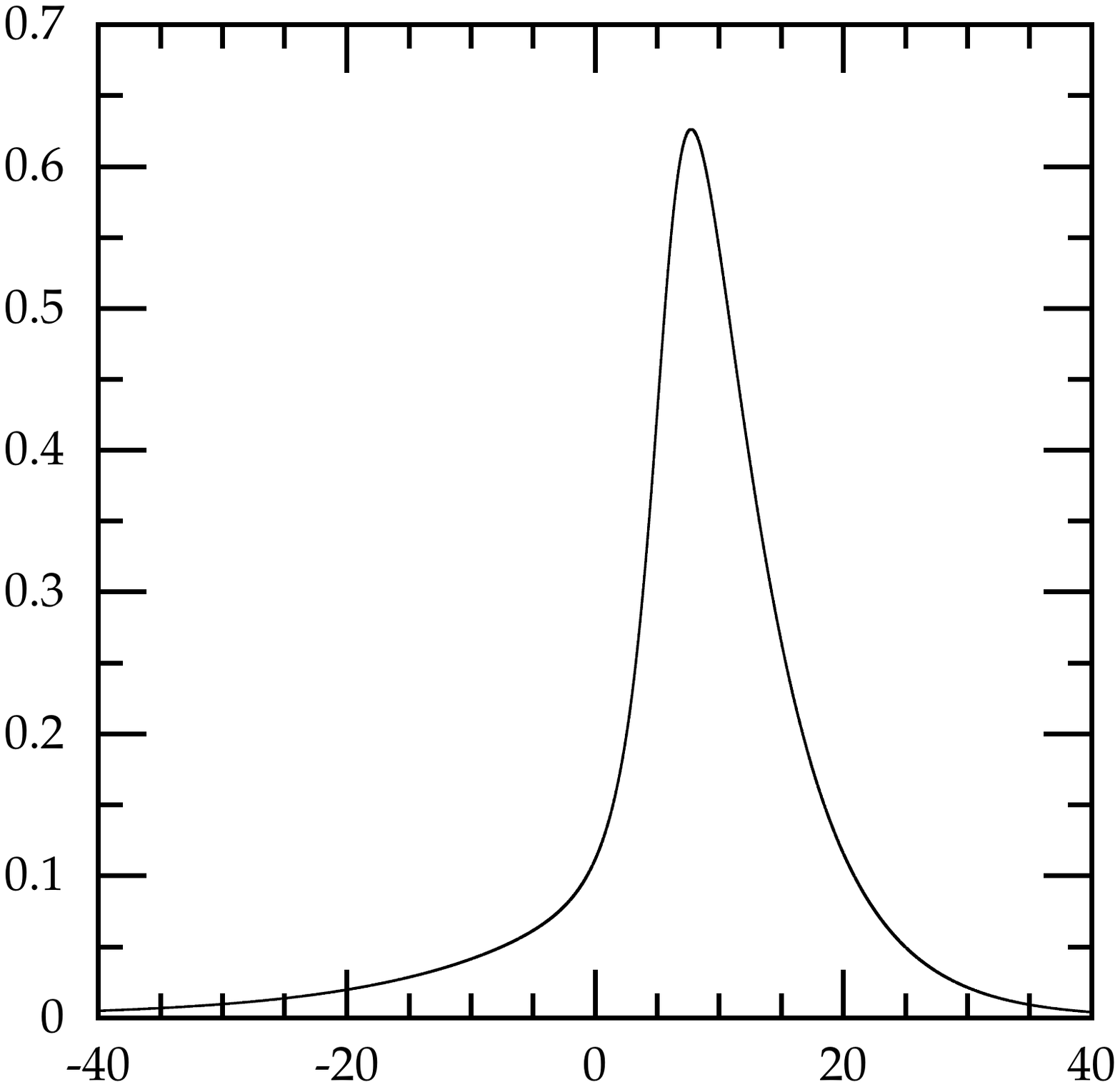}}\subfigure[]{\includegraphics[width=0.32\textwidth,height=0.38\textwidth, angle =0]{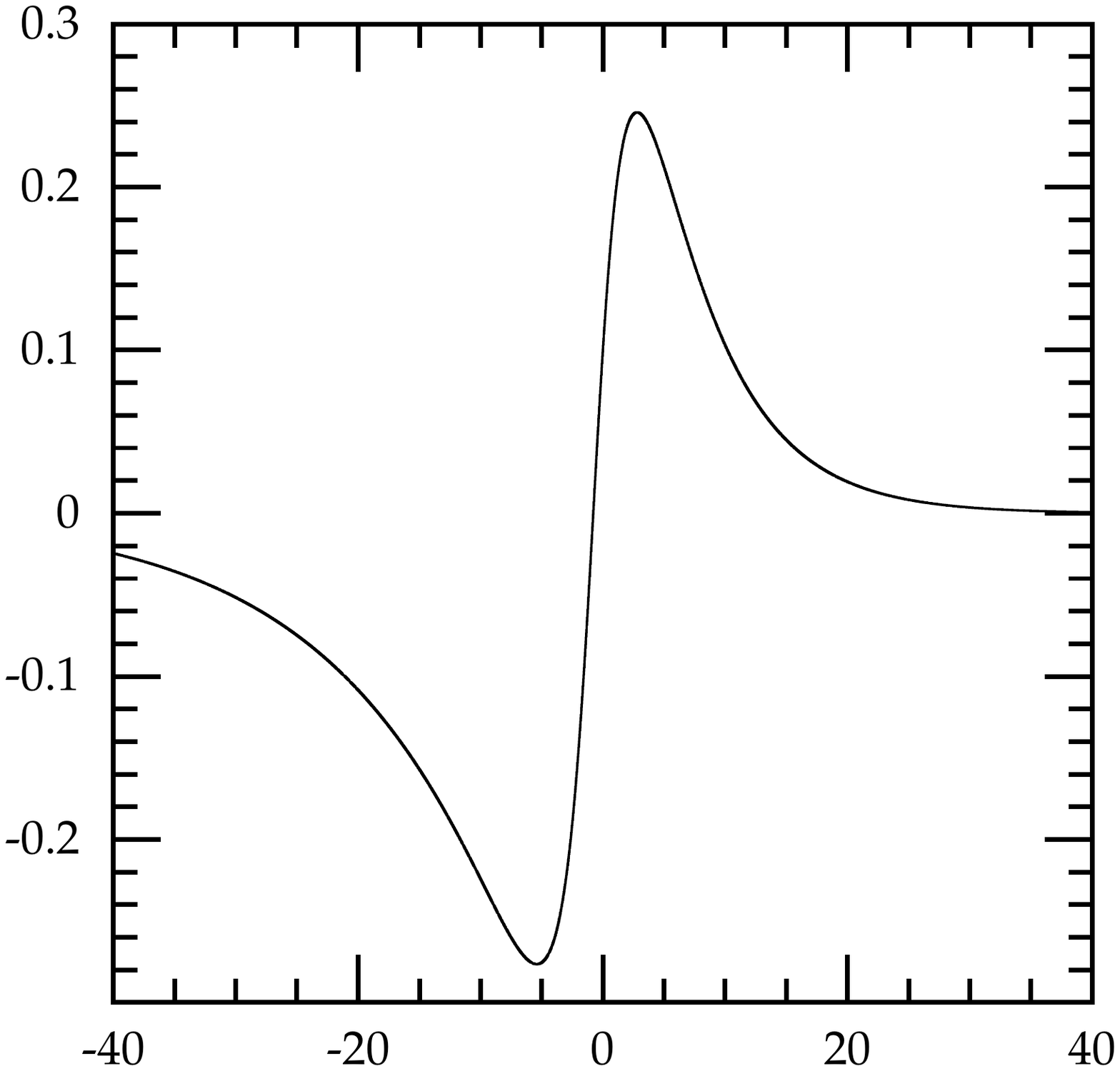}}\subfigure[]{\includegraphics[width=0.32\textwidth,height=0.38\textwidth, angle =0]{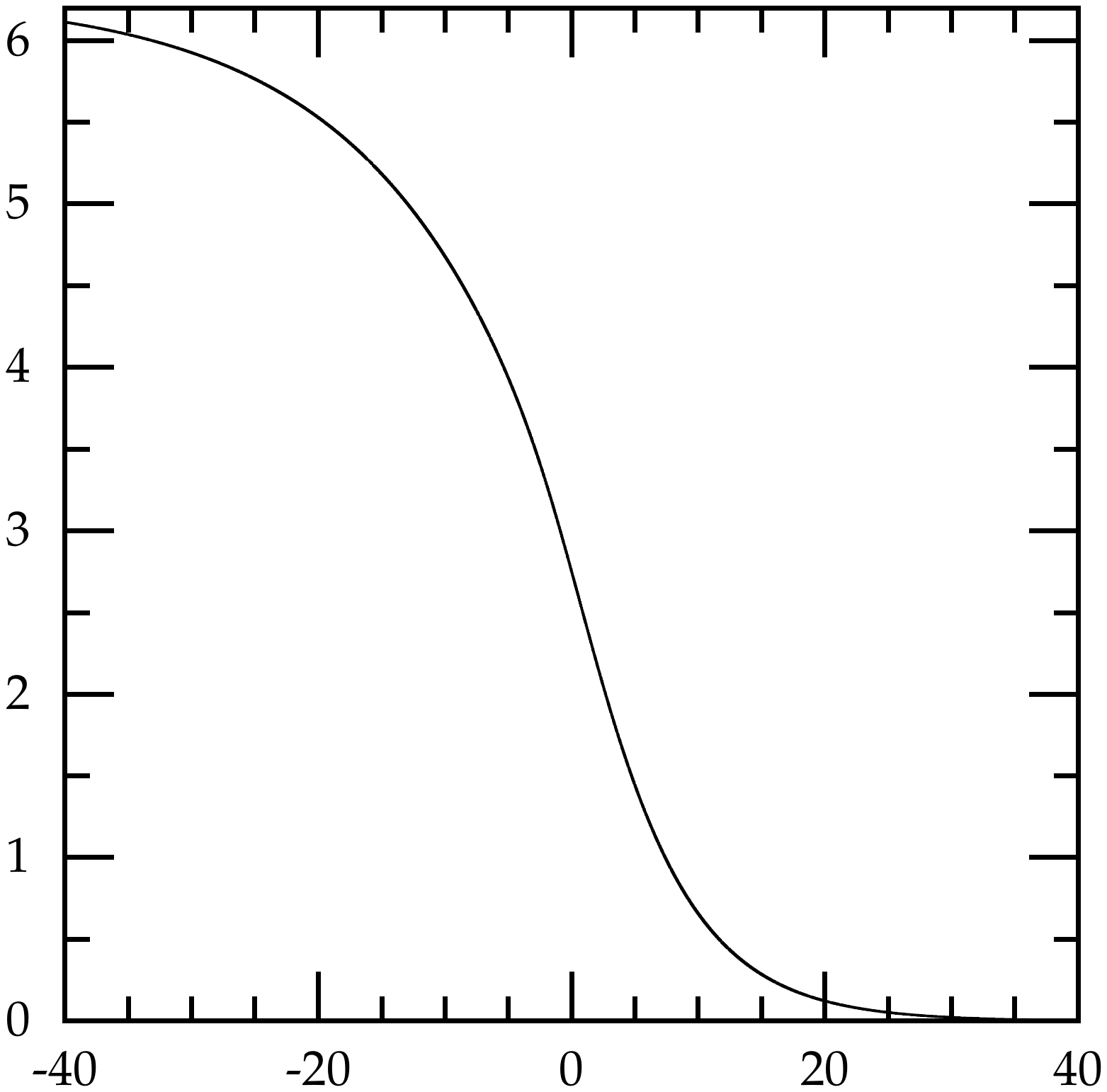}}
\subfigure[]{\includegraphics[width=0.32\textwidth,height=0.36\textwidth, angle =0]{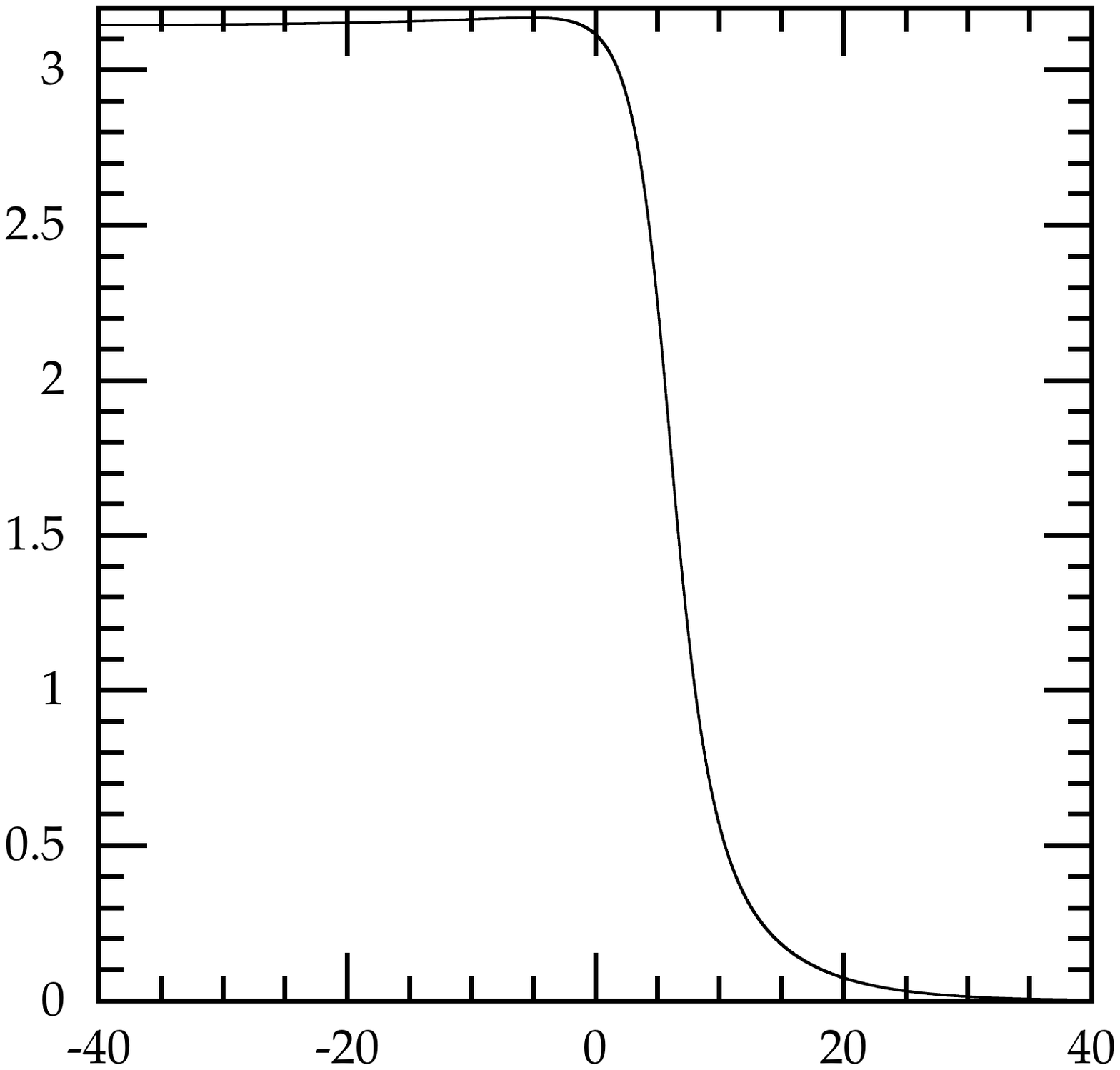}}\subfigure[]{\includegraphics[width=0.32\textwidth,height=0.38\textwidth, angle =0]{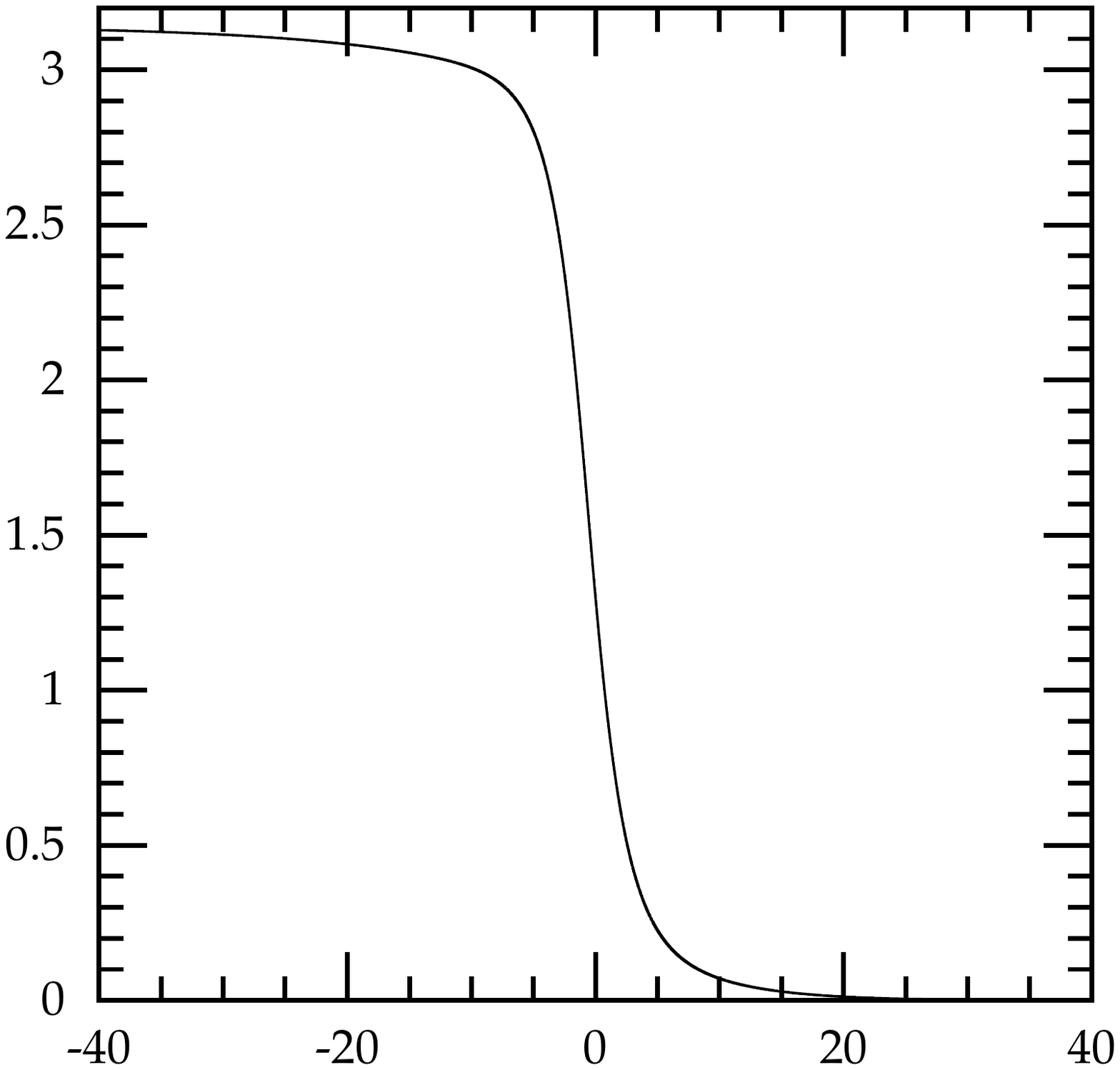}}   
\subfigure[]{\includegraphics[width=0.32\textwidth,height=0.38\textwidth, angle =0]{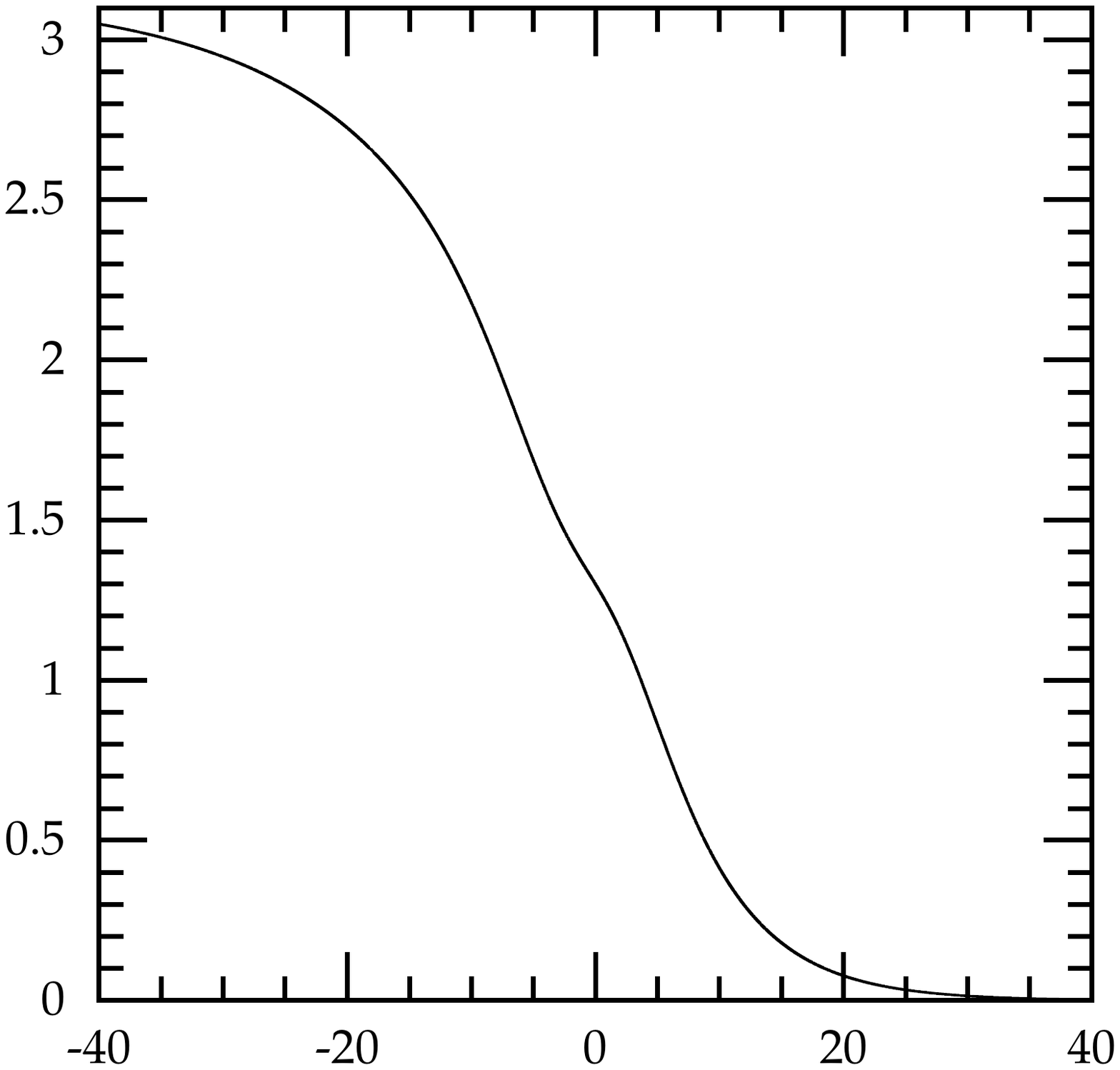}}
\caption{$SU(3)$ fields - case III; (a), (b), (c) field $\vp_1$, (d), (e), (f) field $\vp_2$. Plots (a), (d) were obtained for the initial conditions $\vp_1(0)=0.1$, $\vp_2(0)=3.1095$, (b) and (e) for $\vp_1(0) =0.1$,  $\vp_2(0) = 1.3$ and (c) and (f) for $\vp_1(0) = 2.75$, $\vp_2(0)= 1.3$.}\label{fig:su3case3}
\end{figure}

%%%%%%%%%%%%%%%%%%%%%%%%%%%%%%%%%%%%%%%%%%%%%
%							FIGURE 10 							       %
%%%%%%%%%%%%%%%%%%%%%%%%%%%%%%%%%%%%%%%%%%%%%

\begin{figure}[h!]
\centering
\subfigure[]{\includegraphics[width=0.47\textwidth,height=0.47\textwidth, angle =0]{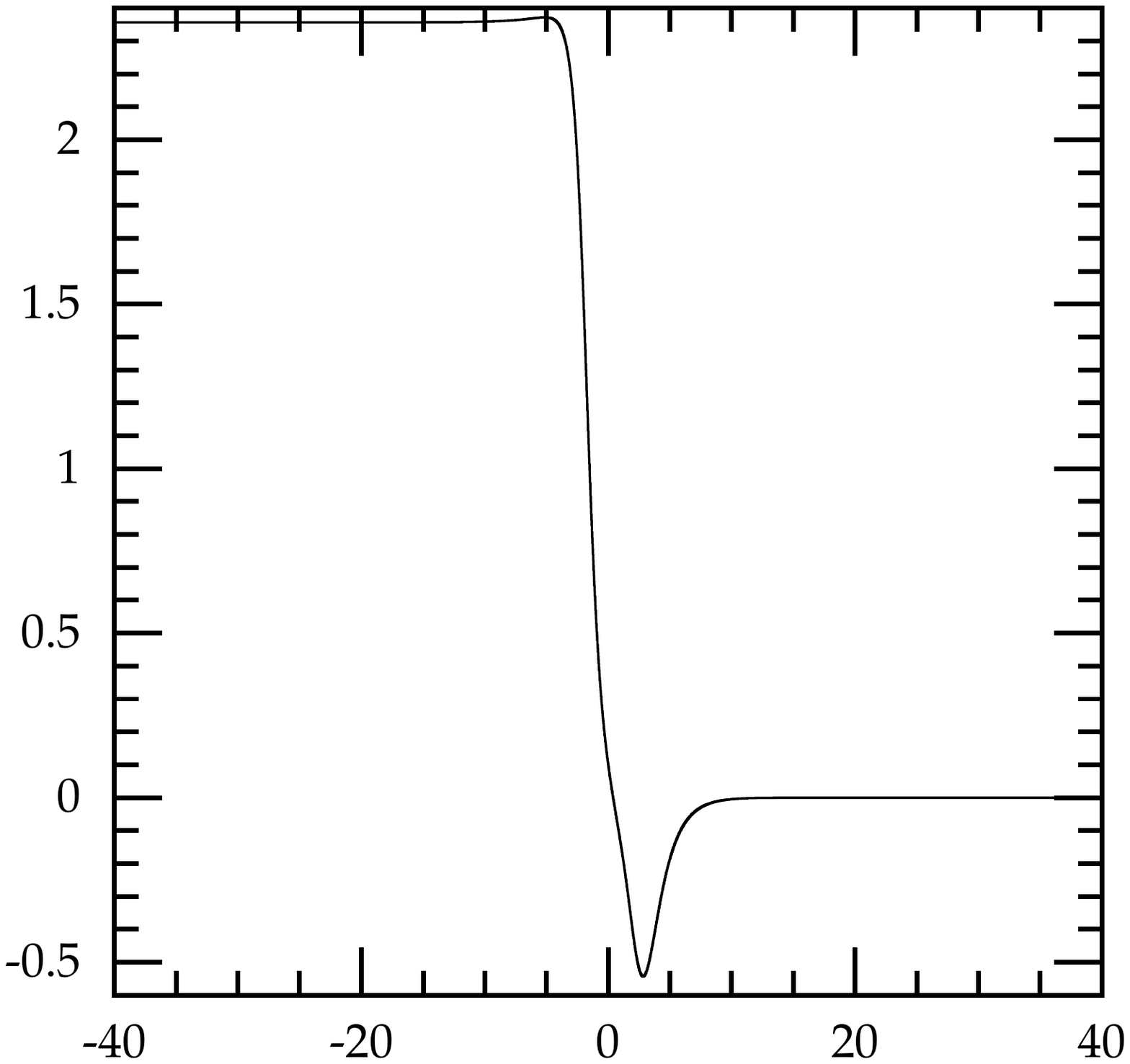}}   
\subfigure[]{\includegraphics[width=0.47\textwidth,height=0.47\textwidth, angle =0]{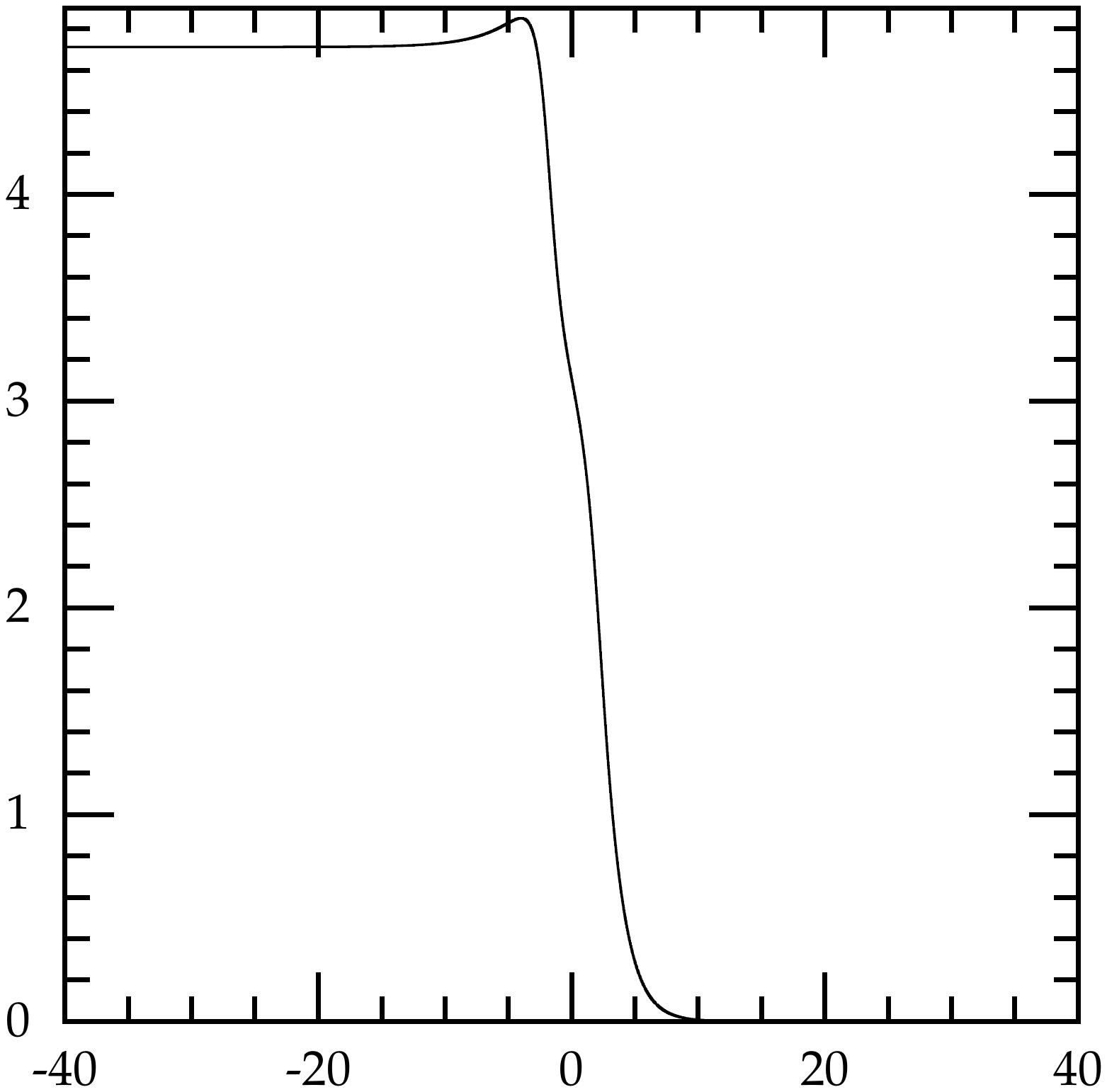}} 
\caption{$SU(3)$ fields - case IV; a) fields $\vp_1$ and $\vp_2$. This case is similar to I but $\lambda=1.8$}\label{fig:su3case4}
\end{figure}

Note looking at all the plots shows very clearly that the fields always go from one vacuum to the vacuum.
In all the cases, as $x$ tends to $\infty$ the fields both tends to the vacua $\vp_1(\infty)=0$ and $\vp_2(\infty)=0$.  As $x$ gets smaller and smaller the fields go to various
vacua. In the cases I, II and IV $\gamma_1=\gamma_3=1$ and so from \rf{vacuasu3b} we see that $\vp_2(-\infty)=2\vp_1(-\infty)$.
The exact values of $\vp_1(-\infty)$ in cases I and II are different and they depend on the value of $\gamma_2$.
The cases of I and IV differ by the value of $\lambda$ and their plots are completely different but the curves go to the same asymptotic values; compare Fig.\ref{fig:su3case1} and Fig.\ref{fig:su3case4}.
In fact, the values of the energies in cases I and IV are the same; this is not surprising as the value of the energy
is determined by the asymptotic values of the fields and these values are the same in these two cases so the values of the energies are also the same.

The case III is special as it presents the plots of three numerical solutions obtained for the identical values of the parameters $\lambda$ and $\gamma_i$. In this case the potential $V$ has only minima of the first kind \rf{firstkindvacuasu3} at $\vp_1=n_1\pi$ and $\vp_2=n_2\pi$, where $n_1$ and $n_2$ are integer and this is seen from our results.  Each solution was obtained for different initial values of $\vp_i(0)$. We present here the results of the studies of the following initial data $\vp_1(0)=0.1$, $\vp_2(0))=3.1095$ shown in Figure \ref{fig:su3case3} (a),(d), $\vp_1(0)=0.1$, $\vp_2(0)=1.3$ shown in Figure \ref{fig:su3case3}(b), (e) and $\vp_1(0)=2.75$, $\vp_2(0)=1.3$ shown in Figure \ref{fig:su3case3}(c), (f).  For other values of the initial conditions the obtained plots have always been similar to one of the three cases shown here.

Of course, the BPS equations do not `know about the topology' and they are just responsible for the evolution to the `nearest' vacuum.
Hence in the III case the field $\vp_1$ evolved in both directions of $x$ to the same value of the vacuum, namely 0, while the field
$\vp_2$ went to $\pi$ and 0. The plots of the numerically determined curves in the space $(\vp_1,\vp_2)$ and the potentials $U$ and $V$ are shown in Fig.\ref{fig:su3case3UV}.

%%%%%%%%%%%%%%%%%%%%%%%%%%%%%%%%%%%%%%%%%%%%%
%							FIGURE 11							               %
%%%%%%%%%%%%%%%%%%%%%%%%%%%%%%%%%%%%%%%%%%%%%

%%%%%%%%%%%%%%%%%%%%%%%%%%%%%%%%%%%%%%%%%%%%%
%							FIGURE 12							               %
%%%%%%%%%%%%%%%%%%%%%%%%%%%%%%%%%%%%%%%%%%%%%
\begin{figure}[h!]
\centering
\subfigure[]{\includegraphics[width=0.45\textwidth,height=0.48\textwidth, angle =0]{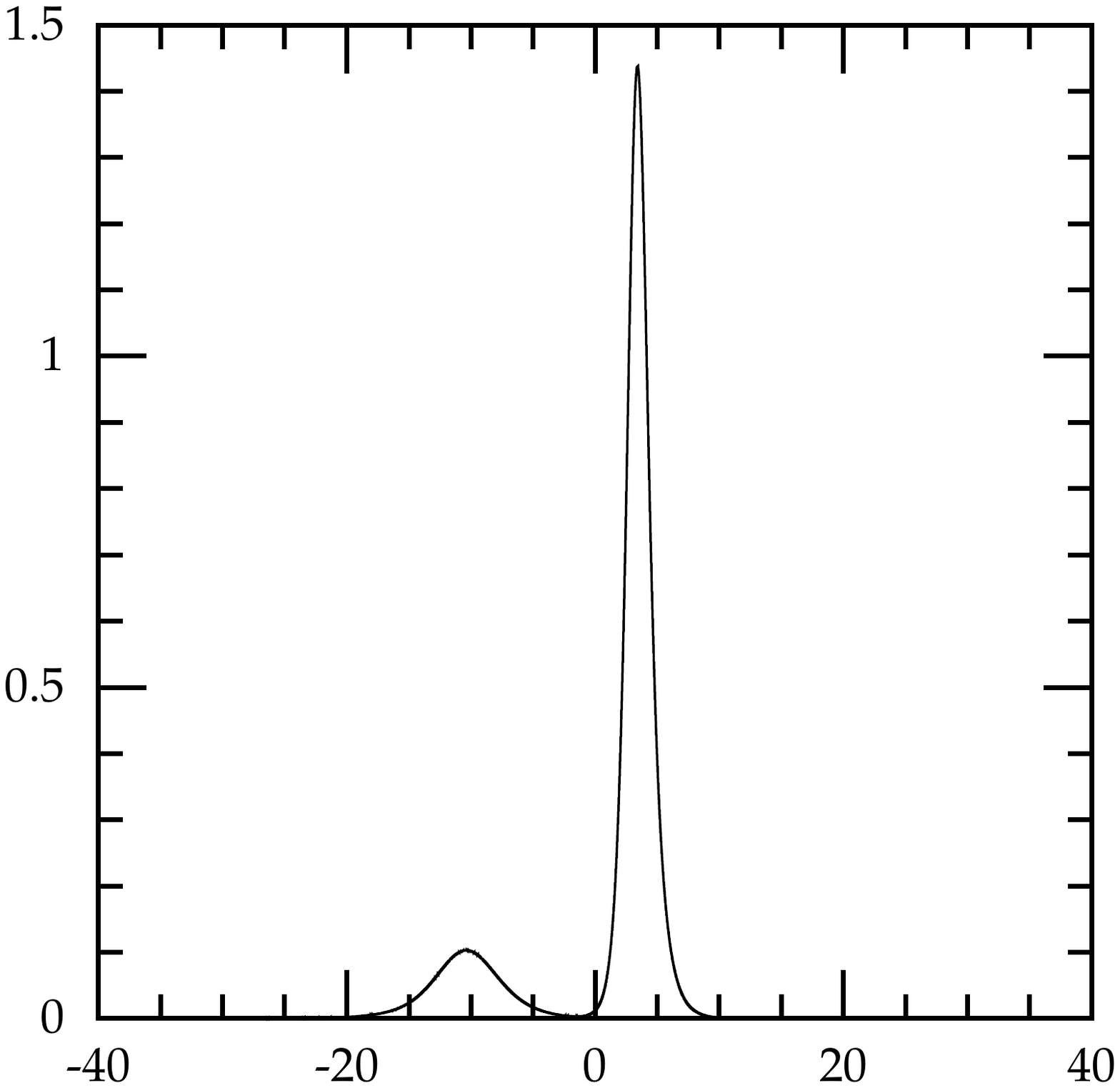}}   
\subfigure[]{\includegraphics[width=0.45\textwidth,height=0.45\textwidth, angle =0]{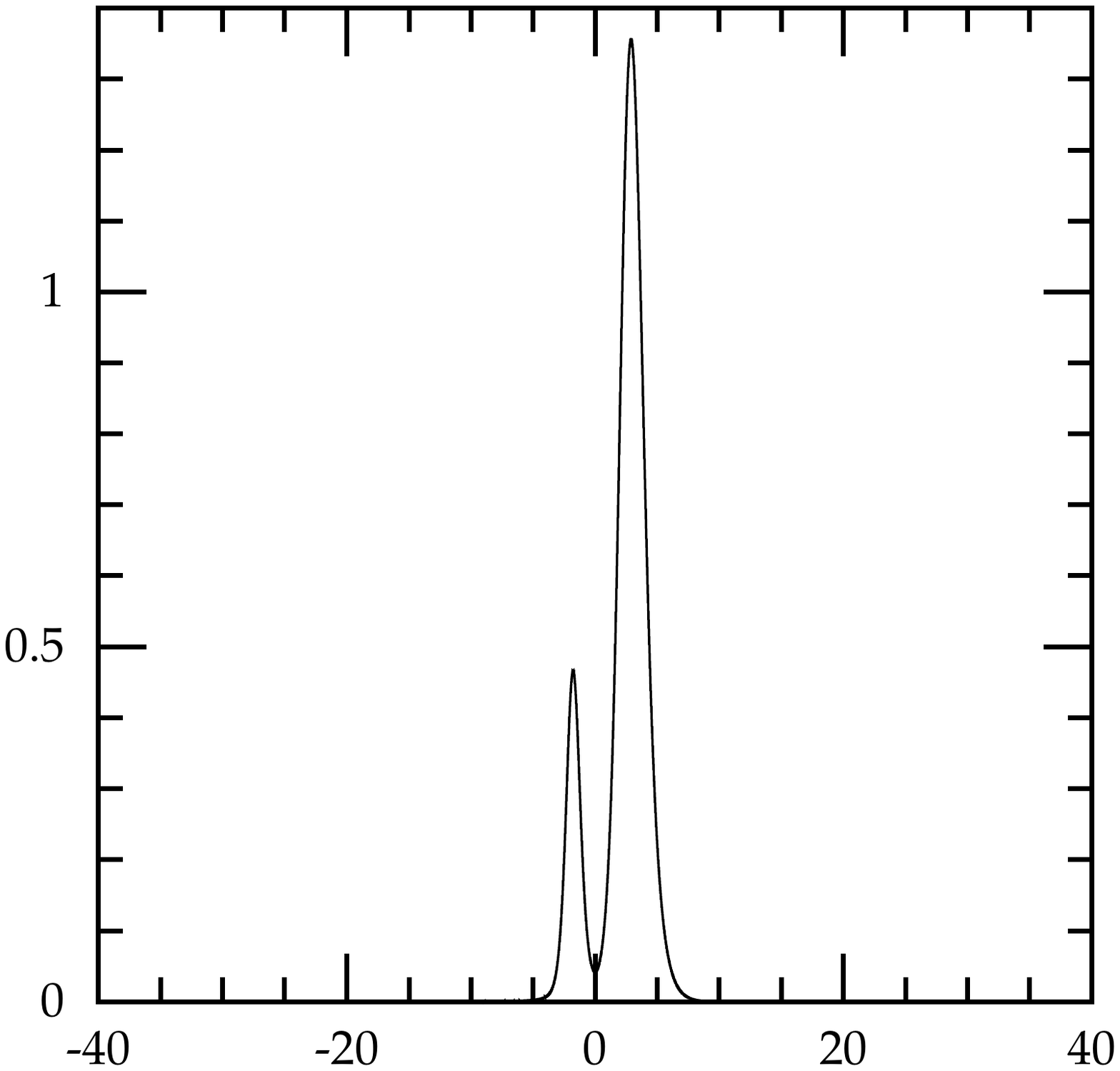}} 
\caption{Energy densities (and topological charge densities) of the $SU(3)$ field configurations seen in cases I (a) and IV (b).}\label{fig:su3case4energy}
\end{figure}

The case IV is similar to the case I but they correspond to the different values of $\lambda$.
The shapes of the curves are very different but it is clear that they go to the same asymptotic values of the fields as $x\to \pm \infty$. {Fields $\vp_1$ and $\vp_2$ for the case IV are plotted in Fig.\ref{fig:su3case4}.}

In Fig.\ref{fig:su3case4energy} we have also plotted the energy density of the field configurations for cases I and IV above. We see that in both cases we have two peaks of the energy density. As the total value is the same in both cases the whole effect of $\lambda$ corresponds to the change of the relative heights of the two peaks and their positions. As we can see 
from the plots when $\lambda$ is larger the peaks are also a little closer together. For smaller values of $\lambda$ these effects
are less visible.

%%%%%%%%%%%%%%%%%%%%%%%%%%%%%%%%%%%%%%%%%%%%%
\subsection{$SO(5)$ Simulations}
%%%%%%%%%%%%%%%%%%%%%%%%%%%%%%%%%%%%%%%%%%%%%

In the case of $SO(5)$ we have solved numerically the self-duality equations \rf{bpseqso5adj} corresponding to the adjoint representation of that group, and we have checked
 the stability of the resultant solutions.  Our simulations confirmed that the self-dual solutions were indeed the static solutions of the full field equations and that 
these solutions were stable. 
This time (as described in detail in the previous sections) the BPS equations were more complicated, as in addition to the coupling parameter $\lambda$ our equations 
depended on 4 $\gamma_i$ parameters.
We present here the results of our simulations for the following sets of values for the $\gamma_i$ and $\lambda$ parameters:
\begin{itemize}
\item Case I): $\(\gamma_1\,,\,\gamma_2\,,\,\gamma_3\,,\,\gamma_4\,,\, \lambda\)=\(1\,,\,1\,,\,1\,,\, 0.4\,,\,0.5\)$

\item Case II): $\(\gamma_1\,,\,\gamma_2\,,\,\gamma_3\,,\,\gamma_4\,,\, \lambda\)=\(1\,,\,1\,,\,1\,,\, 0.2\,,\,0.5\)$
%\item Case III): $\(\gamma_1\,,\,\gamma_2\,,\,\gamma_3\,,\,\gamma_4\,,\, \lambda\)=\(1\,,\,1\,,\,1.2\,,\, 0.5\,,\,0.5\)$
\item Case III): $\(\gamma_1\,,\,\gamma_2\,,\,\gamma_3\,,\,\gamma_4\,,\, \lambda\)=\(1\,,\,-2\,,\,0\,,\, 1\,,\,1.0\)$
\item Case IV): $\(\gamma_1\,,\,\gamma_2\,,\,\gamma_3\,,\,\gamma_4\,,\, \lambda\)=\(\pi\,,\,\sqrt{2}\,,\,\sqrt{3}\,,\, \pi/2\,,\,0.5\)$
\end{itemize} 

We have performed many simulations for other values of parameters but the obtained results were always
similar and not qualitatively different from the results presented here.

\begin{figure}[h!]
\centering
\subfigure[]{\includegraphics[width=0.45\textwidth,height=0.47\textwidth, angle =0]{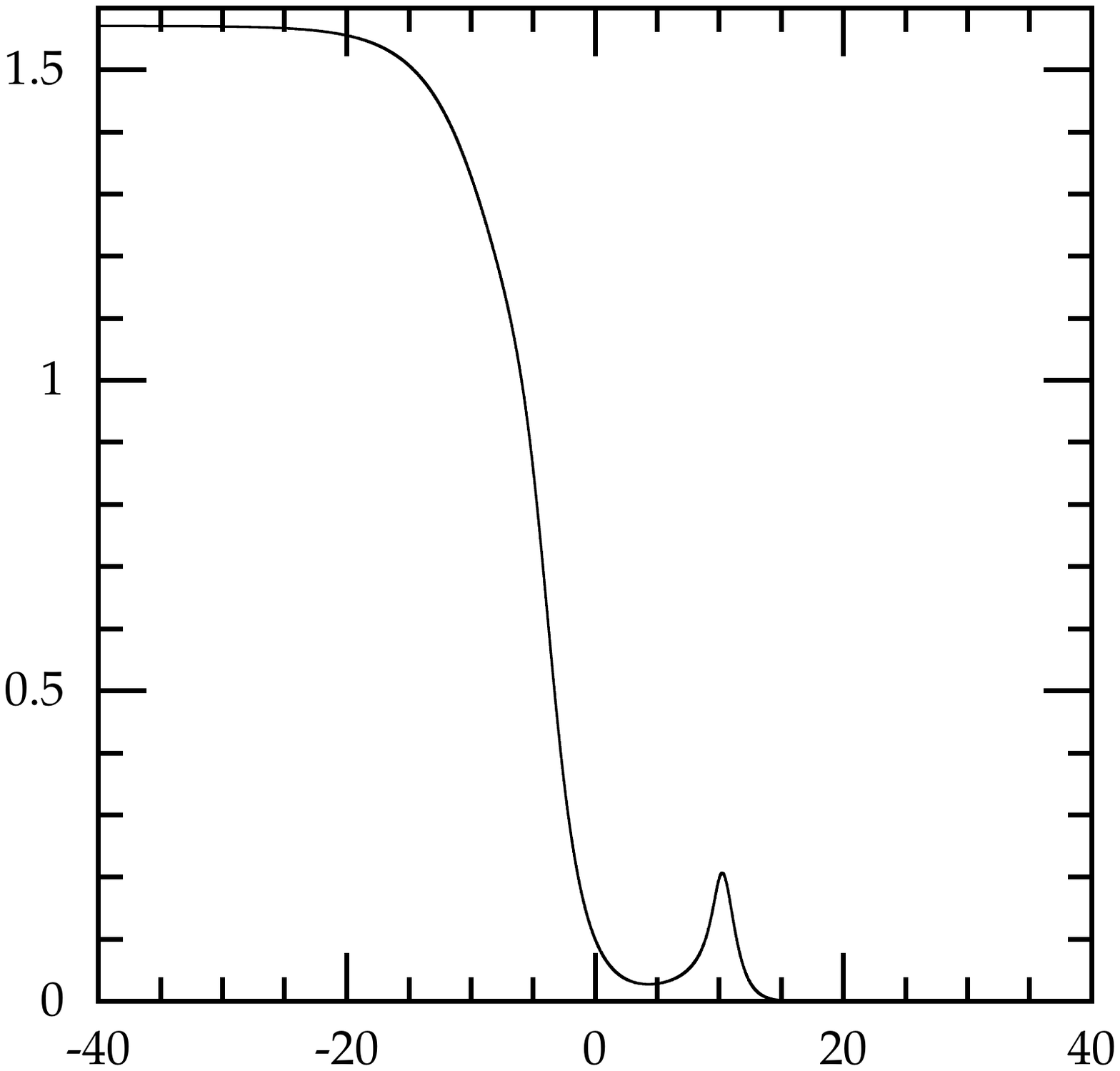}}   
\subfigure[]{\includegraphics[width=0.45\textwidth,height=0.47\textwidth, angle =0]{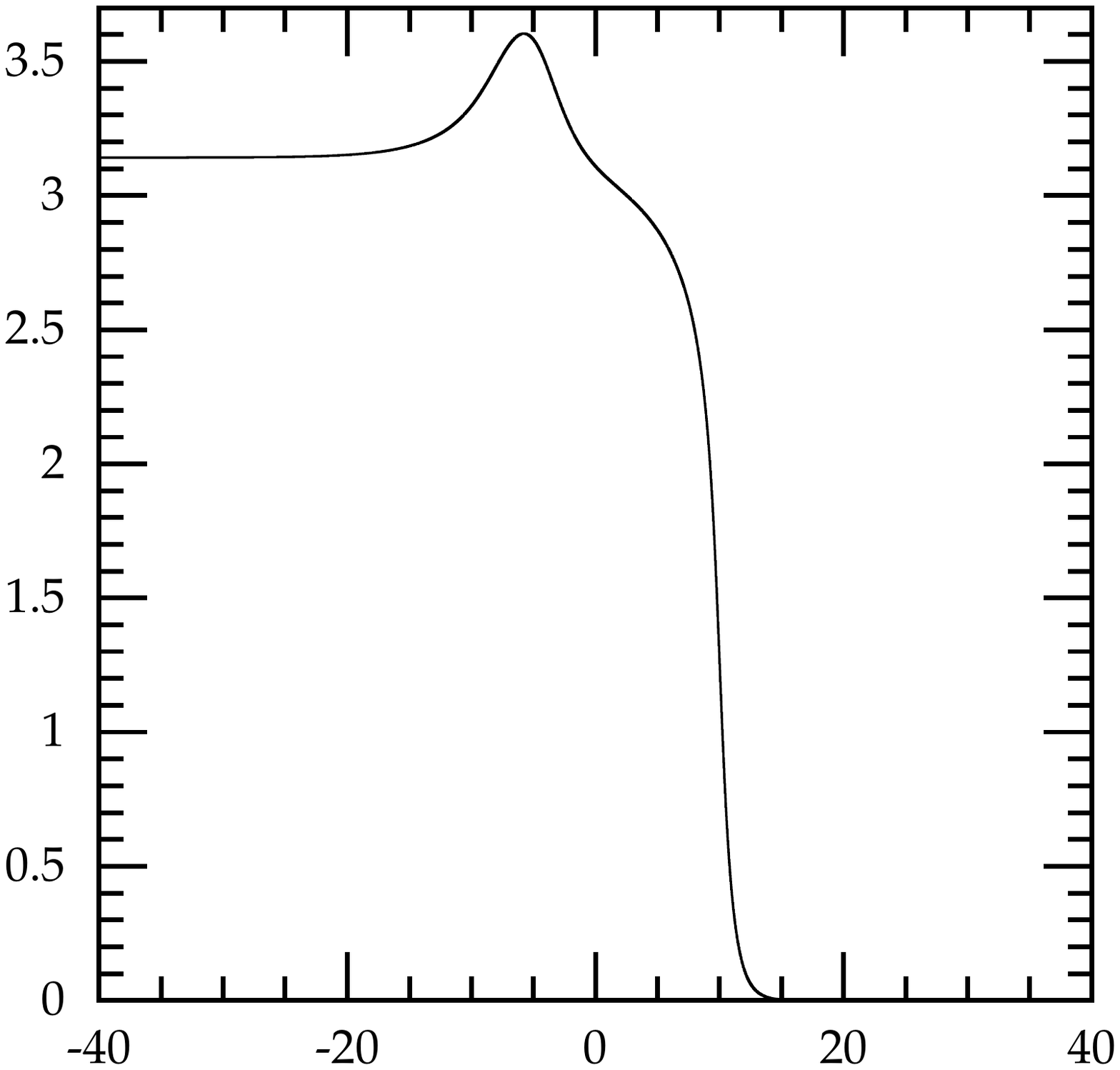} }
\subfigure[]{\includegraphics[width=0.45\textwidth,height=0.4\textwidth, angle =0]{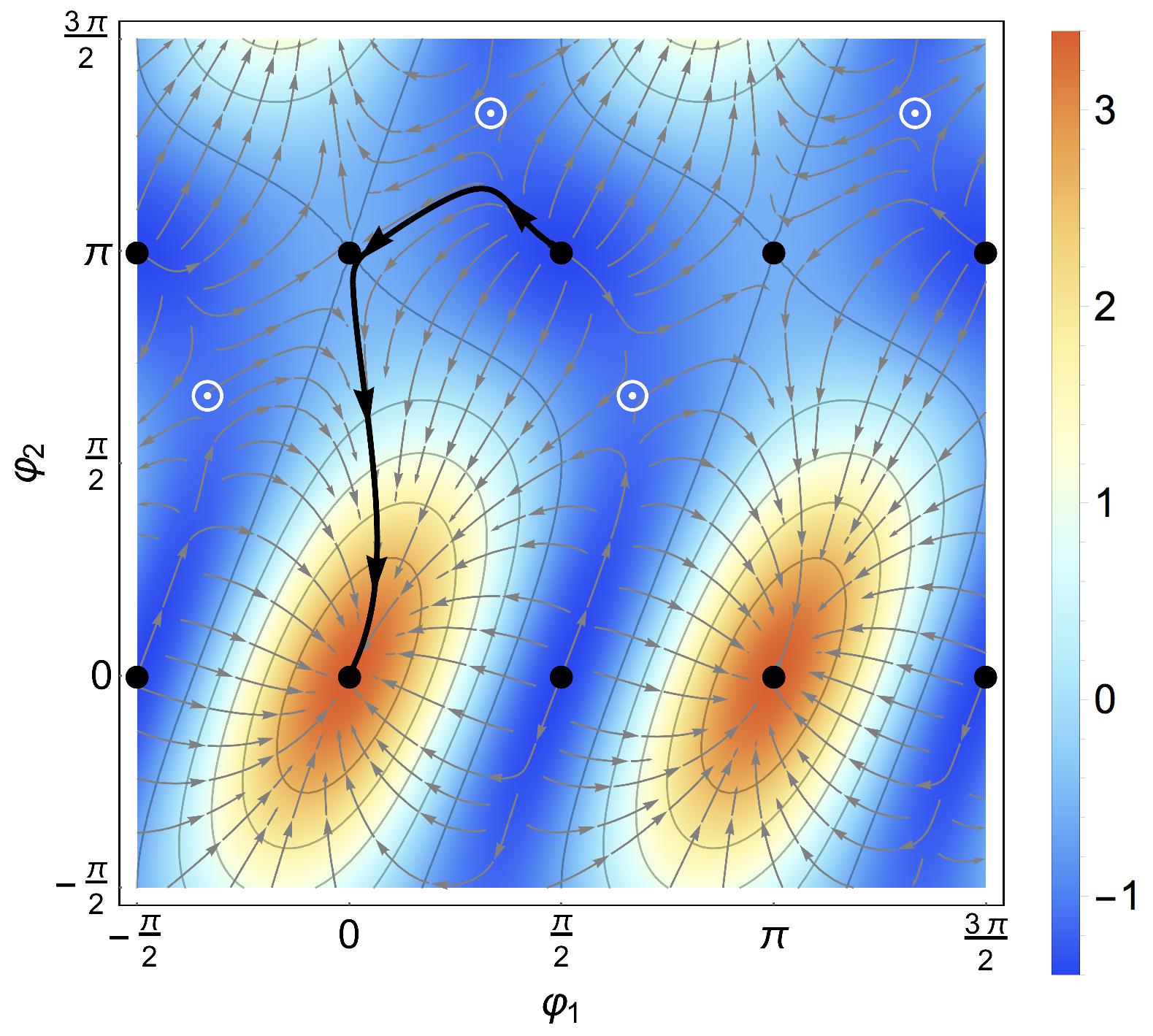} }
\caption{$SO(5)$ fields - case I; (a) $\vp_1$, (b) $\vp_2$  and (c) $\vec\nabla_{\eta} U-$flow. Vacua \eqref{maxmin1} are denoted by dots whereas vacua \eqref{saddle1} are marked by $\odot$.}\label{fig:so5case1}
\end{figure}

The case I corresponds to the potential which has minima at
\begin{align}
(\vp_{1}^{\rm (vac)},\vp_{2}^{\rm (vac)})&=\Big(\frac{\pi}{2}n_1, \pi\, n_2\Big),\label{maxmin1}\\
(\vp_{1}^{\rm (vac)},\vp_{2}^{\rm (vac)})&=\Big(\mp\frac{\pi}{3}+n_1\pi,\pm\frac{2\pi}{3}+n_2\pi \Big),\label{saddle1}
\end{align}
where $n_1, n_2\in \mathbb Z$. Expressions \eqref{maxmin1} give maxima of the pre-potential $U_{\rm max}=\frac{17}{5}$ when $n_1$ and $n_2$ are simultaneously even, minima $U_{\rm min}=-\frac{7}{5}$ when $n_1$ is odd (independently on the value of $n_2$) and saddle points $U_{\rm s1}=-\frac{3}{5}$ for $n_1$ even and $n_2$ odd. On the other hand \eqref{saddle1} always corresponds to the saddle points $U_{\rm s2}=-\frac{11}{10}$ of the pre-potential.  In Fig.\ref{fig:so5case1}(a), (b) we present plots of the fields obtained in the simulation in which the kink $\vp_1$ 
connects the vacua at $\pi/2$ and 0 and the kink $\vp_2$ connects $\pi$ and 0.
%%%%%%%%%%%%%%%%%%%%%%%%%%%%%%%%%%%%%%%%%%%%%
%							FIGURE 13							               %
%%%%%%%%%%%%%%%%%%%%%%%%%%%%%%%%%%%%%%%%%%%%%
We note the familiar ``bumps'' on the kinks. These are the most common solutions of the self-dual
equations found in our simulations. In Fig.\ref{fig:so5case1}(c) we plot the pre-potential, its $\vec\nabla_{\eta} U$-flow, and the numerical BPS solution. The BPS curves in the space of fields follow tightly the $\vec\nabla_{\eta} U$-flows whose form is determined by the existence of the saddle points in this region. Clearly, in this case the existence of ``bumps'' is directly connected  with the presence of saddle points of the pre-potential $U$.

\begin{figure}[h!]
\centering
\subfigure[]{\includegraphics[width=0.45\textwidth,height=0.47\textwidth, angle =0]{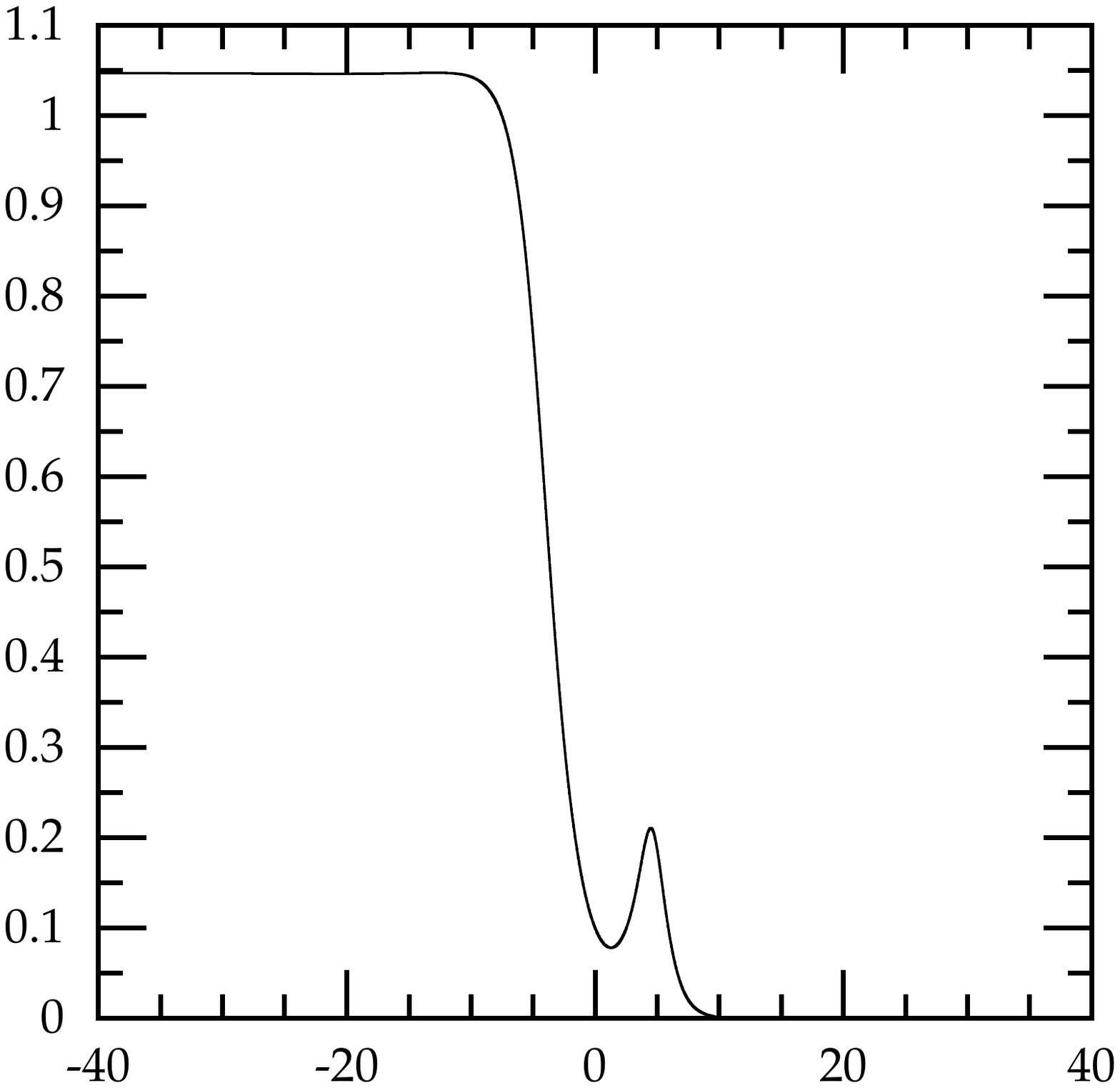}}   
\subfigure[]{\includegraphics[width=0.45\textwidth,height=0.47\textwidth, angle =0]{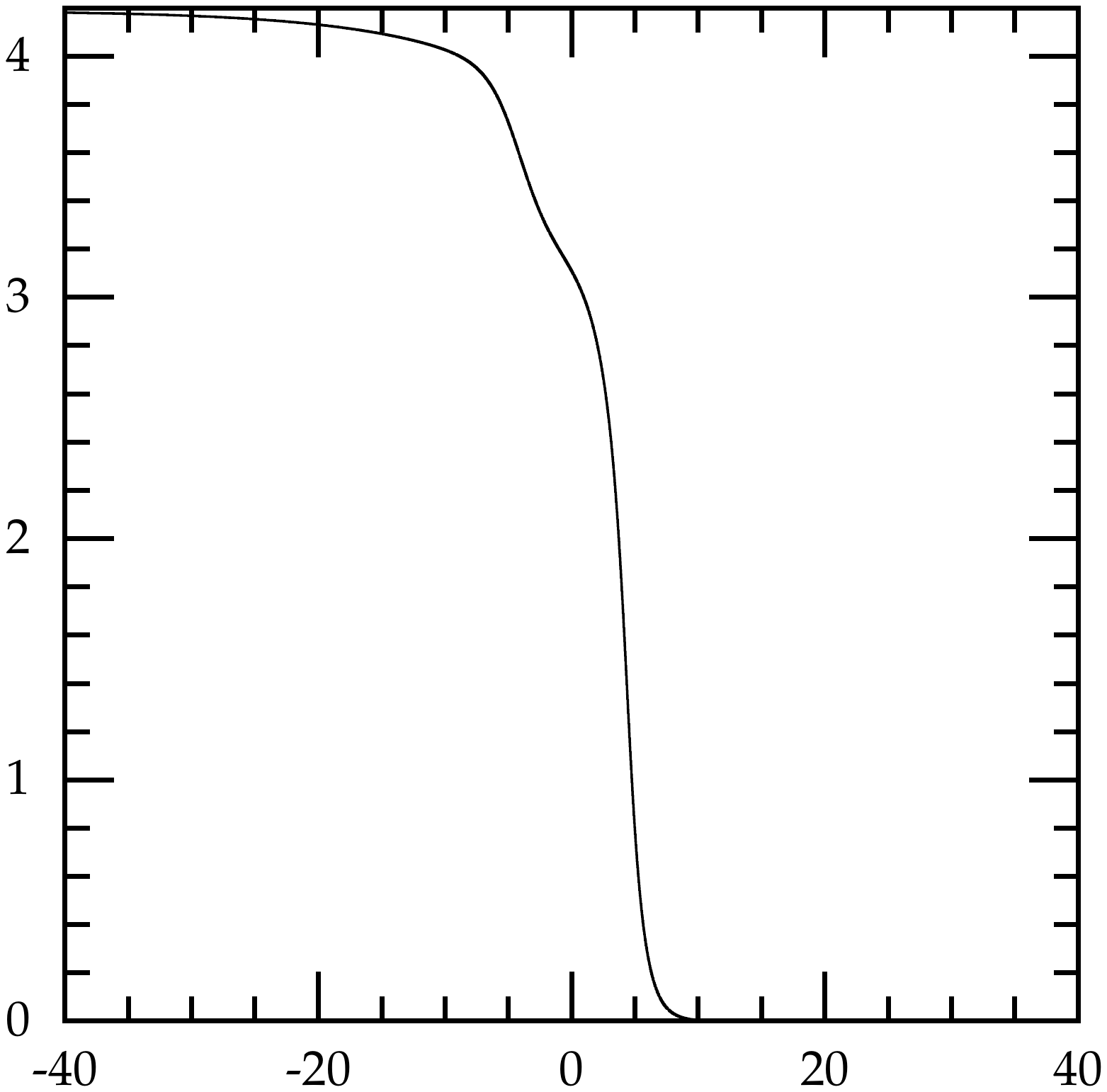} }
\subfigure[]{\includegraphics[width=0.45\textwidth, ,height=0.4\textwidth,angle =0]{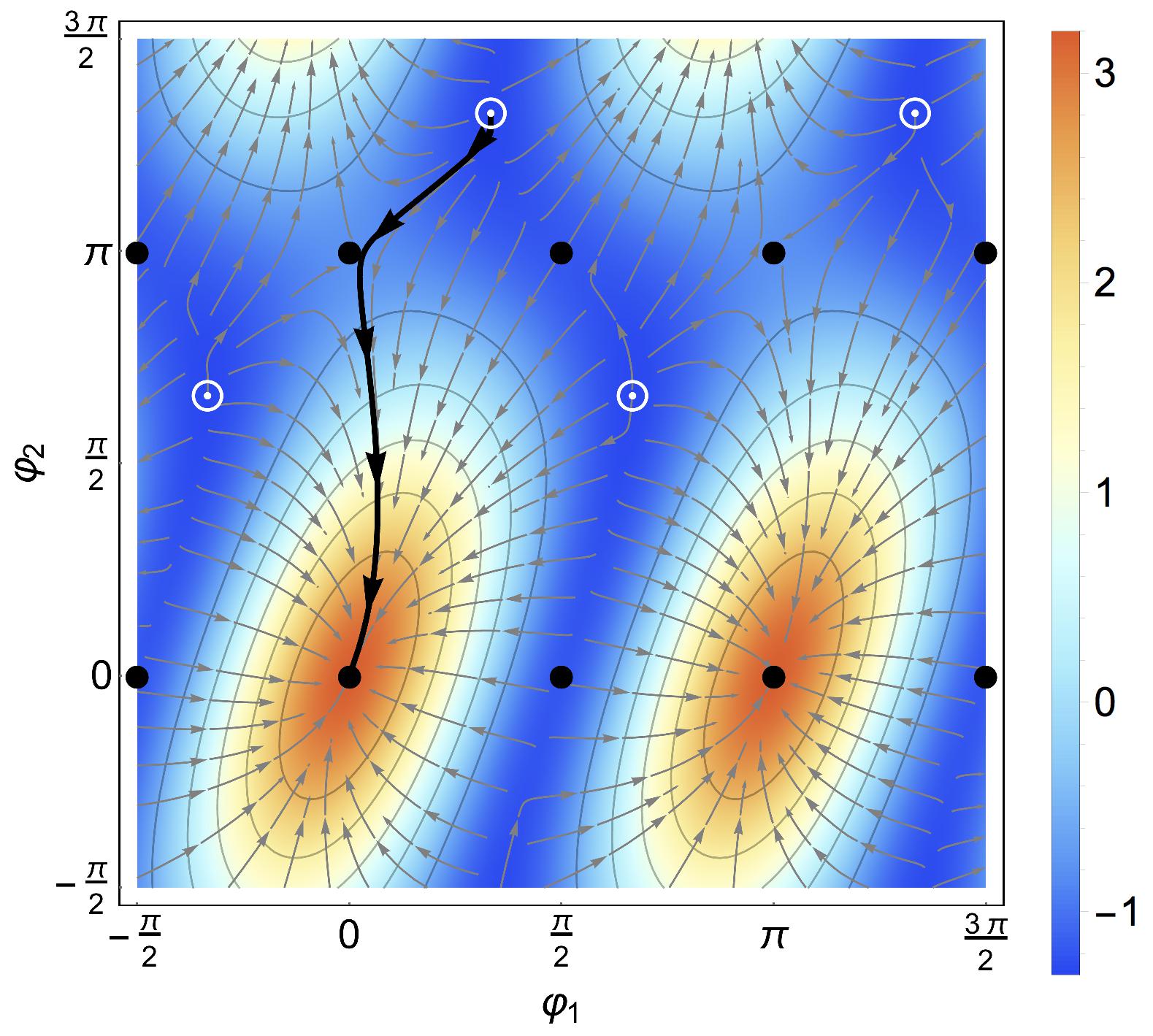} }
\caption{$SO(5)$ fields - case II; (a) $\vp_1$, (b) $\vp_2$  and (c) $\vec\nabla_{\eta} U-$flow. Black dots denote the maxima and saddle points of the pre-potential that correspond to the vacua \eqref{maxmin1} whereas minima of the pre-potential are denoted by $\odot$. They are localized at \eqref{saddle1}.}\label{fig:so5case2}
\end{figure}

The case II shares the vacua \eqref{maxmin1} and \eqref{saddle1} with the case I; however, in this case the extrema of the pre-potential have a different nature.  Expressions \eqref{maxmin1} give maxima $U_{\rm max}=\frac{16}{5}$ for $n_1$ and $n_2$ being simultaneously even, saddle points $U_{\rm s1}=-\frac{6}{5}$ for $n_1$ odd and $n_2$ arbitrary, and different saddle points  $U_{\rm s2}=-\frac{4}{5}$ for $n_1$ even and $n_2$ odd.  Minima of the pre-potential $U_{\rm min}=-\frac{13}{10}$ are given by \eqref{saddle1}. This case looks superficially similar but this time the kinks of $\vp_1$ and $\vp_2$ connect the 
vacua at ($\pi/3$, $4\pi/3$)  to (0,0) (with the first numbers referring to the value of $\vp_1(-\infty)$).  Again, one can see from Fig.\ref{fig:so5case2} that the presence of the saddle point $(0,\pi)$ is tightly related to the ``bump'' in $\vp_1$.

%%%%%%%%%%%%%%%%%%%%%%%%%%%%%%%%%%%%%%%%%%%%%
%							FIGURE 14							               %
%%%%%%%%%%%%%%%%%%%%%%%%%%%%%%%%%%%%%%%%%%%%%

The case III is quite different from the other cases discussed in this paper. In the case III the potential $V(\vp_1,\vp_2)$ has minima $V_{\rm min}=0$ at 
\[
(\vp^{({\rm vac})}_1,\vp^{({\rm vac})}_2)=\left(\frac{2n_1+1}{2}\pi,2n_2\pi\right)
\]
 that correspond to the local minima of the pre-potential, $U_{\rm min}=-4$, and $V$ has local minima at 
 \[
 (\vp^{({\rm vac})}_1,\vp^{({\rm vac})}_2)=(n_1\pi,(2n_2+1)\pi)
 \]
  that correspond to the local maxima, $U_{\rm max}=4$, of the pre-potential. What is different in this case is the presence of the vacua along the straight lines 
 \[
  2\vp^{({\rm vac})}_1-\vp^{({\rm vac})}_2=2n\pi.
  \]
  %%%%%%%%%%%%%%%%%%%%%%%%%%%%%%%%%%%%%%%%%%%%%
%							FIGURE 15							       %
%%%%%%%%%%%%%%%%%%%%%%%%%%%%%%%%%%%%%%%%%%%%%
\begin{figure}[h!]
\centering
\subfigure[]{\includegraphics[width=0.32\textwidth,height=0.33\textwidth, angle =0]{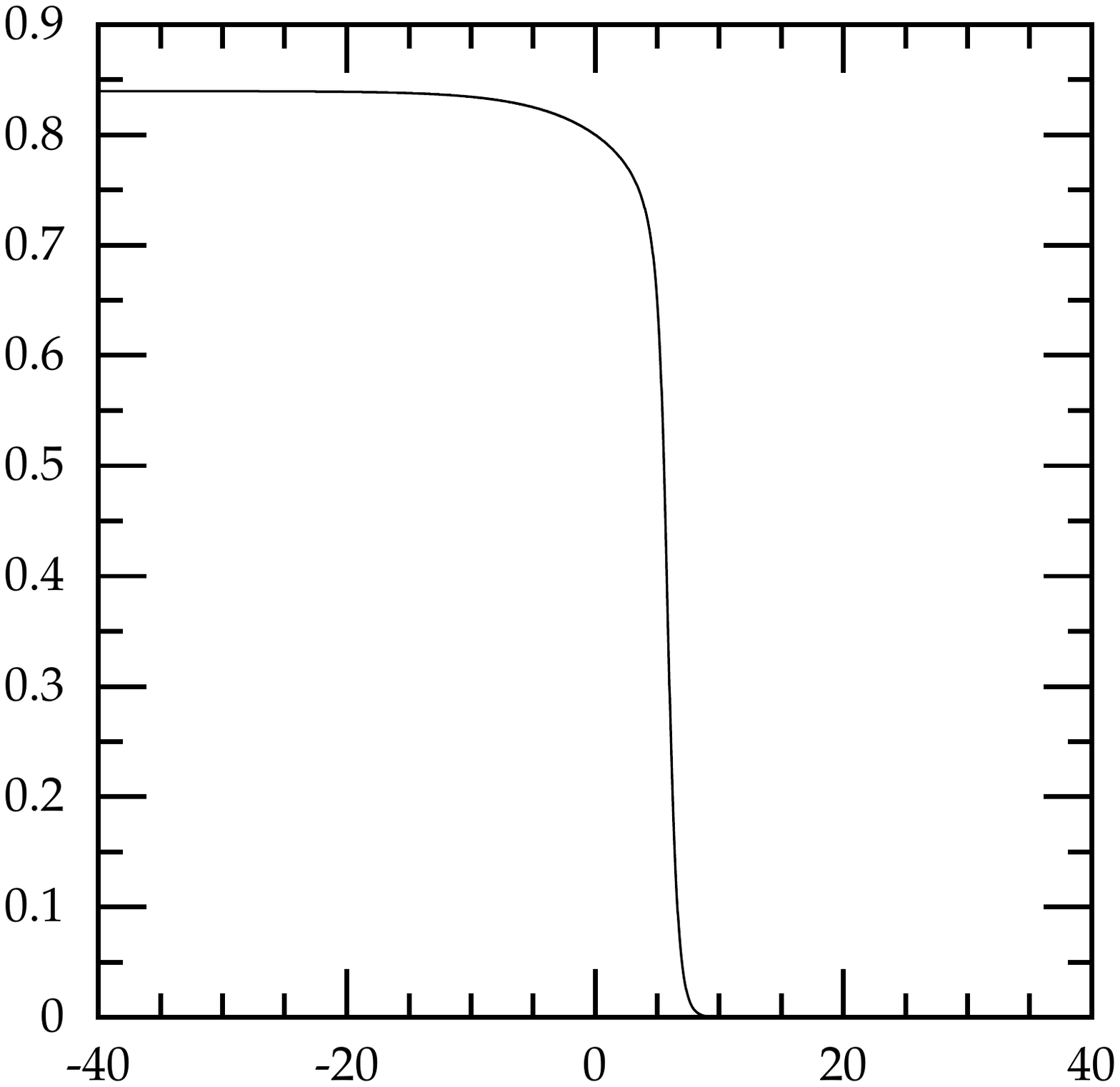}}  
\subfigure[]{\includegraphics[width=0.32\textwidth,height=0.33\textwidth, angle =0]{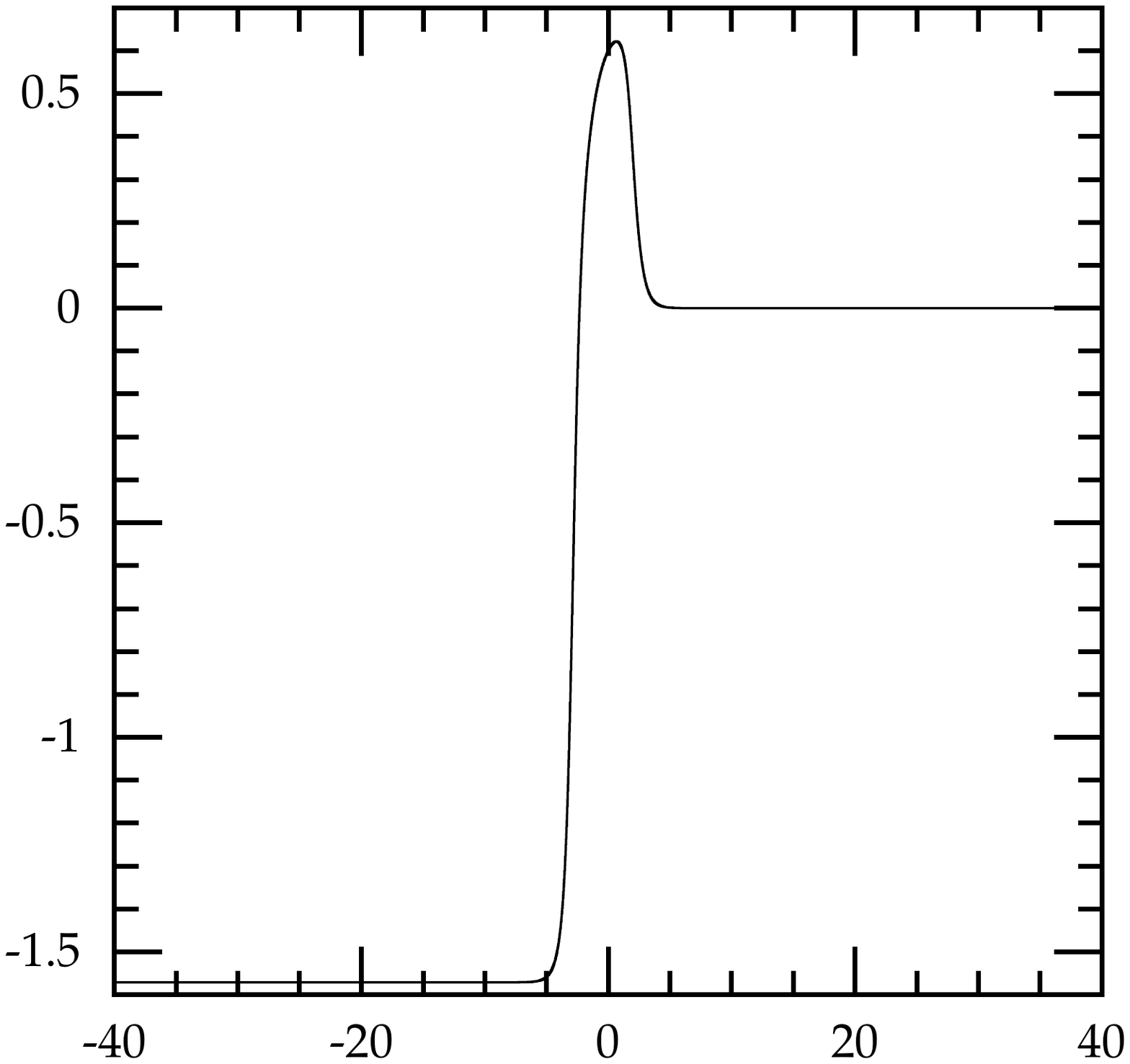}}
\subfigure[]{\includegraphics[width=0.32\textwidth,height=0.33\textwidth, angle =0]{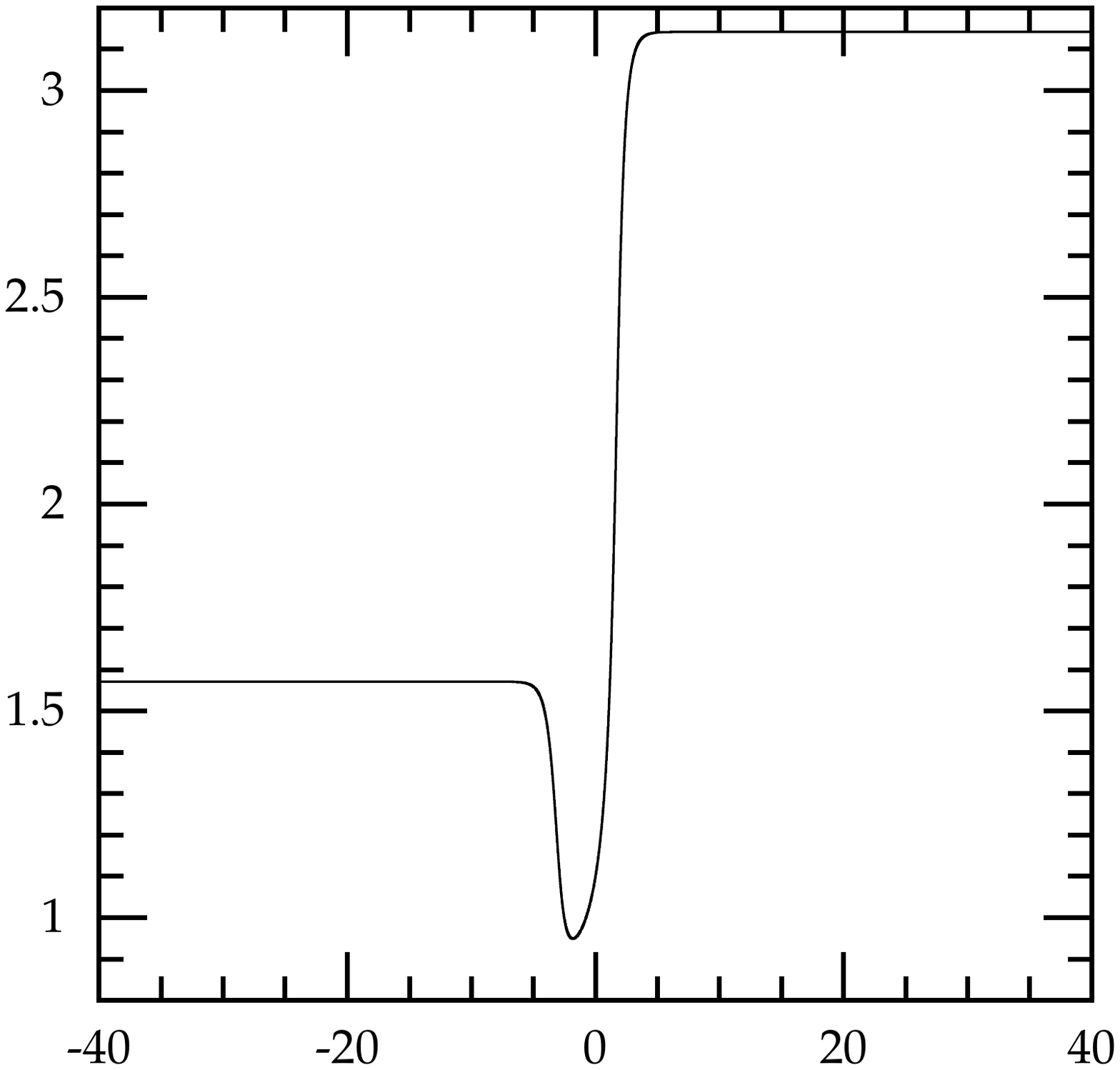}}
\subfigure[]{\includegraphics[width=0.32\textwidth,height=0.33\textwidth, angle =0]{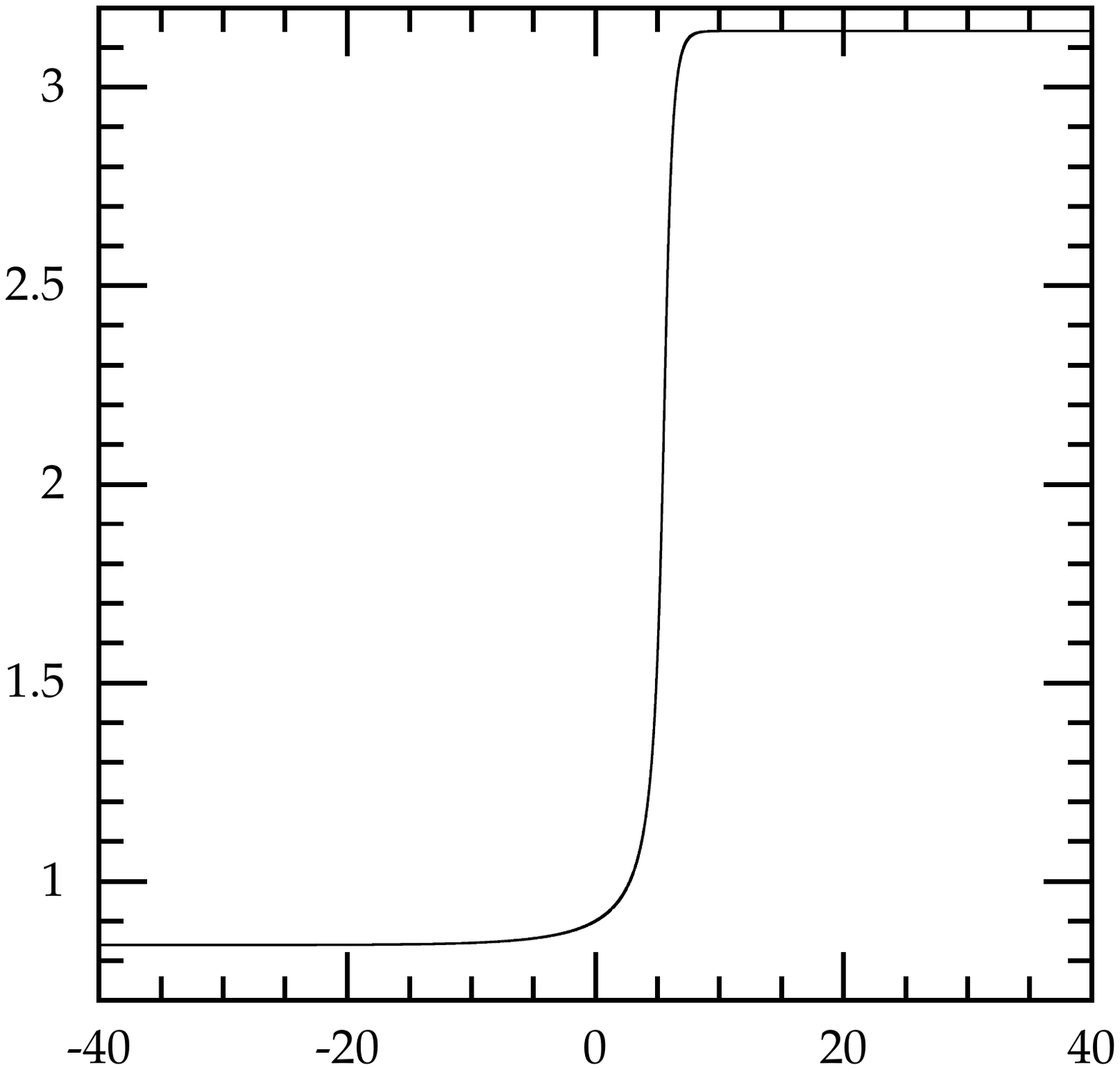}}
\subfigure[]{\includegraphics[width=0.32\textwidth,height=0.33\textwidth, angle =0]{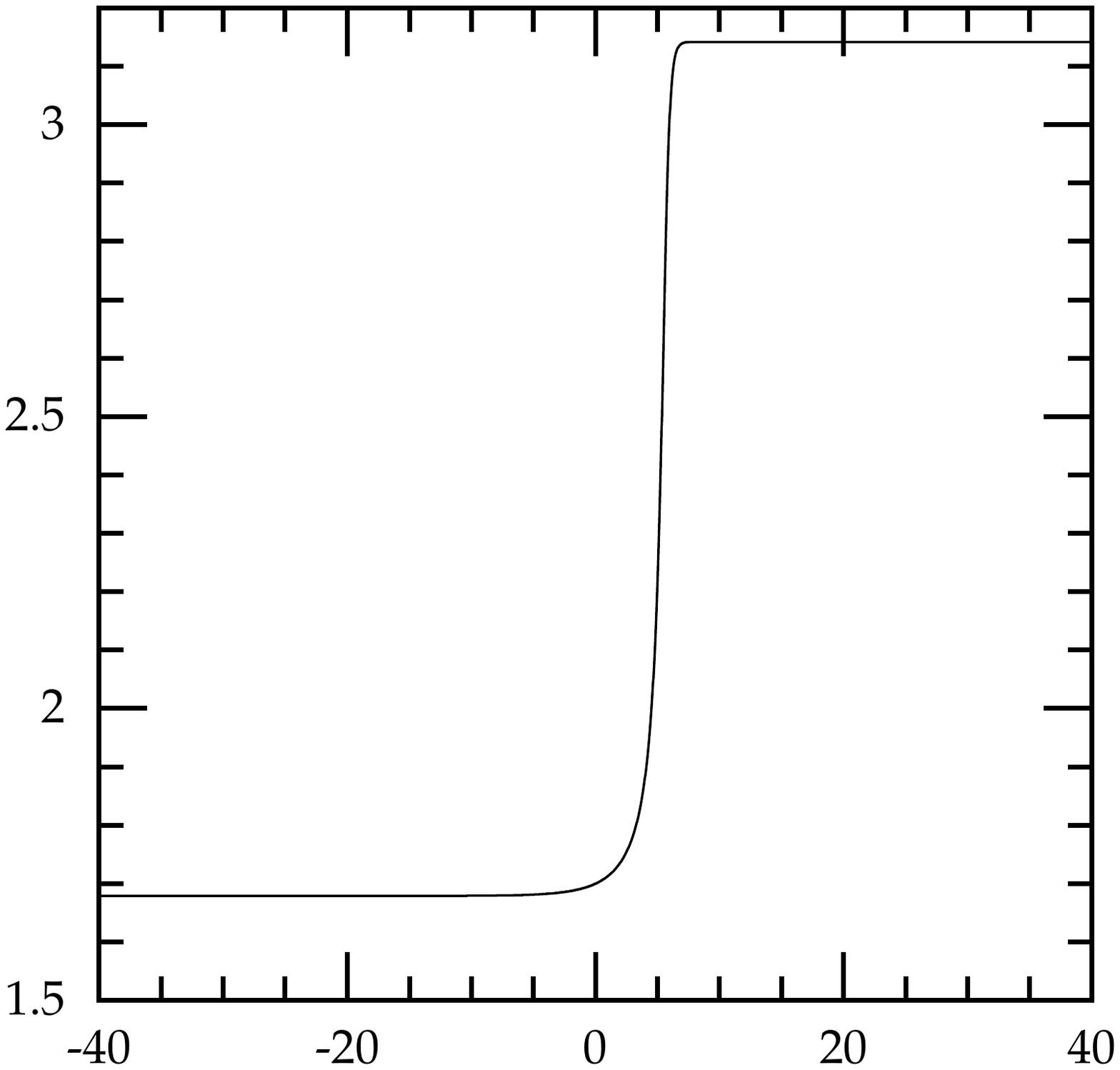} }
\subfigure[]{\includegraphics[width=0.32\textwidth,height=0.33\textwidth, angle =0]{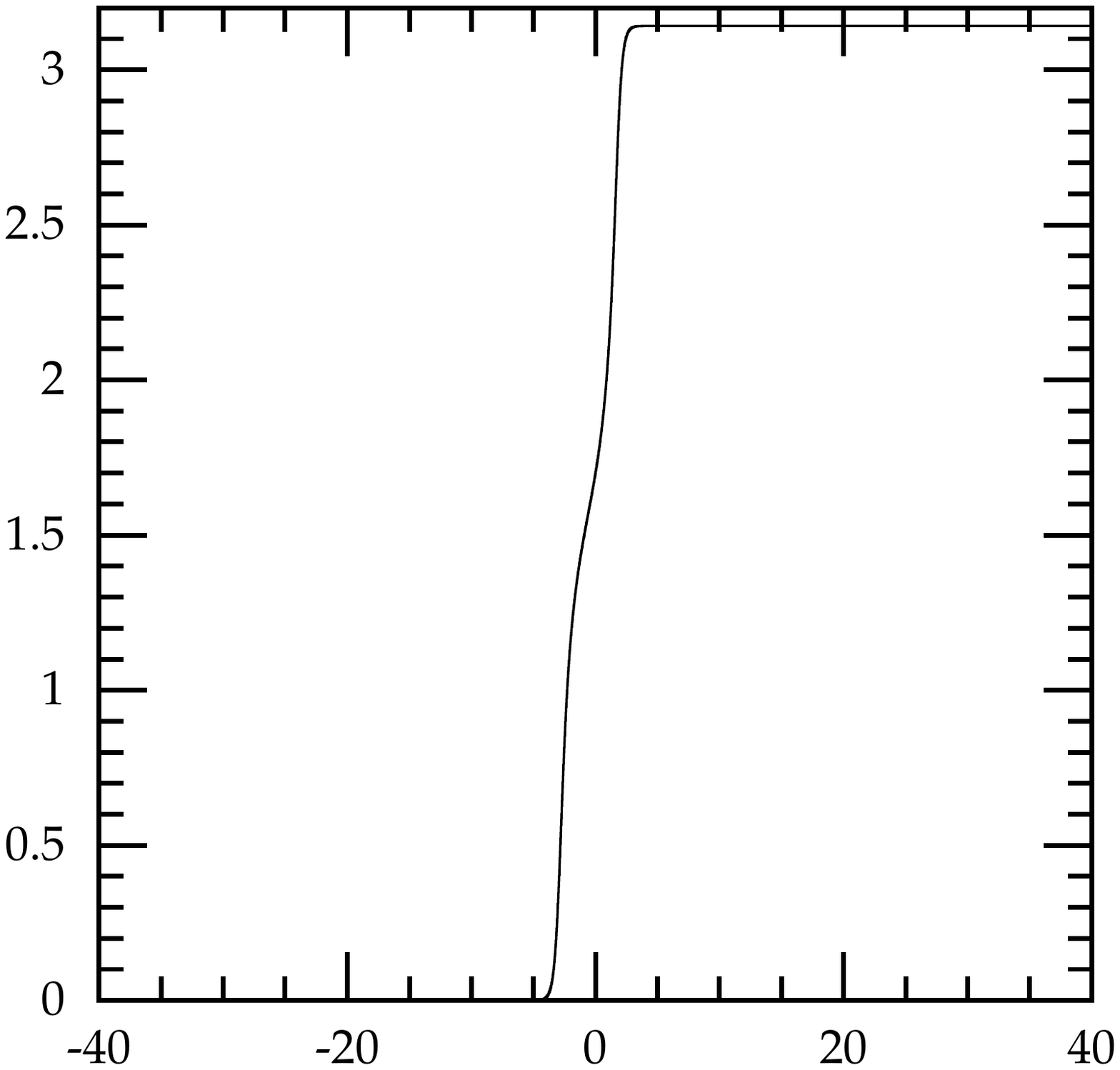} }
\subfigure[]{\includegraphics[width=0.32\textwidth,height=0.33\textwidth, angle =0]{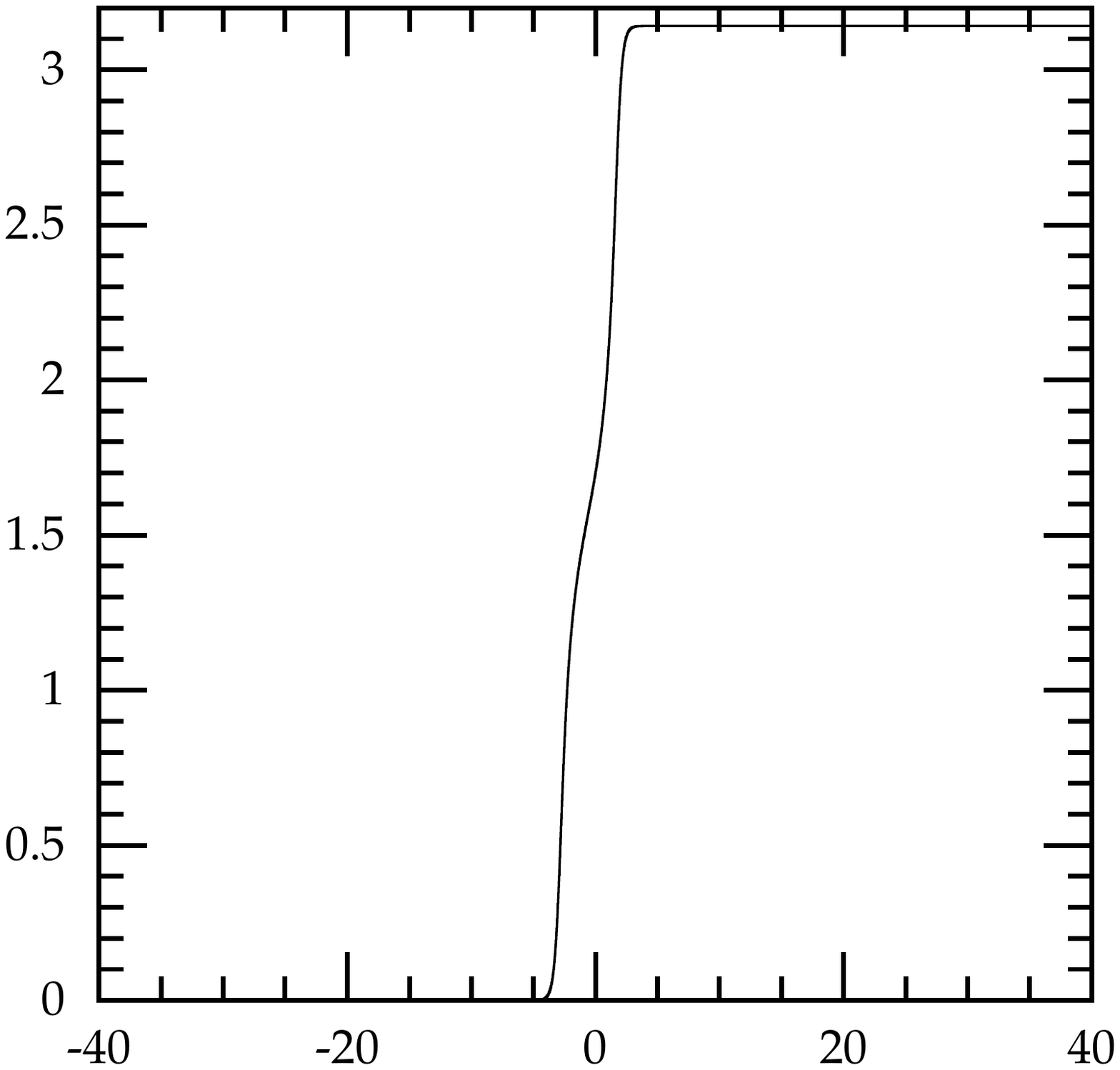} }
\subfigure[]{\includegraphics[width=0.32\textwidth,height=0.33\textwidth, angle =0]{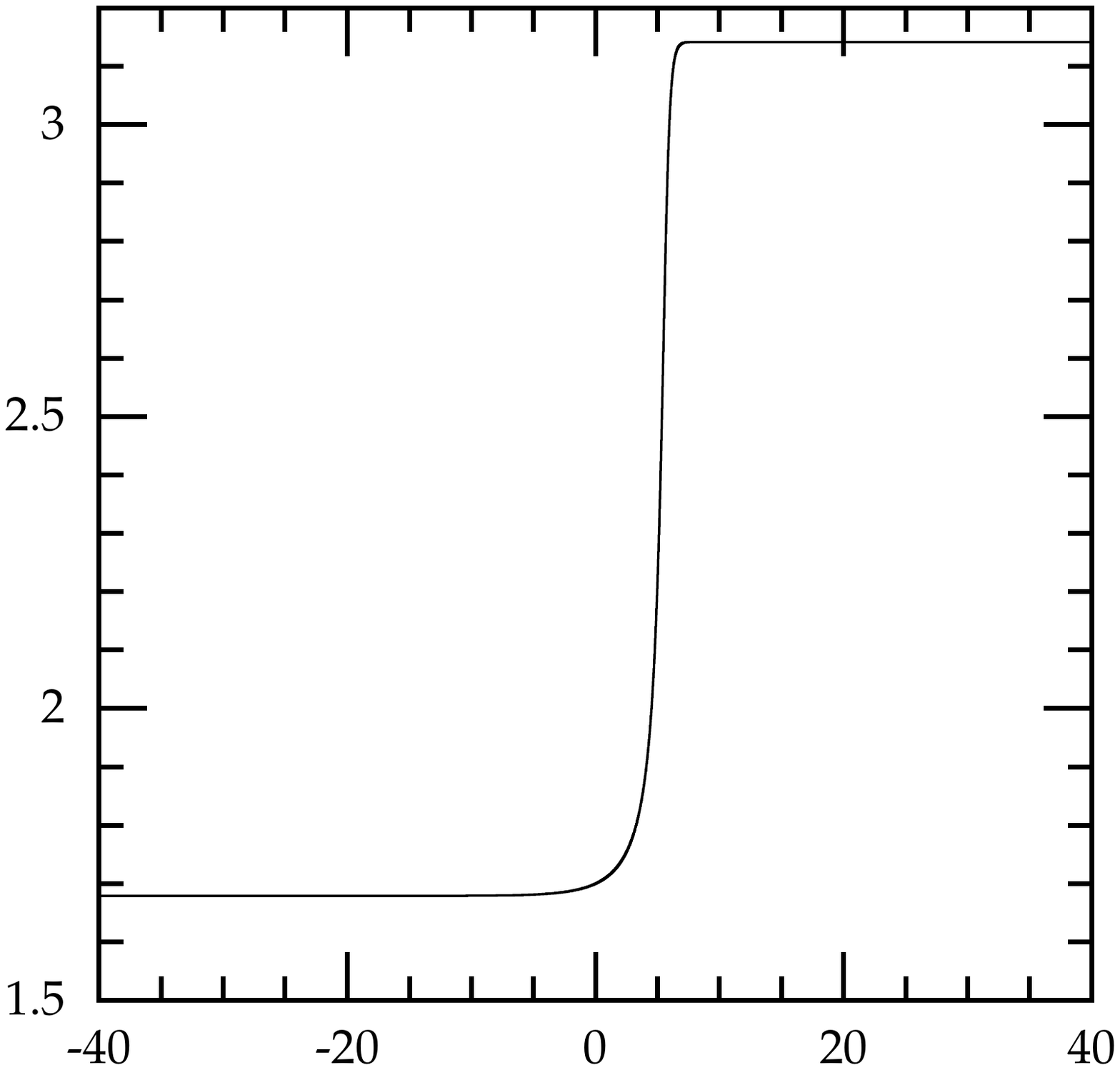} }
\caption{$SO(5)$ fields - case III; (a), (b), (c), (d) field $\vp_1$ and (e), (f), (g), (h) field $\vp_2$. Solutions correspond to simulations started with the initial data $\vp_2(0)=1.7$ and differ by the choice of $\vp_1(0)$. The cases with $\vp_1(0)=0.8$ are shown in plots (a), (e), those with $\vp_1(0)=0.6$ in  (b), (f), $\vp_1(0)=1.1$ in (c), (g) and those with $\vp_1(0)=0.9$ in plots (d) and (h).}\label{fig:so5case4}
\end{figure}

 The pre-potential $U$ and its partial derivatives $\frac{\delta U}{\delta \vp_1}=-4\cos(\vp_2)\sin(2\vp_1-\vp_2)$ and $\frac{\delta U}{\delta \vp_2}=2[\sin(2(\vp_1-\vp_2))+\sin(\vp_2)]$ vanish at these lines.
 %%%%%%%%%%%%%%%%%%%%%%%%%%%%%%%%%%%%%%%%%%%%%
%							FIGURE 16							       %
%%%%%%%%%%%%%%%%%%%%%%%%%%%%%%%%%%%%%%%%%%%%%
\begin{figure}[h!]
\centering
\subfigure[]{\includegraphics[width=0.45\textwidth,height=0.37\textwidth, angle =0]{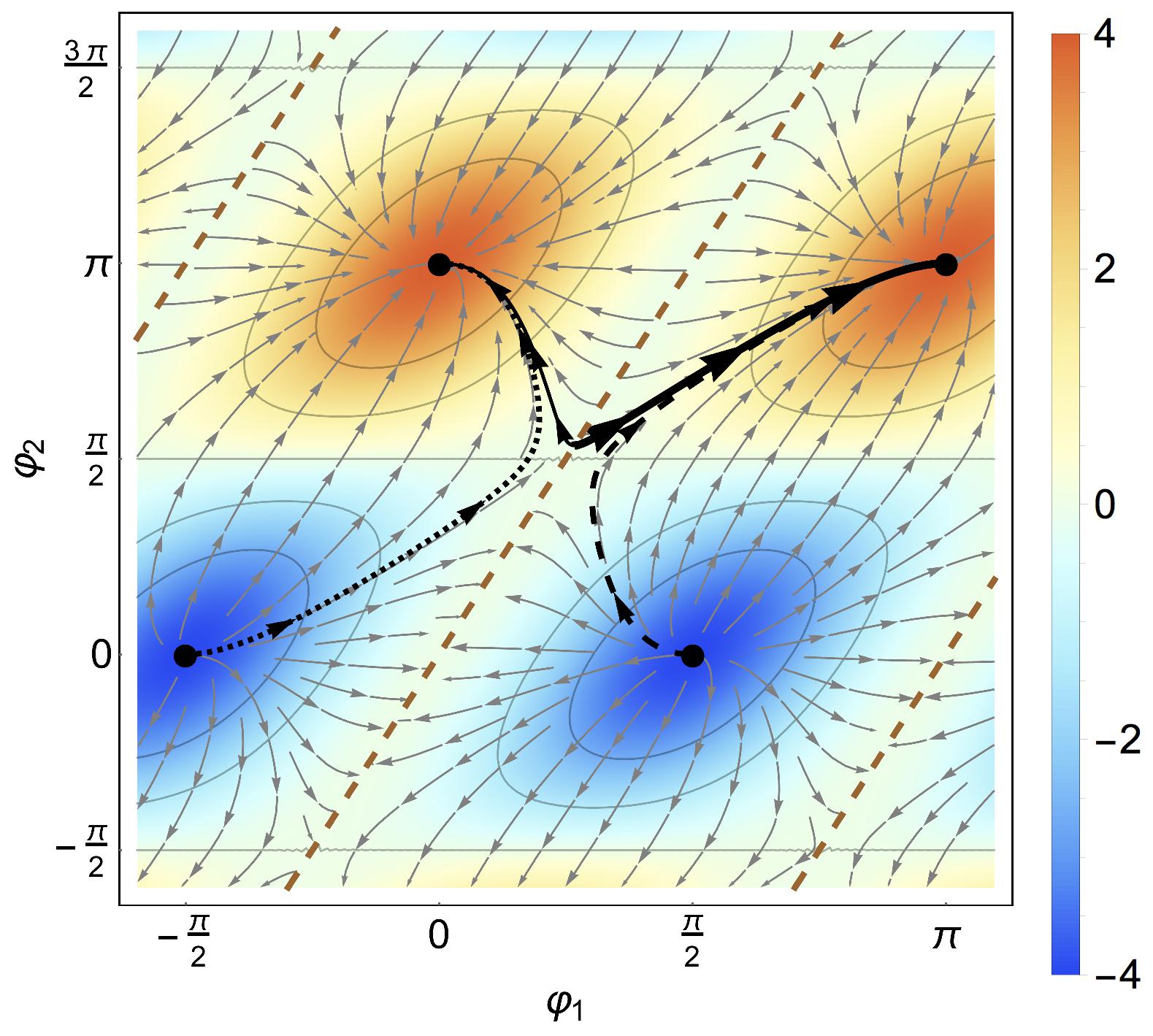}}
\subfigure[]{\includegraphics[width=0.45\textwidth,height=0.37\textwidth, angle =0]{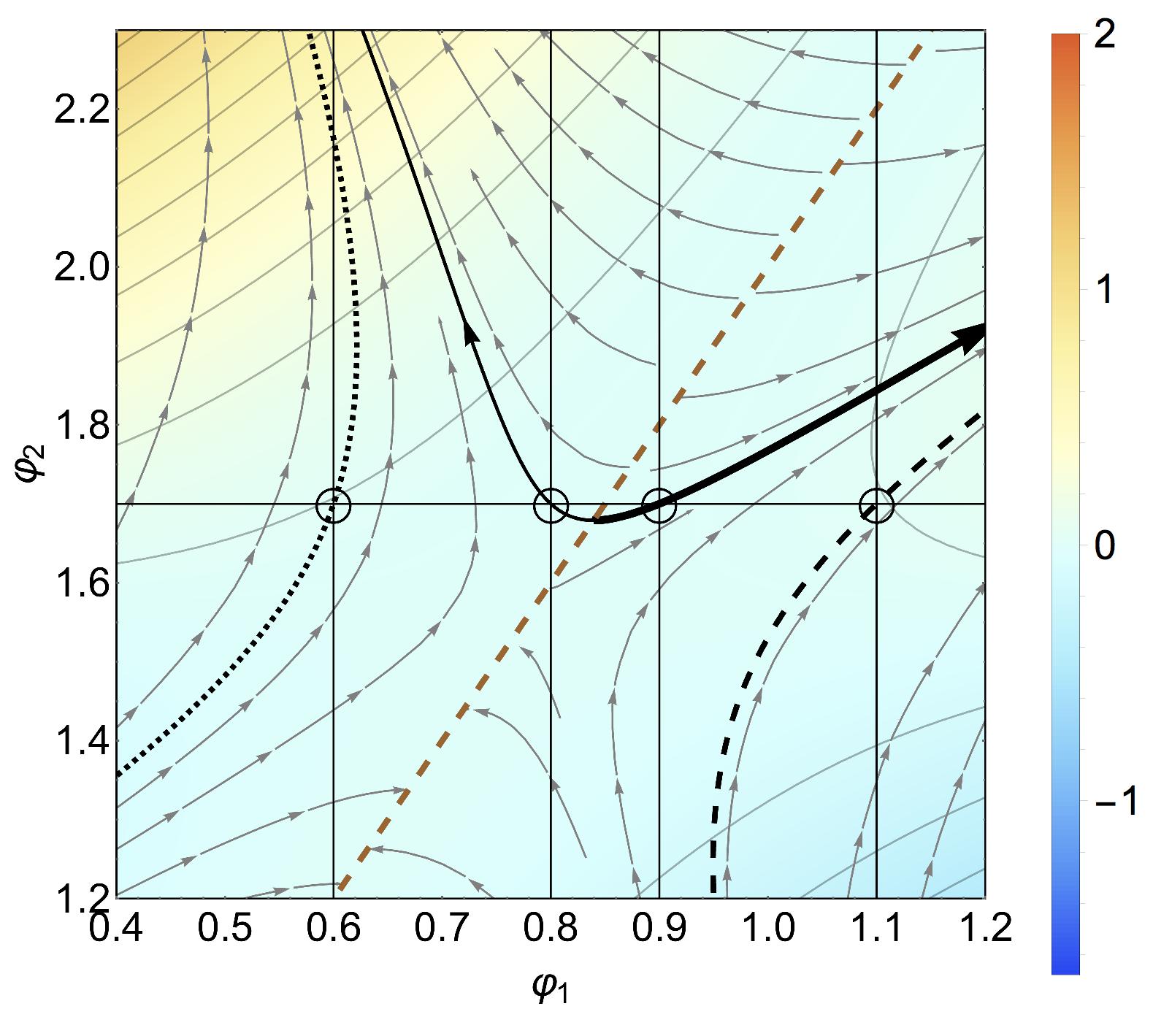}}
\subfigure[]{\includegraphics[width=0.45\textwidth,height=0.37\textwidth, angle =0]{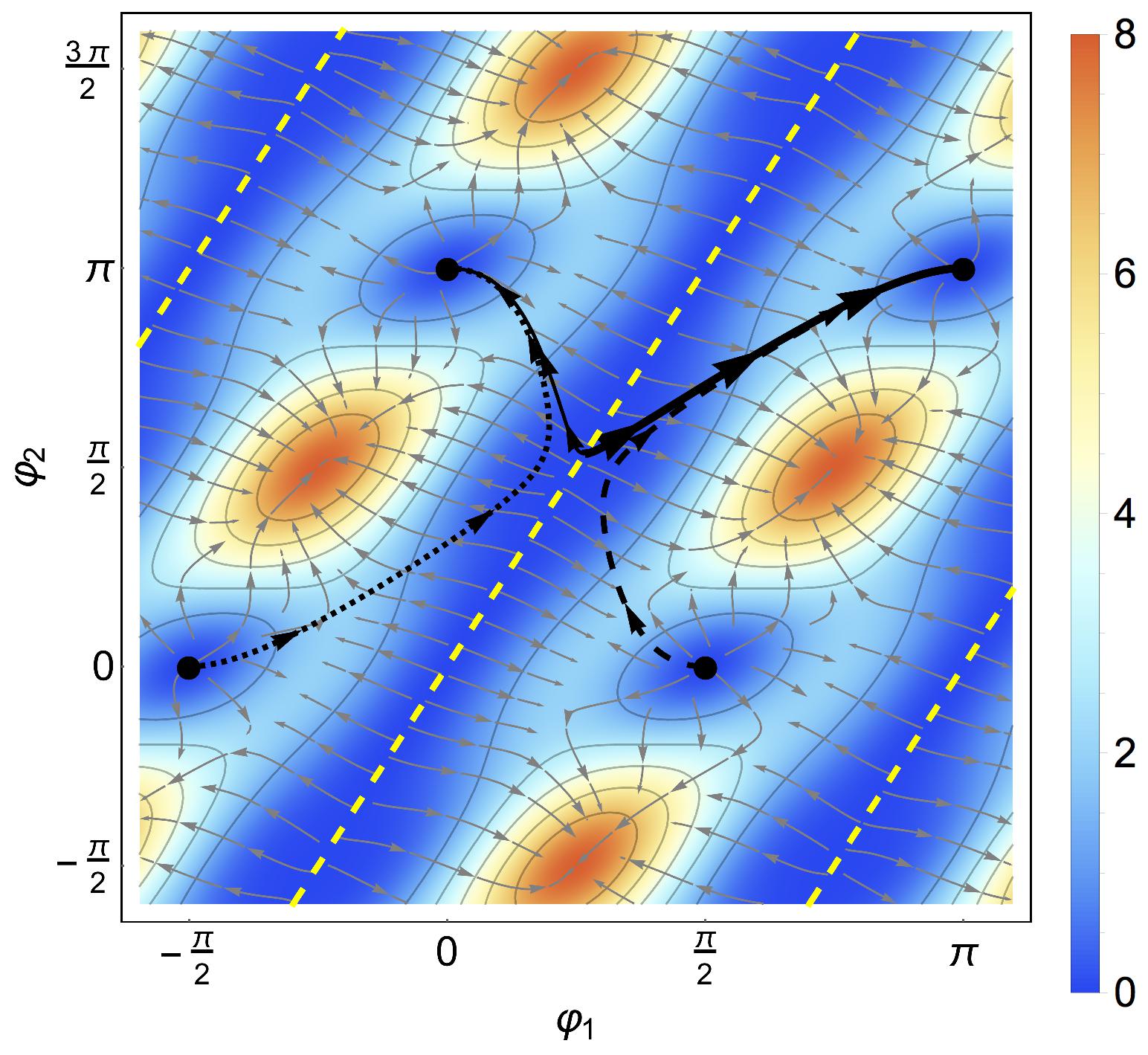}}
\caption{$SO(5)$ fields - case III; (a), (b) pre-potential and $\vec\nabla_{\eta}U-$flow, (c) potential and its gradient flow. The curves correspond to $\vp_2(0)=1.7$ and $\vp_1(0)=0.8$ -solid thin curve, $\vp_1(0)=0.6$ - dotted curve, $\vp_1(0)=1.1$ - dashed curve, $\vp_1(0)=0.9$ - solid thick curve. The initial points for the numerical simulations are shown at figure (b) which is a blow-up of the central region of figure (a).}\label{fig:so5case4UV}
\end{figure}
%%%%%%%%%%%%%%%%%%%%%%%%%%%%%%%%%%%%%%%%%%%%%%%
These vacua can be understood as being the limiting cases of the saddle points. These minima of the potential were denoted by dashed straight lines in  Figure \ref{fig:so5case4UV}. The BPS solutions can interpolate between isolates minima of $V$ (dashed and dotted curves connecting dots at Figure \ref{fig:so5case4UV} (c)) as well as between isolated minima and the valley-shape minima (solid thick and thin curves connecting dashed lines and dots). In  Figure \ref{fig:so5case4UV} (b) we have marked the points that correspond with initial condition for numerical solution. For all four solutions $\vp_2(0)=1.7$ and they differ by the value of $\vp_1(0)$. The analysis of the curves in Fig.\ref{fig:so5case4} allows us to conclude that valley-shape vacua (similarly to the saddle points) are responsible for existence of ``bumps'' in the BPS kinks that connect vacua different from the valley-shape ones. Such ``bumps'' exist for the kink $\vp_1$, see Fig.\ref{fig:so5case4}(b) that is a part of the BPS solution that connects $(-\frac{\pi}{2},0)$ and $(0,\pi)$ (dotted curve in Fig.\ref{fig:so5case4UV}). Similarly the BPS solution that connects vacua $(\frac{\pi}{2},0)$ and $(\pi,\pi)$ (dashed curve in Fig.\ref{fig:so5case4UV}) has a kink in $\vp_1$ with the familiar ``bump'', see Fig.\ref{fig:so5case4}(c).

In the case IV the vacua of the potential $V(\vp_1,\vp_2)$ take the form:
\begin{align}
(\vp_{1}^{\rm (vac)},\vp_{2}^{\rm (vac)})&=\Big(\frac{\pi}{2}n_1, \pi\, n_2\Big),\label{maxmin4}\\
(\vp_{1}^{\rm (vac)},\vp_{2}^{\rm (vac)})&=\Big(\pm a_1+\pi \,n_1,\pm a_2 +\pi\, n_2\Big),\label{saddle4a}\\
(\vp_{1}^{\rm (vac)},\vp_{2}^{\rm (vac)})&=\Big(\pm  b_1+\pi \,n_1,\pm  b_2 +\pi\, n_2\Big),\label{saddle4b}
\end{align}
where $ n_1,n_2\in{\mathbb Z}$ and where $a_1$, $a_2$ are given by the expressions:
\begin{align}
a_1&=\frac{1}{2}\arctan\left[\sqrt{3}\frac{(\pi^2-1)\sqrt{3(2\pi^2-3)\delta^{(+)}}+(2\pi^2-3)\sqrt{(\pi^2-1)\delta^{(+)}}}{-6(2\pi^4-5\pi^2+3)+(8\pi^4-15\pi^2+6)\sqrt{\Delta}}\right]-\frac{\pi}{2},\nonumber\\
a_2&=\pi-\arctan\left[\frac{\sqrt{3(\pi^2-1)\delta^{(+)}}}{9(\pi^2-1)-\sqrt{\Delta}}\right].\nonumber
\end{align}
Similarly, $b_1$ and $b_2$ are given by
\begin{align}
b_1&=\frac{1}{2}\arctan\left[\sqrt{3}\frac{(\pi^2-1)\sqrt{3(2\pi^2-3)\delta^{(-)}}-(2\pi^2-3)\sqrt{(\pi^2-1)\delta^{(-)}}}{6(2\pi^4-5\pi^2+3)+(8\pi^4-15\pi^2+6)\sqrt{\Delta}}\right],\nonumber\\
b_2&=\pi-\arctan\left[\frac{\sqrt{3(\pi^2-1)\delta^{(-)}}}{9(\pi^2-1)+\sqrt{\Delta}}\right],\nonumber
\end{align}
where we have defined
\begin{align}
\Delta&:=3(\pi^2-1)(2\pi^2-3),\nonumber\\
\delta^{(\pm)}&:=16\pi^4-45\pi^2+30\pm\sqrt{\Delta}.\nonumber
\end{align}

\begin{figure}[h!]
\centering
\subfigure[]{\includegraphics[width=0.45\textwidth,height=0.5\textwidth, angle =0]{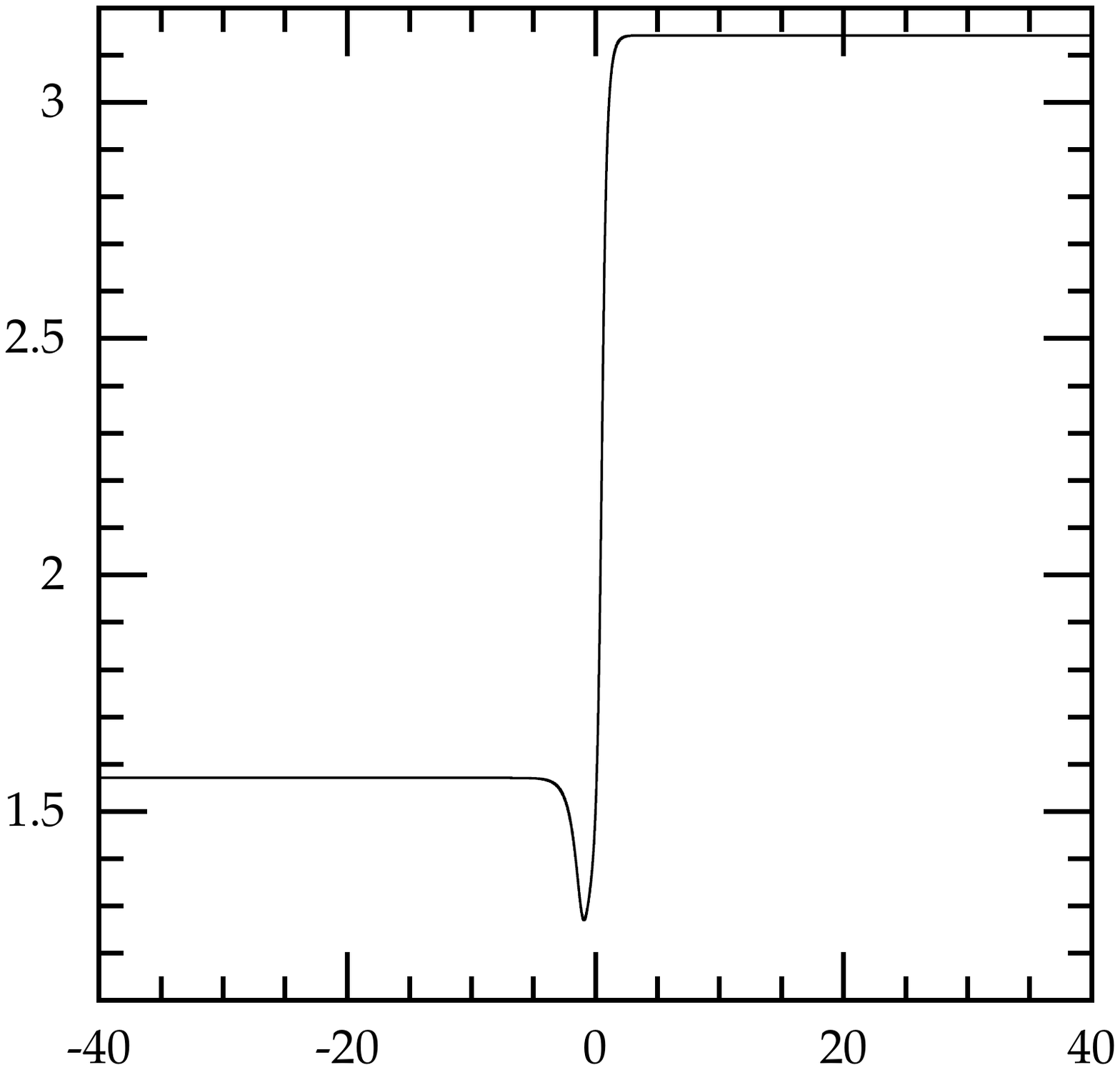}}
\subfigure[]{\includegraphics[width=0.45\textwidth,height=0.5\textwidth, angle =0]{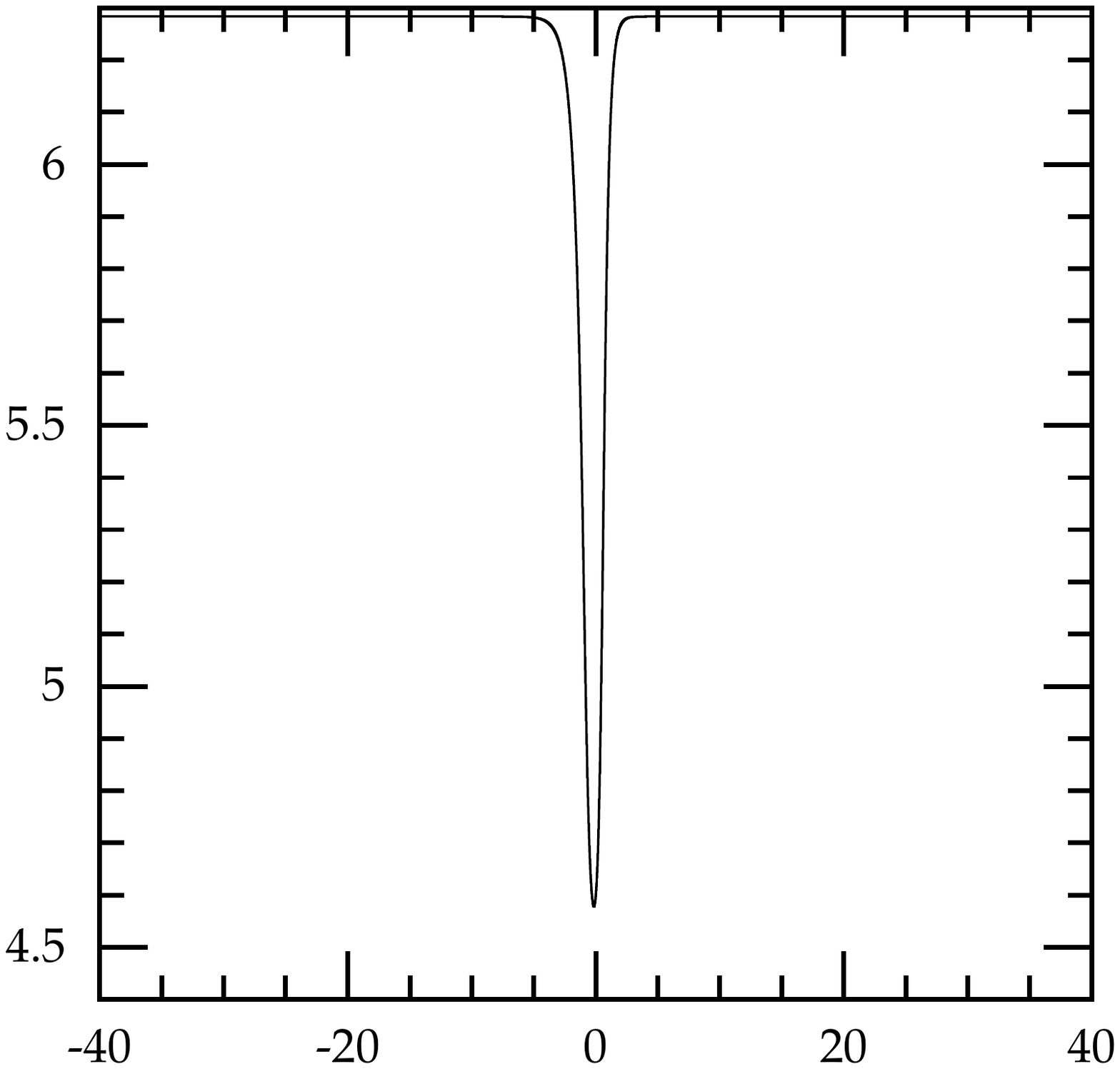}}
\subfigure[]{\includegraphics[width=0.46\textwidth,height=0.4\textwidth, angle =0]{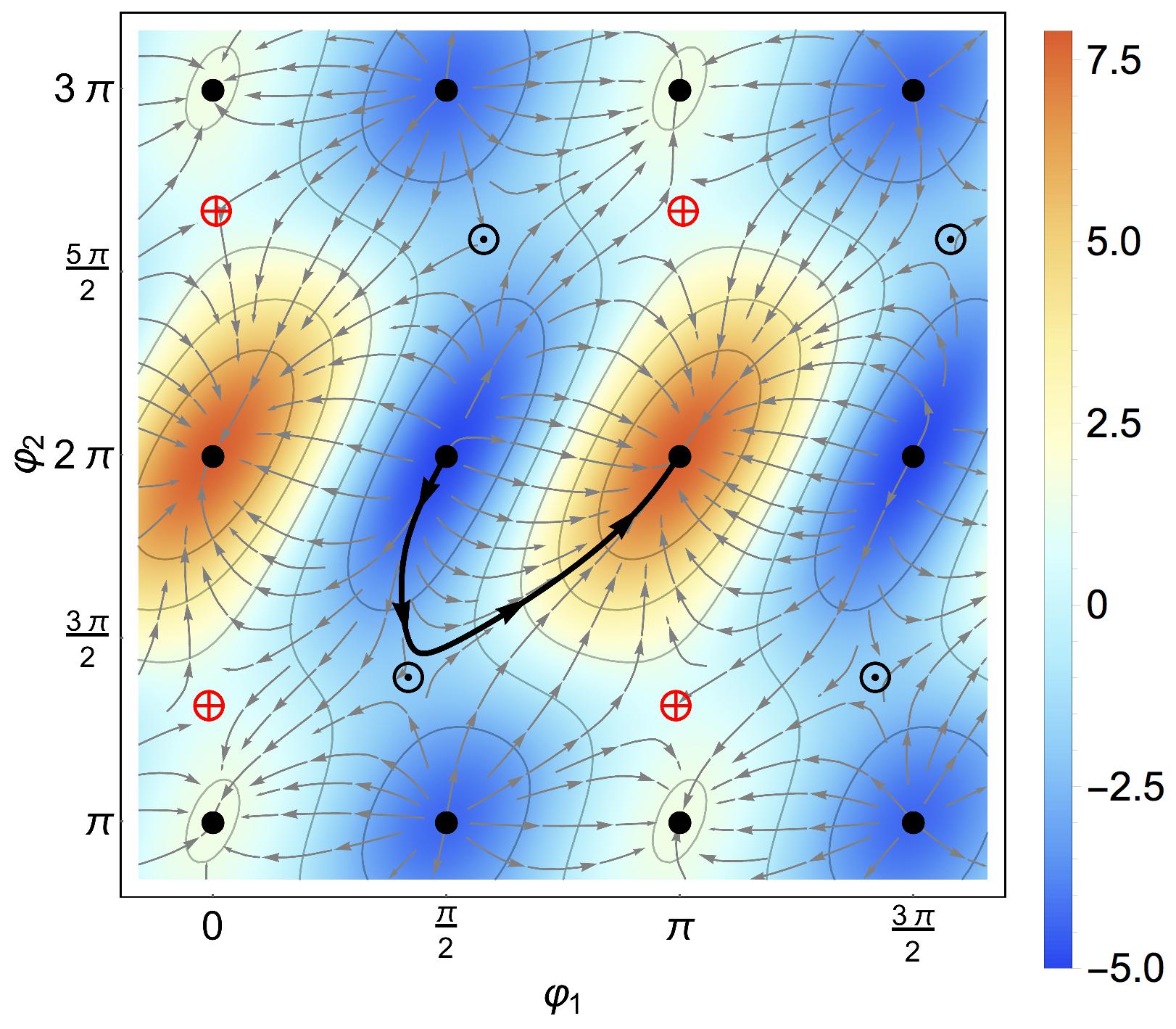}}
\caption{$SO(5)$ fields - case IV; (a) $\vp_1$ and (b) $\vp_2$, and (c) $\vec\nabla_{\eta}U-$flow. Black dots stand for maxima and minima \eqref{maxmin4} of the pre-potential, $\odot$ for its saddle points \eqref{saddle4a} and $\oplus$ for saddle points \eqref{saddle4b}.}\label{fig:so5case5}
\end{figure}

The pre-potential has extrema given by \eqref{maxmin4} which are maxima for $n_1$ even and minima for $n_1$ odd. Global maxima $U_{\rm max1}=\sqrt{2}+\sqrt{3}+\frac{3\pi}{2}$ occur for $n_2$ even and local maxima $U_{\rm max2}=-\sqrt{2}-\sqrt{3}+\frac{3\pi}{2}$ occur for $n_2$ being odd. The minima of the pre-potential become global $U_{\rm min1}=\sqrt{2}-\sqrt{3}-\frac{3\pi}{2}$ for $n_2$ even and are only local $U_{\rm min2}=-\sqrt{2}+\sqrt{3}-\frac{3\pi}{2}$ for $n_2$ being odd. The vacua \eqref{saddle4a} correspond to the saddle points $U_{\rm s1}\approx -1.95$ of the pre-potential (marked by $\odot$ in Fig.\ref{fig:so5case5}) and the vacua \eqref{saddle4b} to the saddle points (marked by $\oplus$ in Fig.\ref{fig:so5case5}) at which the pre potential takes the value $U_{\rm s2}\approx 0.78$.

The case IV is also somewhat unusual in that the fields $\vp_1$ and $\vp_2$ connect the vacua at 
($\pi/2$, 0)  to the vacuum at ($\pi$, 0). In this case the field $\vp_2$ goes from 0 to 0 but the ${\vec \nabla}_{\eta}U$-flow induces a rather complicated path in the ($\vp_1, \vp_2$) space ({\it i.e.}
both fields vary to decrease the overall flow). Superficially, we may have expected $\vp_2$  to remain
constant but the flow shows that this is not the best path.

%%%%%%%%%%%%%%%%%%%%%%%%%%%%%%%%%%%%%%%%%%%%%
%							FIGURE 17							      %
%%%%%%%%%%%%%%%%%%%%%%%%%%%%%%%%%%%%%%%%%%%%%

\section{Conclusions}
\label{sec:conclusions}
\setcounter{equation}{0}

We have presented a method of constructing real scalar field theories in $(1+1)$-dimensions with exact self-dual sectors based on the ideas of a generalized self-duality put forward in \cite{bps}. This methods involves considering a topological charge $Q$ with an integral representation in terms of a pre-potential $U$. The self-duality equations are then obtained by a procedure which involves splitting the topological charge density into a sum of products of terms and a further introduction of an arbitrary matrix $\eta$. This matrix plays the role of a target space metric in the kinetic energy of the discussed $(1+1)$-dimensional theory. The potential energy, in turn, becomes quadratic in the first functional field derivatives of the pre-potential $U$, with the inverse of the $\eta$ matrix playing again the role of the metric contracting these functional derivatives. The constructed theories possess very nice properties when the eigenvalues of the matrix $\eta$ are all positive, and the energy becomes positive definite in such cases. 

We have also given an algebraic construction of the pre-potential $U$ based on representations of Lie groups that lead in a quite natural way to an infinite number of degenerate vacua, allowing topologically non-trivial self-dual solutions to exist. Some concrete examples have been given, based on the groups $SU(2)$, $SU(3)$ and $SO(5)$, and the numerically obtained solutions of these equations have been presented. We have also studied in detail solutions of the corresponding  self-duality equations in these theories. With the exception of the relatively well known $SU(2)$ case for which analytic solutions can be easily found the solutions
of other theories are more complicated and they were obtained numerically. They possess many interesting properties: kink-like solutions with `bumps', some without them and some being even more complicated.
We have also looked at their stability and have found that they all were stable, at least with respect
of small oscillations. The detailed analysis of their properties brought out the importance of the 
pre-potential in determining their properties. The reason for this is that all such solutions follow
the ${\vec \nabla}_{\eta}U$-flow in the space of the fields of the models. This was discussed 
in section V and with many details provided in the numerical section. 

An obvious next step in our investigations of the models presented in this paper is to study time dependent solutions that can perhaps be constructed numerically by taking as the initial configuration two self-dual solutions well separated from each and then evolving them under the full equations of motion. 

This could lead to two-soliton like solutions and it would give information on how such solutions behave during the scattering process.  In addition, it would allow us to investigate whether some of the models presented here are quasi-integrable in the sense of \cite{quasi1,quasi2}.

In our construction we have chosen an approach in which the potential energy is obtained from the given pre-potential. We have not addressed the reversed problem, namely, of finding a pre-potential for a given potential. This inverse problem is certainly very important to study since many well known scalar fields in $(1+1)$ dimensions are known, but it is not clear if they possess self-dual sectors. One example is given by the infinite class of models known by Affine Toda field theories. Except for the simplest example from that class, {\it i.e.} the sine-Gordon model, it is not known if any exact static one or multi-soliton solutions of such exactly integrable  theories are solutions of a self-dual equation or not. Such an investigation involves  solving the equation \rf{potdef} for the pre-potential $U$ for a given explicit potential $V$. This equation is highly non-linear in the field space. We have not managed to  solve it and may even not have solutions for some potentials $V$.

\vspace{.5cm}

{\bf Acknowledgements:} 
The work described here was supported by a Royal Society grant and a Durham/FAPESP SPRINT grant
and it has involved some visits of WJZ to S\~ao Carlos and of LAF to Durham.
Both authors would like to thank the corresponding Institutions for their hospitality.
We would like to thank the organizers of the {\em Workshop on Solitons: Integrability, Duality and Applications}, held in April/2017 at ICTP/SAIFR in S\~ao Paulo, where the authors had the opportunity to develop part of this work. 
Some preliminary aspects of this work were mentioned by WJZ at a meeting in Leeds in July 2018 and
WJZ would like to thank Derek Harland for the invitations of LAF and WJZ to that meeting and the support.
We also would like to thank Nick Manton for his interest and helpful suggestions. LAF is partially supported by CNPq-Brazil.

\end{document}